\documentclass[prb,aps,amssymb,twocolumn,superscriptaddress,notitlepage]{revtex4-1}
\usepackage{amsmath}
\usepackage{amssymb}
\usepackage{amsthm}
\usepackage{amsfonts}
\usepackage{listings}
\usepackage{physics}
\usepackage{enumerate}
\usepackage{latexsym}
\usepackage{psfrag}
\usepackage{bm}
\usepackage{graphicx}
\usepackage[caption=false]{subfig}
\usepackage{blkarray}
\usepackage{array}
\usepackage{color}
\usepackage[normalem]{ulem}
\usepackage{hyperref}
\hypersetup{
     colorlinks = true,
     linkcolor = magenta,
     citecolor = magenta
}

\def\beq{\begin{equation}}
\def\eeq{\end{equation}}
\def\bal{\begin{aligned}}
\def\eal{\end{aligned}}

\begin{document}
\title{Simulating Higher-Order Topological Insulators in Density Wave Insulators}
\author{Kuan-Sen Lin}
\author{Barry Bradlyn}
\affiliation{Department of Physics and Institute for Condensed Matter Theory, University of Illinois at Urbana-Champaign, Urbana, IL, 61801-3080, USA}

\date{\today}
\begin{abstract}
    Since the discovery of the Harper-Hofstadter model, it has been known that condensed matter systems with periodic modulations can be promoted to non-trivial topological states with emergent gauge fields in higher dimensions. 
    In this work, we develop a general procedure to compute the gauge fields in higher dimensions associated to low-dimensional systems with periodic (charge- and spin-) density wave modulations. 
    We construct two-dimensional (2D) models with modulations that can be promoted to higher-order topological phases with $U(1)$ and $SU(2)$ gauge fields in 3D. 
    Corner modes in our 2D models can be pumped by adiabatic sliding of the phase of the modulation, yielding hinge modes in the promoted models. 
    We also examine a 3D Weyl semimetal (WSM) gapped by charge-density wave (CDW) order, possessing quantum anomalous Hall (QAH) surface states. 
    We show that this 3D system is equivalent to a 4D nodal line system gapped by a $U(1)$ gauge field with a nonzero second Chern number. 
    We explain the recently identified interpolation between inversion-symmetry protected phases of the 3D WSM gapped by CDWs using the corresponding 4D theory.
    Our results can extend the search for (higher-order) topological states in higher dimensions to density wave systems.
\end{abstract}

\maketitle

\section{Introduction}
Topological crystalline phases in non-interacting, clean structures have attracted a great deal of recent theoretical and experimental attention\cite{Fu2011,Hsieh2012,ando2015topological,Hughes11,Turner2010,Turner2012}. 
From the discovery of helical edge states in $\mathbb{Z}_2$ topological insulators\cite{Kane04,bernevig2006quantum,konig2007quantum}, to surface Dirac cones protected by time-reversal or crystal symmetries\cite{fukanemele,xia2009observation,Hsieh2012,Hourglass}, the experimental manifestations of band topology have come primarily through the exploration of surface states. 
The recent theoretical prediction of higher-order topological insulators\cite{hotis,Po2017,khalaf2018higher,benalcazar2017quantized} has triggered a wave of materials predictions\cite{NaturePaper,bigmaterials,bigmaterials-china,ashvin-materials,xu2020high} and experimental efforts to observe their predicted gapped surfaces but gapless corners (in 2D) or hinges (in 3D). 
At a theoretical level, topological band insulators can be classified by exploiting the constraints of symmetry, relating the topology of bands to the transformation properties of Bloch functions under crystal symmetries\cite{Kruthoff2016,NaturePaper,Po2017,Fu2007,po2020symmetry,cano2020band,MTQC,watanabe2018structure,Zhang_TMCI_corep_RPB,Slager_magnetic_spacegroup_rep}. 
In the simplest cases, the symmetry eigenvalues of occupied electronic wavefunctions at different crystal momenta in the Brillouin zone can be used to deduce the absence of an exponentially localized, position space description of the occupied states, and hence the presence of non-trivial topology. 
Progress along these lines has led to a full, predictive classification of topological band structures both with and without time-reversal symmetry. 
Essential to these efforts is the presence of discrete translation symmetry, which ensures that localized electronic functions are identical in each unit cell, and hence allows the symmetry properties of the system to be described as a function of momentum. 

At the same time, the interplay between topological bands and symmetry-breaking order has started to attract a great deal of attention. 
It has been argued theoretically that in topological systems with charge-density wave (CDW) order, the collective phason mode of the CDW may inherit topological properties from the Fermi sea, such as an induced axion coupling to electromagetic fields\cite{wang2013chiral,you2016response,BurkovCDW,zyuzin2012weyl,maciejko2014weyl}. 
Signatures of this axion coupling have been recently experimentally detected in (TaSe$_4$)$_2$I\cite{gooth2019evidence,shi2019charge}. 
Additionally, the quantum anomalous Hall phase in the Dirac semimetal ZrTe$_5$ can be understood as originating from a magnetic-field induced CDW transition\cite{tang2019three,qin2020theory,song2017instability,zhang2017transport}. 
Because CDW order is in general incommensurate with the underlying lattice, a full understanding of the interplay between mean-field CDW order and band topology requires us to examine topology of incommensurately modulated electronic systems. 
Such a study would also yield insights into topology in artificially modulated photonic\cite{ozawa2018topological,ozawa2016synthetic}, metamaterial\cite{grinberg2020robust}, and cold-atomic lattice systems\cite{Ane1602685}, which have become a focus of recent research due to their tunability and experimental accessibility.

Naively, the breaking of translational and point group symmetries implied by incommensurate modulation would seem to prohibit the application of symmetry-based tools which have been so successful in identifying and classifying topological crystalline systems. 
However, it is often possible to view the single-particle dynamics in an incommensurately modulated system as describing the behavior of a particle in a larger number of dimensions, the phase offsets of the incommensurate modulations playing the role of momenta in the extra ``synthetic'' dimensions. 
The canonical example of this mapping is the 1D Harper (Aubry-Andr\'{e}) model with incommensurate on-site potential. 
As was shown some time ago\cite{hofstadter-original}, the Hamiltonian for the Harper model is equivalent to the Hamiltonian for a 2D square-lattice system coupled to a background magnetic field\cite{equivalence_Fibo_Harper_2012,Kraus_1D_QC_to_2D_QHE}. 
The phase of the on-site modulation plays the role of the momentum in the second, synthetic dimension, while the wavevector of the modulation plays the role of the magnetic flux per plaquette in the 2D lattice. 
Bands in the enhanced, 2D system can be characterized by a Chern number, which mandates the presence of gapless chiral modes at the edges of the system. 
Reducing back to 1D, these two-dimensional edge states manifest as boundary states of a 1D wire which appear and disappear as a function of the phase of modulation, thus realizing a Thouless pump\cite{Thouless_pump_original_paper,niu1984quantised,Topologically_quantized_current_PRR}. 
\textcolor{black}{Recent studies also show that certain generalization of the 1D Harper model allows for the investigation of higher-order topological phases~\cite{Zeng_generalized_AAH_model_HOTI_PRB}.} 

In this work, we will extend the connection beyond 1D, to show how modulated systems in 2D and 3D can be related to topological crystalline phases in higher dimensions. 
We will first review a general method for representing a modulated system as a higher dimensional system coupled to a background gauge field\cite{RiceMele,Thouless_pump_original_paper,Kraus_1D_QC_to_2D_QHE,equivalence_Fibo_Harper_2012}.
For systems with negligible spin-orbit coupling and spin-independent modulation, the gauge field will be a $U(1)$ magnetic field; for spin-dependent modulations we will show that there can also be induced $SU(2)$ gauge fields. 
We will exploit the fact that both $U(1)$ and $SU(2)$ gauge fields with constant field strength preserve inversion symmetry to show that 2D modulated systems can realize higher-order chiral ($U(1)$) and helical ($SU(2)$) topological phases in one extra synthetic dimension. 
We show how the hinge states of these synthetic higher-order topological insulators (HOTIs) manifest as corner modes in 2D, with energies that can be tuned by changing the phase of the modulation. 
Going further, we use the mapping to synthetic dimensions to bring order to the complex landscape of eigenstates of the modulated system, showing how the states can be interpreted as bulk and surface Landau level (LL) wavefunctions in synthetic dimensions. Finally, we also revisit a 3D minimal model for a Weyl semimetal (WSM) with (generally incommensurate) CDW order\cite{dynamical_axion_insulator_BB}, and show how it realizes a 4D nodal line semimetal gapped into a phase with a non-trivial second Chern number. We will verify our conclusions with a combination of exact numerical results and approximate low-energy analytic calculations.
We will also exploit the fact that the phase of a (charge- or spin-) density wave (DW) order parameter can be shifted with an applied electromagnetic field, by exciting the (nominally gapless, but sometimes pinned) sliding mode\cite{gruner1988dynamics}. 
This will allow us to make predictions about topological pumping of boundary states in modulated structures, driven by the sliding mode of the DW.
In contrast to other recent proposals for topological pumping in synthetic dimensions, the coupling of the DW sliding mode to electromagnetic fields allows for tunability of synthetic dimensions in modulated structures. 
We will comment on potential experimental realizations in condensed matter, photonic, and cold-atom systems throughout. 
This work will enable new avenues for exploring higher-order topological phenomena which, with the exception of some promising results in Bismuth\cite{hsu2019topology,nayak2019resolving,schindler2018higher}, have not been unambiguously identified in crystalline electronic systems.

The rest of the paper is organized as follows. 
In Sec.~\ref{sec_review_thouless_pump_Rice_Mele}, we review how the Thouless pump in a 1D Rice-Mele [Su-Schrieffer-Heeger (SSH)] chain is realized by the sliding of a CDW, and we review its connection to topology by promoting the model to a 2D $\pi$-flux lattice. 
In Sec.~\ref{sec_Dimension_promotion} we next develop a general method to compute the $U(1)$ gauge fields that are coupled to a higher dimensional models promoted from a low-dimensional modulated system. 
In Secs.~\ref{sec_chiral_HOTI} and~\ref{sec_helical_HOTI_sliding_modes}, we construct 2D modulated systems that can be promoted to 3D chiral and helical HOTIs coupled to $U(1)$ and $SU(2)$ gauge fields, respectively. 
We demonstrate the pumping of corner modes by the sliding of DWs in these systems via numerical calculations of the energy spectra. 
We examine the properties of wave functions in these 2D modulated systems by constructing low energy theories coupled to gauge fields in 3D. 
We show how the evolution of bulk, edge, and corner states in 2D can be understood from the perspective of the low energy theory in 3D. 
In Sec.~\ref{sec:Weyl_CDW}, we turn to a model for a 3D WSM gapped by a CDW. We show that this model can be promoted to a 4D nodal line system gapped by a $U(1)$ gauge field. 
We derive the corresponding low energy theory in 4D, and use it to explain both the existence of QAH surface states and the interpolation between topologically distinct QAH phases at the two inversion-symmetric values of the CDW phase in this 3D system.
Finally, in Sec.~\ref{sec:outlook}, we give an outlook as to how our work may extend the search of (higher-order) topological insulators in higher dimensions and enable simulations of $SU(2)$ gauge physics in higher dimensions. 
Some details of our models, further numerical results, and detailed derivations of the low energy theories are presented in the Supplementary Material (SM)\cite{SM}. 

Throughout this paper, we use units where $\hbar = c = |e| = 1$, and where the electron has charge $-|e| = -1$. 
Furthermore, the Einstein summation convention will not be used; whenever there is a summation over an index, we will write the summation explicitly.

\section{\label{sec_review_thouless_pump_Rice_Mele}Review - Thouless Pump as Sliding Mode }

In this section, we review the CDW picture of the Rice-Mele (SSH) model\cite{RiceMele,ssh1979}, and the interpretation of the Thouless pump\cite{Thouless_pump_original_paper} as a CDW sliding mode. 
Consider the following Hamiltonian for a 1D chain
\begin{align}
    H_{\text{Rice-Mele}} = \sum_{n} &  \left( t + \delta t (-1)^{n} \cos{\phi} \right)c^{\dagger}_{n+1}c_{n} + \text{h.c.} \nonumber\\
    & + \sum_{n} (-1)^{n+1} \Delta  \sin{\phi} c^{\dagger}_{n}c_{n}, 
    \label{eq:1DRiceMele}
\end{align}
where $c^{\dagger}_{n}$ is the creation operator for an electron at site $n$. 
The nearest-neighbor hopping and on-site potential are modulated with periodicity $2$, and their relative strength is related to the phase $\phi$ of the modulation. 
We thus identify $\phi$ as the phase of this CDW modulation. In this paper, we use the terms "CDW sliding phase" and "phases of the mean-field CDW order parameter" interchangeably to refer to $\phi$. 
For suitable choices of $t$, $\delta t$ and $\Delta$, the spectrum of this Hamiltonian is gapped for all $\phi \in [0,2\pi)$. 
Focusing on the half-filled insulating ground state in this parameter regime, the occupied-band Wannier centers\cite{Kohn59,Brouder2007,Marzari2012,NaturePaper,shockley1939surface} will be pumped by a length of one unit cell (two sites) as the phase $\phi$ adiabatically slides from $0$ to $2\pi$, leading to a quantized change of bulk polarization\cite{xiao2010berry,Aris2014,RiceMele,ksv}. 
This quantization has a topological origin: If we regard $\phi$ as a crystal momentum along a second, synthetic dimension which we call $y$, Eq.~(\ref{eq:1DRiceMele}) is equivalent to a 2D square lattice model with a $U(1)$ $\pi$-flux (equivalent to half flux quantum $\Phi_{0} = 2\pi \hbar / |e|$ where electron has charge $-|e|$) per plaquette, and with a fixed crystal momentum $k_{y}$ along $y$. 
The quantized polarization change is then identified as the Chern number\cite{tknn,niu1984quantised,niu1985quantized,Aris2014,bernevigbook} of the occupied bands in 2D.
We provide further details, including numerical verification of charge pumping, and the explicit construction of the dimensional promotion to 2D, in the SM\cite{SM}.

We see from this example that promoting the dimension of a modulated system to a higher dimensional lattice coupled to gauge fields can help explain the topological origin of low-dimensional properties, including charge transport and boundary modes. 
A general method for dimensional promotion will thus be helpful in dealing with various topological modulated systems in more than 1D. 
In what follows, we will show that the dimensional promotion approach can be extended to higher dimensions, and to cases where the modulation is incommensurate with the underlying lattice periodicity.

\section{\label{sec_Dimension_promotion}Dimensional Promotion Procedure}

In this section, we will generalize the $1$D-to-$2$D dimensional promotion of the Rice-Mele chain to general dimensions. 
To begin, let us consider a $d$-dimensional ($d$D) electronic model on a cubic lattice with $N$ mutually 
incommensurate on-site modulations\cite{Earliest_dimension_promotion_superspace,Kraus_1D_QC_to_2D_QHE,equivalence_Fibo_Harper_2012,2D_QC_4D_QHE,time_periodic_1,time_periodic_2}, described by the Hamiltonian
\begin{equation}
    H_{\text{low-dim}} = \sum_{\vec{n},\vec{m}} {\psi}^{\dagger}_{\vec{n}+\vec{m}} \left[{H}_{\vec{m}}\right] {\psi}_{\vec{n}} + \sum_{\vec{n}} \sum_{i=1}^{N}{\psi}^{\dagger}_{\vec{n}} \left[{V}^{(i)}_{\vec{n}}\right] {\psi}_{\vec{n}}.\label{eq:h_low_dim}
\end{equation}
Here both $\vec{n} = (n_{1},\cdots,n_{d})$ and $\vec{m} = (m_{1},\cdots,m_{d}) \in \mathbb{Z}^{d}$ are vectors in the $d$D cubic lattice, and ${\psi}^{\dagger}_{\vec{n}}$ is the electron creation operator for an electron at position $\vec{n}$ with a given set of spin and orbital degrees of freedom.
We denote by $\left[{H}_{\vec{m}}\right]$ the hopping matrix connecting position $\vec{n}$ to $\vec{n}+\vec{m}$, and by $\left[{V}^{(i)}_{\vec{n}}\right]$ the matrix representing $i^{\text{th}}$ modulated on-site energy at position $\vec{n}$ ($i = 1 ,\ldots, N$), with matrix indices encoding the spin and orbital dependence of the hopping\footnote{throughout this work, we will use square brackets to denote matrices and matrix-valued functions}. 
Note that hermiticity of the Hamiltonian requires that $\left[{H}_{\vec{m}}\right] = \left[{H}_{-\vec{m}} \right]^\dag$ {{and $\left[{V}^{(i)}_{\vec{n}}\right]^{\dagger}=\left[{V}^{(i)}_{\vec{n}}\right]$}}. 
We further assume that $\left[{V}^{(i)}_{\vec{n}}\right] = \left[f^{(i)}\left(2\pi\vec{q}^{(i)}\cdot\vec{n} + \phi^{(i)} \right)\right]$ with $\left[f^{(i)}(x)\right] = \left[f^{(i)}(x+2\pi)\right]$, where $\vec{q}^{(i)}$ is the $i^{\text{th}}$ modulation wave vector and $\phi^{(i)}$ is the sliding phase associated with the $i^{\text{th}}$ modulation.
For the cubic system with unit lattice vectors we are discussing here, each component $q^{(i)}_{j}$, ($j = 1 ,\ldots, d$) of $\vec{q}^{(i)}$ is defined within $[0,1)$; that is, each $2\pi \vec{q}^{(i)}$ lies within the primitive Brillouin zone of the unmodulated system.
Since each $\left[{V}^{(i)}_{\vec{n}}\right]$ is a periodic function, they can be expanded in terms of Fourier series as
\begin{equation}
     \left[{V}^{(i)}_{\vec{n}}\right] =\sum_{p_{i}\in \mathbb{Z}} \left[{V}^{(i)}_{p_{i}}\right] e^{ip_{i}\left( 2\pi\vec{q}^{(i)}\cdot\vec{n} + \phi^{(i)} \right)}, \label{eq:expand_FT_V}
\end{equation}
where $\left[{V}^{(i)}_{p_{i}}\right]$ is the matrix-valued ${p_{i}}^{\text{th}}$ Fourier component of $\left[{V}^{(i)}_{\vec{n}} \right]$. 
Note that $\left[{V}^{(i)}_{p_{i}}\right] = \left[{V}^{(i)}_{-p_{i}} \right]^{\dagger}$ due to hermiticity of the Hamiltonian.

To perform the enhancement of dimensions, we first insert the expansion Eq.~($\ref{eq:expand_FT_V}$) into the Hamiltonian Eq.~(\ref{eq:h_low_dim}). 
We then regard each $\phi^{(i)}$ as the $i^{\text{th}}$ crystal momentum $k^{i}$ along one of the additional $N$ synthetic dimensions. 
We then promote the $d$D model to a $(d+N)$D space by summing over $\vec{k} = (k^{1},\cdots,k^{N}) \in \mathbb{T}^{N}$ (where $\mathbb{T}^N$ denotes the $N$-dimensional torus), which yields the Hamiltonian in $(d+N)$D as
\begin{align}
    H_{\text{high-dim}} &= \sum_{\vec{n},\vec{m},\vec{k}} {\psi}^{\dagger}_{\vec{n}+\vec{m},\vec{k}} \left[{H}_{\vec{m}}\right] \psi_{\vec{n},\vec{k}} \nonumber  \\
    & + \sum_{\vec{n},\vec{k},i,p_{i}} {\psi}^{\dagger}_{\vec{n},\vec{k}} \left[{V}^{(i)}_{p_{i}}\right] e^{ip_{i}k^{i}} e^{i2\pi p_{i} \vec{q}^{(i)}\cdot \vec{n}}{\psi}_{\vec{n},\vec{k}}. \label{eq:d_n_Bloch}
\end{align}
Each physically distinct configuration of $\{\phi^{(i)}\}$ can be recovered by restricting the Hamiltonian Eq.~(\ref{eq:d_n_Bloch}) to a single $\vec{k}$-point. 
Once we sum over $\vec{k}$, however, we can reinterpret the Hamiltonian in a $(d+N)$D space. 
As we will see below, adiabatic pumping of the phases $\phi^{(i)}$ by an external field will allow us to explore dynamics in the full $d+N$ dimensional space. 

To obtain the $(d+N)$D model in position-space, we perform an inverse Fourier transform of ${\psi}^{\dagger}_{\vec{n},\vec{k}}$, yielding
\begin{align}
    H_{\text{high-dim}} &= \sum_{\vec{n},\vec{m},\vec{\nu}} {\psi}^{\dagger}_{\vec{n}+\vec{m},\vec{\nu}} \left[{H}_{\vec{m}}\right] \psi_{\vec{n},\vec{\nu}} \nonumber  \\
    & + \sum_{\vec{n},\vec{\nu},i,p_{i}} {\psi}^{\dagger}_{\vec{n},\vec{\nu}-p_{i}\hat{\nu}_{i}} \left[{V}^{(i)}_{p_{i}}\right] e^{i2\pi p_{i}\vec{q}^{(i)}\cdot \vec{n} }\psi_{\vec{n},\vec{\nu}}, \label{eq:general_n_plus_d_model}
\end{align}
where $\vec{\nu} = (\nu_{1},\cdots,\nu_{N}) \in \mathbb{Z}^{N}$ and $\hat{\nu}_{i}$ is the unit vector along the $i^{\text{th}}$ additional dimension, such that $\vec{\nu}-p_{i}\hat{\nu}_{i}= (\nu_{1},\cdots,\nu_{i}-p_{i},\cdots,\nu_{N})$. 
Eq.~(\ref{eq:general_n_plus_d_model}) can be viewed as the Hamiltonian for a system on a $(d+N)$D {{cubic lattice}} whose lattice sites are located at $(\vec{n},\vec{\nu}) = (n_{1},\cdots,n_{d},\nu_{1},\cdots,\nu_{N}) \in \mathbb{Z}^{d+N}$. 
The system is coupled to a continuous $U(1)$ gauge field
\begin{align}
    \vec{A} = (\underbrace{\vec{0}}_\text{$d$D},\underbrace{2\pi \vec{q}^{(1)}\cdot \vec{r},\cdots,2\pi \vec{q}^{(N)}\cdot \vec{r}}_\text{$N$D}) \label{eq:expression_A}
\end{align}
through a Peierls substitution\cite{Peierls_substitution}, explaining the appearance of the phase factors multiplying $\left[{V}^{(i)}_{p_{i}}\right]$ in Eq.~(\ref{eq:general_n_plus_d_model}). 
Note $\vec{r} \in \mathbb{R}^{d}$ is a vector in the original $d$D space.

As the vector potential in Eq.~(\ref{eq:expression_A}) is linear in position $\vec{r}$, the antisymmetric field strength $F_{\mu \nu} = \partial_{\mu}A_{\nu} - \partial_{\nu}A_{\mu}$ is constant in space. 
In particular, Eq.~(\ref{eq:expression_A}) implies that the nonzero components of $F_{\mu \nu}$ are given by
\begin{align}
    F_{i,j+d} = \partial_{i}A_{j+d} - \partial_{j+d}A_{i} = \partial_{i}A_{j+d} = 2\pi q^{(j)}_{i}, \label{eq:F_i_j_plus_d}
\end{align}
where $i = 1 ,\ldots, d$ and $j = 1 ,\ldots, N$. 
Due to the antisymmetry of the field strength, $F_{i+d,j}$ with $i = 1 ,\ldots, N$, $j = 1 ,\ldots, d$ is also nonzero and given by $F_{i+d,j} = -2\pi q^{(i)}_{j}$. 
Therefore the (nonzero) constant field strength is proportional to the magnitude of the modulation wave vectors.

This shows that that a $d$D modulated system with phase offset $\{\phi^{(i)}\}$ is equivalent to the Bloch Hamiltonian (see Eq.~(\ref{eq:d_n_Bloch})) of the promoted $(d+N)$D lattice model with fixed crystal momenta $\vec{k}$, once we identify $\phi^{(i)}$ as $k^{i}$.
In practice, the modulation $\left[{V}^{(i)}_{\vec{n}}\right]$ can be induced by a set of DW modulations. 
The phase offset $\{\phi^{(i)}\}$ is then regarded as the phason degrees of freedom, namely the phase of the  $i^{\text{th}}$ mean-field DW order parameter. 
By applying electric fields that depin the DWs and make them slide\cite{gooth2019evidence,Zakphase,gruner1988dynamics}, we may sample the whole spectrum of the $(d+N)$D model. 
In particular, and as we will explore in subsequent sections, non-trivial topology in the $(d+N)$D lattice model--which may support localized boundary states--will manifest in the response of the $d$D model to adiabatic sliding of the DW phase mode(s). 
\textcolor{black}{We emphasize here that in our dimensional promotion procedure for a DW system, there are no emergent electric fields in the promoted $(d+N)$D space. The electric fields mentioned here are external and serve as a way to depin the DW in order to vary $\{\phi^{(i)}\}$ adiabatically. This allows for the sampling of the entire spectrum of the $(d+N)$D model as a function of $\{\phi^{(i)}\}$, namely the additional crystal momenta.}

Before we move on to consider the band topology of promoted lattice models, let us make a few general comments about our dimensional promotion procedure. 
First, note that the dimensional promotion procedure places no constraints on the modulation vectors $\vec{q}^{(i)}$; in particular, they need not be commensurate with the underlying lattice. 
In the case of incommensurate modulation, the dimensional promotion procedure allows us to write the $d$D incommensurate model in terms of a periodic $(d+N)$D model with an irrational $U(1)$ flux per plaquette. 
We will see below how we can use this to explore the topology of systems with incommensurate modulation. 
\textcolor{black}{We emphasize that the dimensional promotion procedure is independent of whether in the original $d$D space the system is infinite or finite. When we promote the dimension of a $d$D system to $(d+N)$D space, the $(d+N)$D system is inherently infinite along the additional $N$ dimensions, as it allows a Fourier transformation to obtain the Bloch Hamiltonian with fixed $N$ additional crystal momenta. From this viewpoint there are two ways to utilize the dimensional promotion procedure. If we promote the dimension of an infinite $d$D system, we will obtain an infinite $(d+N)$D system that allows us to discuss the non-trivial bulk topology in the promoted $(d+N)$D space. If we instead promote the dimension of a finite $d$D system, we will obtain a $(d+N)$D system which is finite along the original $d$ dimensions and infinite in the additional $N$ dimensions. This allows us to compute the energy spectrum to examine whether there are boundary states protected by the non-trivial bulk topology in $(d+N)$D space.}

Second, although here we consider only dimensional promotion of a $d$D cubic lattice model with only on-site modulations \textcolor{black}{and all orbitals located at the lattice points labelled by $\vec{n} \in \mathbb{Z}^{d}$} to a $(d+N)$D cubic lattice model, we may generalize our method to $d$D models with modulations in both on-site and hopping matrix elements, together with non-orthogonal lattice vectors \textcolor{black}{and arbitrary orbital positions}. 
We show how to systematically promote the dimensions of such $d$D models to $(d+N)$D and compute the corresponding $U(1)$ gauge fields in the SM\cite{SM}. 
We also give several examples in the SM\cite{SM}, including the dimensional promotion of: (1) the 1D Rice-Mele chain in Sec.~\ref{sec_review_thouless_pump_Rice_Mele} to a 2D square lattice with $\pi$-flux, (2) 1D lattices with modulation in both on-site energies and hopping terms to 2D hexagonal lattices under a perpendicular magnetic field, and (3) 2D modulated systems with hexagonal lattice to 3D systems also with hexagonal lattices coupled to a $U(1)$ gauge field. 
The $U(1)$ gauge fields will take a slightly different form from Eq.~(\ref{eq:expression_A}) when we consider a system with non-orthogonal lattice vectors. 
However, the vector potentials will still be linear in $\vec{r} \in \mathbb{R}^{d+N}$, and hence will still produce constant field strengths $F_{\mu \nu}$. 
Furthermore, note that although we considered for simplicity models where the electrons were localized to the origin of each unit cell, this is not essential for the application of our formalism.

Third, we emphasize that no additional parameters are used in the above derivation. 
The hopping matrices connecting $(\vec{n},\vec{\nu})$ to $(\vec{n}+\vec{m},\vec{\nu})$ and $(\vec{n},\vec{\nu}-p_{i}\hat{\nu}_{i})$ are given by $\left[{H}_{\vec{m}}\right]$ and $\left[{V}^{(i)}_{p_{i}}\right]$, respectively, in the $(d+N)$D model. 
The phase $\phi^{(i)}$ corresponds to the $i^{\text{th}}$ crystal momentum along the $i^{\text{th}}$ additional dimensions. 
Further, the modulation wave vectors $\vec{q}^{(i)}$ specify the strength of the $U(1)$ gauge field in $(d+N)$D, see Eq.~(\ref{eq:expression_A}). 
Notice that the on-site modulations $\left[{V}^{(i)}_{\vec{n}}\right]$ only lead to hopping parallel to $\hat{\nu}_{i}$ in $(d+N)$D. 
If we also consider modulated hopping matrices in $d$D, upon dimensional promotion we will get hopping along $\vec{m} + p_{i}\hat{\nu}_{i}$ in $(d+N)$D\cite{equivalence_Fibo_Harper_2012,Oded_4D_CI_to_2D_HOTI}, which we show in the SM\cite{SM}. 
Notice that the index $i$ is not summed over in $p_{i}\hat{\nu}_{i}$. 
Recall also that $\vec{m}$ and $\hat{\nu}_{i}$ are vectors in the original $d$D and additional $N$D space, respectively. 
An example that demonstrates this is the 1D Rice-Mele model\cite{RiceMele} in Sec.~\ref{sec_review_thouless_pump_Rice_Mele}. 
In the SM\cite{SM} we promote Eq.~(\ref{eq:1DRiceMele}) to a 2D lattice with $\pi$-flux per plaquette in which the electrons can hop along $\hat{x}+\hat{y}$ (where $\hat{x}$ and $\hat{y}$ are in the original 1D and additional 1D space, respectively).
 
Next, our construction provides a way to compute the promoted $(d+N)$D model and the $U(1)$ gauge field to which it is coupled. 
As a $U(1)$ gauge field breaks time-reversal-symmetry (TRS), this dimensional promotion procedure is suitable to investigate non-trivial topological phases in $(d+N)$D space without TRS. 
Below we will also consider a dimensional promotion to $(d+N)$D space with an $SU(2)$ gauge field, which preserves TRS and allows us to explore non-trivial topological phases protected by TRS\cite{Kane04,bernevig2006quantum,ryu2010topological,KitaevClassify}.
In order to construct a low dimensional modulated model equivalent to a higher dimensional lattice coupled to an $SU(2)$ gauge field, we adopt a top-down approach. 
We will in Sec.~\ref{sec_helical_HOTI_sliding_modes} present a 2D modulated model which is obtained from a 3D model coupled to one $SU(2)$ gauge field with a fixed crystal momentum.

In the following sections, we explore various 2D and 3D modulated systems that admit a dimensional promotion to a higher dimensional topological phases coupled to either $U(1)$ or $SU(2)$ gauge fields. 
We will show how an analysis of the higher-dimensional models can shed light on the eigenstates and boundary state dynamics of incommensurate DWs.

\section{\label{sec_chiral_HOTI}Chiral Higher-Order Topological Sliding Modes}
In this section, we will show how the dimensional promotion procedure can be used to realize 3D chiral HOTIs in 2D density wave (DW) materials. 
We will first construct a Hamiltonian for an insulating 2D modulated system that is inversion-symmetric for special values of the DW sliding phase $\phi$. 
Then, we will show how, after dimensional promotion, the Hamiltonian corresponds to a 3D inversion-symmetric chiral HOTI coupled to a $U(1)$ gauge field. 
We will explore the connection between hinge states of the 3D system and corner states of the 2D system using a combination of numerical diagonalization and a 3D low-energy $\vec{k}\cdot\vec{p}$ theory.

\subsection{Dimensionally Promoted Chiral Model}

Consider the following 2D Hamiltonian for electrons on a square-lattice, with one modulated on-site potential $[V(\vec{q},\vec{n},\phi)]$:
\begin{equation}
\begin{aligned}
H = {} & \sum_{\vec{n}} \psi^{\dagger}_{\vec{n}+\hat{x}} [H_{+\hat{x}}]\psi_{\vec{n}} + \psi^{\dagger}_{\vec{n}+\hat{y}} [H_{+\hat{y}}]\psi_{\vec{n}} + \text{h.c.}\\
&+ \sum_{\vec{n}} \psi^{\dagger}_{\vec{n}}\left( [H_{\text{on-site}}] + [V(\vec{q},\vec{n},\phi)] \right)\psi_{\vec{n}},
\end{aligned}\label{chiral_modulated_2}
\end{equation}
where the unmodulated hoppings and on-site energies are
\begin{align}
    & [H_{+\hat{e}_{i}}] = \frac{J_{i}}{2}\tau_{z}\sigma_{0} - \frac{\lambda_{i}}{2i}\tau_{x}\sigma_{i}, \label{eq:chiral_xy_hopping}\\
    & [H_{\text{on-site}}] = M \tau_{z}\sigma_{0}+\tau_{0} \vec{B_0} \cdot \vec{\sigma} . \label{eq:chiral_on_site}
\end{align}
We use $\hat{e}_{i}$ to denote the unit vector along the $i^{\text{th}}$ ($i = 1, 2$) direction. 
The Pauli matrices $\vec{\tau} = (\tau_{x},\tau_{y},\tau_{z})$ and $\vec{\sigma} = (\sigma_{x},\sigma_{y},\sigma_{z})$ denote respectively  orbital (for example $s$ and $p$ orbitals) and spin degrees of freedom. 
This Hamiltonian is inversion-symmetric, with inversion symmetry represented by $\tau_z$. 
Furthermore, when $\vec{B}_0=0$ the model is also time-reversal (TR) symmetric, with the TR operator represented as $i\sigma_y\mathcal{K}$ (where $\mathcal{K}$ is the complex conjugation operator).

We assume that both orbital degrees of freedom are located at the lattice sites.
The hopping matrices $[H_{+\hat{e}_{i}}]$, and $M \tau_{z}\sigma_{0}$ give rise to, at low energy, four-component massive Dirac fermions, allowing us to access various topological phases\cite{ryu2010topological,haldanemodel,bernevigbook}.
Physically, we can interpret $M \tau_{z}\sigma_{0}$ as the on-site energy difference for different orbitals, and $\tau_{0}\vec{B_0} \cdot \vec{\sigma}$ as a ferromagnetic potential which splits the spin degeneracy of bands\cite{wieder2018axion}. 
The modulated on-site potential, which can arise from a density wave modulation, is
\begin{align}
    [V(\vec{q},\vec{n},\phi)] =&  J_{z}\cos\theta_{\vec{q},\vec{n},\phi}\tau_{z}\sigma_{0} + \lambda_{z}\sin\theta_{\vec{q},\vec{n},\phi}\tau_{x}\sigma_{z}, \label{eq:chiral_modulated_V}
\end{align}
where $\theta_{\vec{q},\vec{n},\phi} = 2\pi \vec{q}\cdot \vec{n} + \phi$, $\vec{q} = (q_{x},q_{y})$ is the modulation wave vector in 2D, $\vec{n} \in \mathbb{Z}^{2}$ is the lattice position, and $\phi$ is the sliding phase. 
The first term in Eq.~(\ref{eq:chiral_modulated_V}) modulates the mass $M \tau_{z}\sigma_{0}$ in Eq.~(\ref{eq:chiral_on_site}), while the second modulation denotes an on-site spin-orbit coupling between $s$ and $p$ orbitals. 
Note that the modulation $J_{z}\cos\theta_{\vec{q},\vec{n},\phi}\tau_{z}\sigma_{0}$ is a TR-even charge ordering, while $\lambda_{z}\sin\theta_{\vec{q},\vec{n},\phi}\tau_{x}\sigma_{z}$ is a TR-odd spin ordering. 
To see this, note that TR maps $(\tau_{0},\tau_{x},\tau_{y},\tau_{z}) \to (\tau_{0},\tau_{x},-\tau_{y},\tau_{z})$ and $(\sigma_{0},\sigma_{x},\sigma_{y},\sigma_{z}) \to (\sigma_{0},-\sigma_{x},-\sigma_{y},-\sigma_{z})$. 
In addition, the modulations $J_{z}\cos\theta_{\vec{q},\vec{n},\phi}\tau_{z}\sigma_{0}$ and $\lambda_{z}\sin\theta_{\vec{q},\vec{n},\phi}\tau_{x}\sigma_{z}$ are both inversion-symmetric when $\phi=0$, $\pi$.

Denoting the third, synthetic dimension as $z$ and identifying $\phi$ as the corresponding crystal momentum $k_{z}$, we may use our general procedure in Sec.~\ref{sec_Dimension_promotion} to promote this 2D modulated system to a 3D lattice model. 
We first expand the modulations in terms of Fourier series as
\begin{align}
    & J_{z}\cos\theta_{\vec{q},\vec{n},\phi}\tau_{z}\sigma_{0} = \frac{J_{z}}{2}\left(e^{i\theta_{\vec{q},\vec{n},\phi}} + e^{-i\theta_{\vec{q},\vec{n},\phi}} \right) \tau_{z}\sigma_{0}, \label{eq:J_expand} \\
    & \lambda_{z}\sin\theta_{\vec{q},\vec{n},\phi}\tau_{x}\sigma_{z} = \frac{\lambda_{z}}{2i}\left( e^{i\theta_{\vec{q},\vec{n},\phi}} - e^{-i \theta_{\vec{q},\vec{n},\phi}} \right)\tau_{x}\sigma_{z}. \label{eq:lambda_z_expand}
\end{align}
According to Eqs.~(\ref{eq:expand_FT_V}) and (\ref{eq:general_n_plus_d_model}), the hopping along $+\hat{z}$ can be identified with the terms associated with $e^{-i \theta_{\vec{q},\vec{n},\phi}}$ in Eqs.~(\ref{eq:J_expand}) (\ref{eq:lambda_z_expand}). 
Therefore, the hopping along $+\hat{z}$ in the promoted 3D space reads
\begin{equation}
[H_{+\hat{z}}]=\frac{J_{z}}{2}\tau_{z}\sigma_{0} -\frac{\lambda_{z}}{2i}\tau_{x}\sigma_{z}.
\end{equation} 
From Eq.~(\ref{eq:expression_A}) we can also identify the vector potential in the promoted 3D space as
\begin{align}
    \vec{A} = (0,0,2\pi \vec{q} \cdot \vec{r}) = (0,0,2\pi q_{x} x + 2\pi q_{y}y), \label{A_U1}
\end{align}
where $\vec{r} = (x,y) \in \mathbb{R}^{2}$. 
Therefore, we have that the lattice Hamiltonian in the promoted 3D space is given by
\begin{widetext}
\begin{align}
    H = \sum_{\vec{n}} & \left[ \left( \psi^{\dagger}_{\vec{n}+\hat{x}} [H_{+\hat{x}}]\psi_{\vec{n}} +\psi^{\dagger}_{\vec{n}+\hat{y}} [H_{+\hat{y}}]\psi_{\vec{n}} + {\psi}^{\dagger}_{\vec{n}+\hat{z}} [H_{+\hat{z}}] e^{-i2\pi (q_{x}n_{x}+q_{y}n_{y})} {\psi}_{\vec{n}} + \text{h.c.} \right)+ \psi^{\dagger}_{\vec{n}} [H_{\text{on-site}}] \psi_{\vec{n}} \right], \label{eq:lattice_model_chiral_sliding}
\end{align}
\end{widetext}
where the vector potential Eq.~(\ref{eq:expression_A}) is coupled to the system through a Peierls substitution\cite{Peierls_substitution}, and
$[H_{+\hat{x}}]$, $[H_{+\hat{y}}]$, $[H_{+\hat{z}}]$ and $[H_{\text{on-site}}]$ are given by Eqs.~(\ref{eq:chiral_xy_hopping}) and (\ref{eq:chiral_on_site}), respectively. 
Hereafter, we will set $J_{x} = J_{y} = J_{z} = J$ for simplicity. 
If we Fourier transform Eq.~(\ref{eq:lattice_model_chiral_sliding}) along $z$ and regard $k_{z}$ (the wavenumber along $z$) as the sliding phase $\phi$, we can obtain the 2D modulated system in Eq.~(\ref{chiral_modulated_2}). 

We will now use Eq.~(\ref{eq:lattice_model_chiral_sliding}) to analyze the topological properties of the higher-dimensional model, in order to infer the properties of the low-dimensional modulated system.
This approach can also be employed in other low-dimensional modulated systems provided the corresponding higher-dimensional models are constructed. 
For $q_{x}=q_{y}=0$ and $\vec{B_0}=0$, Eq.~(\ref{eq:lattice_model_chiral_sliding}) describes a TR and inversion-symmetric insulator whose inversion operation is represented by $\tau_{z}$ (note that inversion symmetry acts to flip the sign on the synthetic momentum $k_z$). 
We can employ the theory of symmetry-based indicators of band topology~\cite{khalaf,Po2017,song2017,xu2020high,MTQC,Wieder_spin_decoupled_helical_HOTI,dynamical_axion_insulator_BB,Kruthoff2016} to compute the $\mathbb{Z}_4$ indicator
\begin{align}
    z_{4} = \frac{1}{4}\sum_{\vec{k}_{a} \in \text{TRIMs}} \left( n^{a}_{+} - n^{a}_{-} \right) \text{ mod }4,
\end{align}
where $n^{a}_{+}$[$n^{a}_{-}$] is the number of positive[negative] parity eigenvalues in the valence band at the time-reversal invariant momentum (TRIM) $\vec{k}_{a} $. 
We find that for $|M/J|>3$, $|M/J|<1$, $1<M/J < 3$, $-3<M/J<-1$, the $\mathbb{Z}_{4}$ symmetry-based indicator is given by $z_{4} = 0$, $0$, $1$ and $3$, respectively. 
The regimes where $z_{4} \mod 2 =1$ give a strong TRS-invariant topological insulator (TI). 
For non-zero $\vec{B_0}$ which breaks TRS but does not induce additional band inversions, the magnetic $\mathbb{Z}_{4}$ symmetry-based indicator~\cite{xu2020high,NaturePaper,MTQC,Po2017,watanabe2018structure,khalaf2018higher,wieder2018axion,Zhang_TMCI_corep_RPB,Slager_magnetic_spacegroup_rep,JiabinZhidaAXIDirac,MurakamiAXI1,MurakamiAXI2,dynamical_axion_insulator_BB} 
is given by 
\begin{align}
    \tilde{z}_{4} = \frac{1}{2}\sum_{\vec{k}_{a} \in \text{TRIMs}} \left( n^{a}_{+} - n^{a}_{-} \right) \text{ mod }4,
\end{align}
such that for $|M/J|>3$, $|M/J|<1$, $1<|M/J| < 3$ we have $\tilde{z}_{4} = 0$, $0$ and $2$. 
The corresponding weak indices are all necessarily trivial.
Therefore, for $1 < |M/J| <3$ with $q_{x}=q_{y}=0$, the system described by Eq.~(\ref{eq:lattice_model_chiral_sliding}) gives a strong TI with $\vec{B_0}=0$ and a chiral HOTI (axion insulator)\cite{MTQC} with $\vec{B_0}\ne0$, where the gapless surface states of the strong TI are gapped by the inversion-preserving ferromagnetic potential $\tau_{0} \vec{B_0} \cdot \vec{\sigma}$. 
Therefore, Eq.~(\ref{eq:lattice_model_chiral_sliding}) with $\vec{q} \ne 0$ describes an inversion-symmetric chiral HOTI\cite{pozo2019quantization} coupled via a Peierls substitution to a 3D $U(1)$ gauge field given by the $\vec{A}$ in Eq.~(\ref{eq:expression_A}).
This $\vec{A}$ produces a constant $U(1)$ magnetic field 
\begin{align}
    \nabla \cross \vec{A}  = (2\pi q_{y},-2\pi q_{x},0), \label{B_U1}
\end{align} 
which preserves the inversion symmetry represented by $\tau_{z}$ in 3D, up to a gauge transformation (see SM\cite{SM}). 
Therefore, for a suitable choice of parameters, as long as the $U(1)$ gauge field does not close the bulk gap in 3D, the insulating ground state will be in the same inversion symmetry-protected non-trivial chiral HOTI phase. 
This implies that our model should exhibit the characteristic boundary modes of a chiral HOTI in 3D. 
In particular, our promoted model will support odd numbers of sample-encircling chiral hinge modes in rod geometries which respect inversion symmetry~\cite{pozo2019quantization,wieder2018axion,dynamical_axion_insulator_BB}.

\begin{figure}[t]
\hspace{-0.5cm}
\includegraphics[width=\columnwidth]{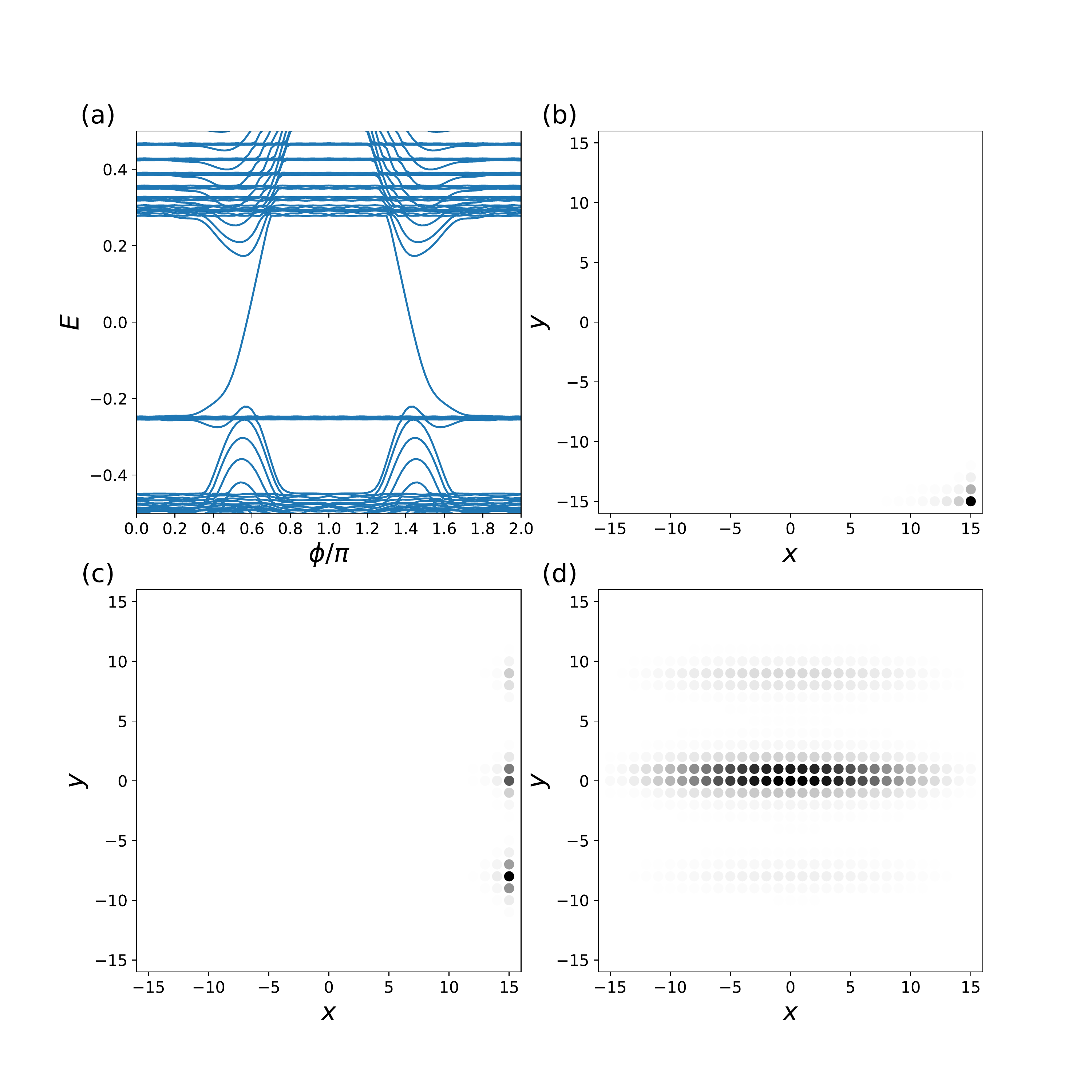}
\caption{(a) $\phi$-sliding spectrum of the chiral 2D model in Eq.~(\ref{chiral_modulated_2}) with parameters given in the text. 
(b) Probability distribution of corner modes in the gap-crossing bands at $\phi = 0.5\pi$ and $E=-0.1368$. 
(c) $\&$ (d) Probability distribution of edge and bulk modes at $\phi = 0.9\pi$ and $E = -0.2508$ and $E = 0.278$, respectively. 
The darker (black) color in (b)--(d) implies higher probability density. \textcolor{black}{(b), (c) and (d) correspond to the corner mode, edge-confined mode and bulk-confined mode discussed in Sec.~\ref{sec:chiral_sliding_corner_modes}, \ref{sec:chiral_sliding_confined_edge_modes} and \ref{sec:chiral_sliding_confined_bulk_modes}, respectively.}
In (b), (c) and (d), the $x$- and $y$-coordinate both range from $-15 ,\ldots, +15$.}
\label{fig:chiral_sliding_main_text}
\end{figure}

\subsection{\label{sec:chiral_sliding_corner_modes}Corner states}

Recalling that in our case the $\hat{z}$ direction is conjugate to the phase $\phi$ of the sliding mode (regarded as the crystal momentum $k_z$), it is natural for us to consider inversion-symmetric rod geometries which are finite in the $\hat{x}$ and $\hat{y}$ directions, and infinite in the $\hat{z}$ direction.
In our 2D system, this corresponds to considering the properties of a finite system as a function of the phase $\phi$.
We can thus compute the energy spectrum of our 2D system in an open geometry with size $L_{x} \times L_{y}$ as a function of $\phi$ to obtain the energy dispersion along $k_{z}$ in the promoted model. 
In the following, we call this kind of calculation the {\it $\phi$-sliding spectrum}, since the variation of $\phi$ can be obtained by electromagnetically exciting the sliding mode of the underlying DW.
Fig.~\ref{fig:chiral_sliding_main_text} (a) shows the $\phi$-sliding spectrum of Eq.~(\ref{chiral_modulated_2}) with parameters $J = 1$, $M =2 $, $\lambda_{i}= 1$, $(\vec{B}_{0})_{i}= 0.5/\sqrt{3}$\cite{pozo2019quantization}, and $\vec{q} = (0,q_{y})$, where $q_{y} = 0.11957$ is comparable with the experimental CDW wave vectors in (TaSe$_4$)$_2$I\cite{shi2019charge} and is incommensurate with the underlying 2D square lattice in Eq.~(\ref{chiral_modulated_2}).
The system size is $31 \times 31$. 
As we can see the spectrum contains modes which, as a function of $\phi$, traverse the bulk spectral gap. 
Examining the wave functions of these ``gap-crossing modes,'' we see that they are localized to the corners of our $2$D sample, as shown in Fig.~\ref{fig:chiral_sliding_main_text} (b). 
The gap-crossing modes with opposite slopes correspond to states at inversion-related corners; in our example one mode is localized at the corner $(x_{\text{corner}},y_{\text{corner}})=(L/2,-L/2)$ (Fig.~\ref{fig:chiral_sliding_main_text} (b)) and the other at $(x_{\text{corner}},y_{\text{corner}})=(-L/2,L/2)$ where $L = 30$.
If we start in a half-filled insulating ground state (with Fermi level $E_{F}=0$), then as $\phi$ slides from $0$ to $2\pi$, we realize charge pumping as one corner mode merges into the occupied-state subspace while the inversion-related counterpart flows into the unoccupied state subspace. 
The ground states at the two inversion-symmetric values $\phi=0,\pi$ differ in electron number by $1$, demonstrating a ''filling anomaly''\cite{benalcazar2018quantization,wieder2020strong}. \tabularnewline
Because these corner modes originate as hinge modes in the $3$D dimensionally promoted system (where, recall, $\phi$ is the momentum $k_z$), their existence is mandated by the non-trivial higher-order topology of the model Eq.~(\ref{eq:lattice_model_chiral_sliding}).

By analyzing the low energy theory of the 3D hinge modes, we will now derive the dynamics of the 2D corner modes as a function of $\phi$. 
In 3D, the corresponding low energy 1D hinge Hamiltonian\cite{hasan2010colloquium,khalaf,hotis} with a chiral mode as a function of $k_{z}$ is given by
\begin{align}
    H_{\text{hinge}} = \xi v_{F} \left( k_{z} + 2\pi \left( q_{x}x_{\text{hinge}} + q_{y}y_{\text{hinge}} \right) \right). \label{eq:chiral_low_energy_hinge_H_1}
\end{align}
We have assumed that for the hinge along $z$ at position $(x_{\text{hinge}},y_{\text{hinge}})$ there is only one chiral mode with Fermi velocity $\xi v_{F}$ where $v_{F} > 0$. 
We have introduced $\xi = \pm 1$ to denote whether the chiral mode has positive or negative velocity. 
Following our dimensional promotion procedure, Eq.~(\ref{eq:chiral_low_energy_hinge_H_1}) is then minimally coupled to a $U(1)$ gauge field in Eq.~(\ref{A_U1}) through the Peierls substitution $k_{z} \to k_{z} + 2\pi (q_{x}x+q_{y}y)$, where $x = x_{\text{hinge}}$ and $y = y_{\text{hinge}}$ are fixed. 

To map Eq.~(\ref{eq:chiral_low_energy_hinge_H_1}) in 3D to the corner mode dispersion in 2D, it is helpful to first compute the $\phi$-sliding spectrum for Eq.~(\ref{chiral_modulated_2}) with $\vec{q} = (0,0)$, as shown in Fig.~\ref{fig:chiral_sliding_low_energy_main_text_temp} (a). 
If we identify $\phi$ as $k_{z}$ in the hinge theory (modulo a constant offset that we will fix later), Fig.~\ref{fig:chiral_sliding_low_energy_main_text_temp} (a) is the $\hat{z}$-directed rod band structure for Eq.~(\ref{eq:lattice_model_chiral_sliding}) without coupling to any vector potential. 
As we can see, there are linear dispersing hinge modes spanning the bulk gap, which cross each other at $k_{z} = \pi$. 
This will be used below in Eq.~(\ref{eq:low_E_corner_chiral}) to complete the mapping from Eq.~(\ref{eq:chiral_low_energy_hinge_H_1}) to 2D. 
Fig.~\ref{fig:chiral_sliding_low_energy_main_text_temp} (a) will also serve as a reference calculation when we examine the response of the $\phi$-sliding spectrum as we increase the magnitude of $\vec{q}$, which will confirm our low energy analysis.

We now use Eq.~(\ref{eq:chiral_low_energy_hinge_H_1}) to construct a low energy description of the corner modes in Fig.~\ref{fig:chiral_sliding_main_text} (a) for Eq.~(\ref{chiral_modulated_2}). 
Upon projecting from 3D to 2D, the fixed hinge mode position $(x_{\text{hinge}},y_{\text{hinge}})$ becomes the fixed corner mode position $(x_{\text{corner}},y_{\text{corner}})$, and the hinge modes become corner modes. 
Since the gap-crossing modes in the $q=0$ system shown in Fig.~\ref{fig:chiral_sliding_low_energy_main_text_temp} (a) intersect at $\phi = \pi$, we replace $k_{z}$ in the hinge theory by $\Delta \phi = \phi - \pi$. 
Thus, we obtain an effective low energy description of the corner modes as
\begin{align}
    H_{\text{corner}} = \xi v_{F} \cdot \left(\Delta \phi+2\pi \left(q_{x}x_{\text{corner}}+q_{y}y_{\text{corner}}\right)\right).
    \label{eq:low_E_corner_chiral} 
\end{align}
We now verify Eq.~(\ref{eq:low_E_corner_chiral}) by numerically computing the $\phi$-sliding spectrum shown in Fig.~\ref{fig:chiral_sliding_low_energy_main_text_temp} (b) with same parameters as Fig.~\ref{fig:chiral_sliding_main_text} (a) but with $q_{y}$ changed to $0.02$. This small value of $q_y$ gives a smooth modulation--and hence a low flux per plaquette in the dimensionally-promoted model--and is thus a suitable platform to examine the low energy theory with minimal coupling. 
We observe gap-crossing modes with negative and positive slopes corresponding to corner modes at $(-L/2,L/2)$ and $(L/2,-L/2)$ where $L =30$, respectively. 
These are shown in Figs.~\ref{fig:chiral_sliding_low_energy_main_text_temp} (c) and (d) at $\phi = 0.4\pi$ and $1.6\pi$, respectively. 
Using Eq.~(\ref{eq:low_E_corner_chiral}), we have the low energy descriptions for these two corner modes governed by the Hamiltonians
\begin{align}
    & H_{\text{corner 1}} = -v_{F}\left( \Delta \phi + \pi q_{y}L \right), \label{eq:eff_corner_chiral_1} \\
    & H_{\text{corner 2}} = +v_{F}\left( \Delta \phi - \pi q_{y}L \right), \label{eq:eff_corner_chiral_2}
\end{align}
where we have used $q_{x} = 0$. 
Thus, if we ramp up $q_{y}$ from $0$ to some non-zero value, we expect to see the corner mode dispersion shift along the $\phi$-axis. 
This is demonstrated in Fig.~\ref{fig:chiral_sliding_low_energy_main_text_temp} (b), which is to be compared with Fig.~\ref{fig:chiral_sliding_low_energy_main_text_temp} (a). 
In fact, a careful examination of Figs.~\ref{fig:chiral_sliding_low_energy_main_text_temp} (a) and (b) shows that the dispersions of the two corner modes shift in opposite directions as a function of $\Delta\phi$, as indicated in Eq.~(\ref{eq:eff_corner_chiral_1}) and Eq.~(\ref{eq:eff_corner_chiral_2}), with the shift given by $\pi q_{y}L \approx 0.6 \pi$ for $q_{y} = 0.02$ and $L = 30$. 
We thus see that the corner mode dispersion in Fig.~\ref{fig:chiral_sliding_low_energy_main_text_temp} (b) can be explained by Eq.~(\ref{eq:low_E_corner_chiral}). 
This demonstrates the origin of the corner modes in the 2D modulated system as higher dimensional hinge modes minimally coupled to a $U(1)$ gauge field. 
If we consider larger $q_{y}$, such as in Fig.~\ref{fig:chiral_sliding_main_text} where we have $q_{y} = 0.11957$, then the shift of the corner mode dispersion is predicted to be $\pi q_{y} L \approx 3.5871 \pi$. This lies outside the first Brillouin zone and needs to be folded back into the range $\phi = [0,2\pi)$. 
This occurs because, in passing from low energy continuum theory to a lattice model, the periodicity of $\phi$--which in the promoted dimension is the continuous wavenumber $k_{z}$--is restored.
Additionally, note that Eq.~(\ref{eq:low_E_corner_chiral}) implies that we may tune the range of $\phi$ where the corner mode energies emerge from the bulk continuum by varying the periodicity of the modulation $\sim 1/|\vec{q}|$. 
As shown in Eq.~(\ref{A_U1}) and Eq.~(\ref{B_U1}), tuning $\vec{q}$ is equivalent to changing the direction and strength of the $U(1)$ gauge field and the corresponding magnetic field in 3D.

\subsection{\label{sec:chiral_sliding_confined_edge_modes}Edge states}
 Having accounted for the low energy description of the corner modes, we observe that in Fig.~\ref{fig:chiral_sliding_main_text} (a), there are additional modes with {\it flat dispersion}. 
 These non-dispersing modes describe states confined either to the bulk or edge of the system, as shown in Figs.~\ref{fig:chiral_sliding_main_text} (c) and (d). 
 We now use low energy theories to demonstrate that these states originate from the $U(1)$ Landau quantization of the surface and bulk electrons in the promoted 3D chiral HOTI. 
 We will revisit Figs.~\ref{fig:chiral_sliding_main_text} (c) and (d) after we complete the low energy theory analysis using relatively small $q_{y}$.

\begin{figure*}[t]
\hspace{-0.5cm}
\includegraphics[width=\linewidth]{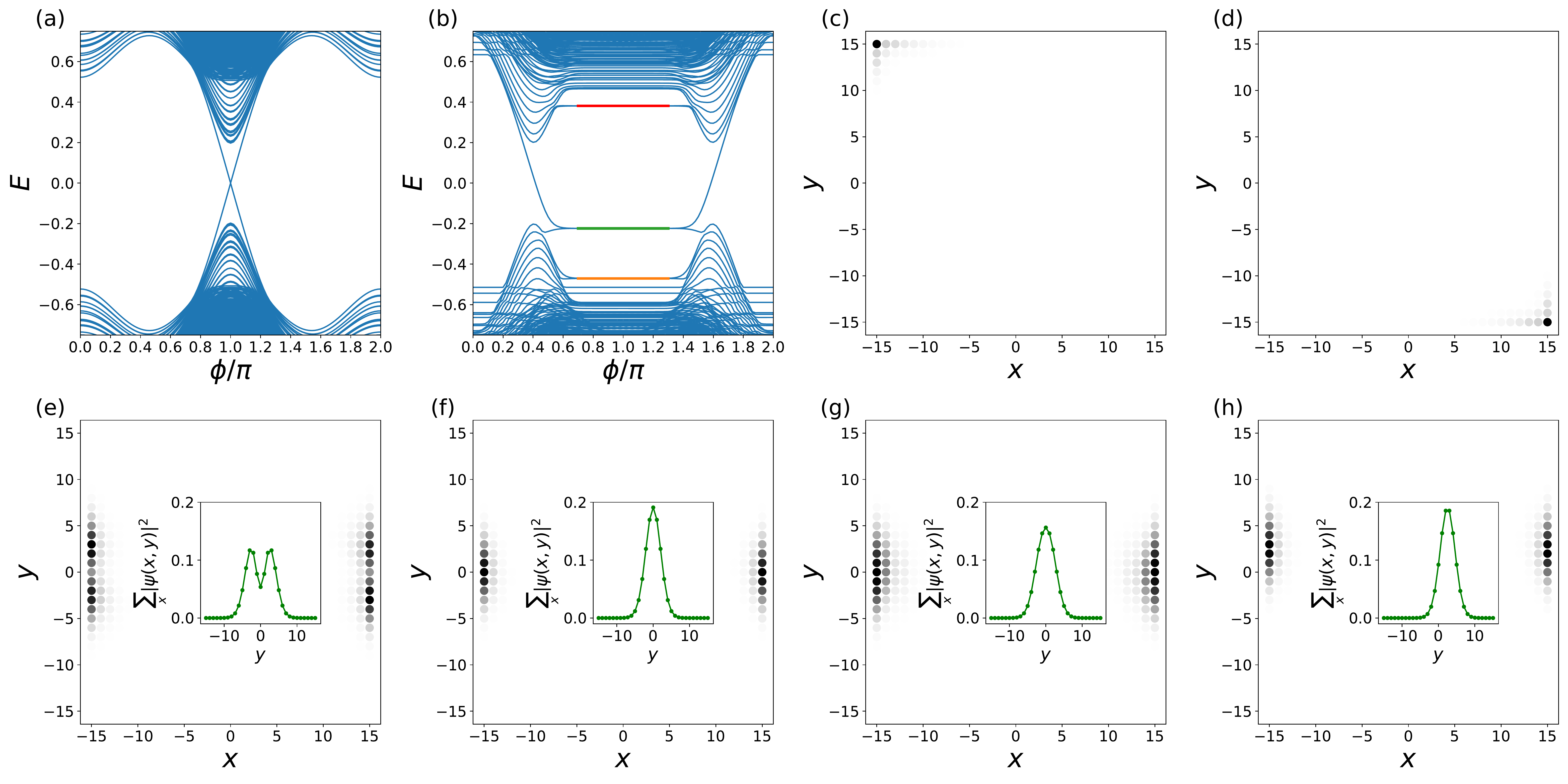}
\caption{(a) $\&$ (b) $\phi$-sliding spectrum of the chiral 2D model with the same parameters as Fig.~\ref{fig:chiral_sliding_main_text} (a) but with $q_{y} = 0$ and $0.02$, respectively. 
(c) $\&$ (d) Probability distribution of the corner modes in (b) at $\phi = 0.4 \pi$ and $1.6 \pi$ and both with $E = 0.0365$. 
(e)--(g) Average of the probability distribution for the doubly degenerate edge-confined modes in the flat bands in (b) at $\phi=\pi$. 
The double degeneracy is due to the pair of opposite edges related by inversion symmetry. 
(e)--(g) are edge-confined modes at $\phi = \pi$ with energies $E=-0.4705$, $-0.2235$ and $0.3811$, which are marked orange, green and red respectively in (b). 
The corresponding energy eigenvalues are $E^{-}_{k_{z}=0,n=1}$, $E^{-}_{k_{z}=0,n=0}$ and $E^{+}_{k_{z}=0,n=1}$ in Eq.~(\ref{eq:U1_surface_EE_EV}). 
(h) Edge-confined mode at $\phi = 0.9 \pi$ with energy $E = -0.2235$ corresponding to $E^{-}_{k_{z}=-0.1\pi,n=0}$ in Eq.~(\ref{eq:U1_surface_EE_EV}). 
The darker (black) color in (c)--(h) implies higher probability density. 
The inset in (e)--(h) is the probability distribution integrated over all $x$ coordinates. 
In (c)--(h), the $x$- and $y$-coordinate both range from $-15 ,\ldots, +15$.}
\label{fig:chiral_sliding_low_energy_main_text_temp}
\end{figure*}

We start with the edge-confined modes. 
Since a chiral HOTI can be obtained by gapping out the surface of a 3D inversion and TR-symmetric TI with a TR-breaking mass term, the generic surface theory reads\cite{khalaf,wieder2018axion,vanderbiltaxion}
\begin{align}
    H_{\text{surf}} = \left( \vec{p} \times \vec{\sigma}' \right)\cdot \hat{n} + m \hat{n} \cdot \vec{\sigma}',
\end{align}
where $\vec{\sigma}'$ are Pauli matrices that act in the basis of low-energy surface states and which capture their spin and orbital texture, $\vec{p}$ is the momentum operator, and $\hat{n}$ is the surface normal vector. 
The time-reversal operator in this surface theory is given by $\mathcal{T} = i\sigma_{y}'\mathcal{K}$ such that $\mathcal{T} \vec{\sigma}' \mathcal{T}^{-1} = -\vec{\sigma}'$. 
The momentum dependent term $\left( \vec{p} \times \vec{\sigma}' \right)\cdot \hat{n}$ describes a helical surface Dirac cone, while $m \hat{n} \cdot \vec{\sigma}'$ is the TR-breaking mass term. 
As shown in Eq.~(\ref{B_U1}), if $q_{x} = 0$, which is the case we consider in Fig.~\ref{fig:chiral_sliding_main_text} and Fig.~\ref{fig:chiral_sliding_low_energy_main_text_temp}, we have that $\nabla \cross \vec{A}$ is parallel to $\hat{x}$. 
We then consider a surface theory on the $yz$-plane coupled to a perpendicular magnetic field $B \hat{x}$ generated by a Landau-gauge $U(1)$ gauge field $\vec{A} = (0,0,By)$.
The corresponding surface Hamiltonian with the $U(1)$ gauge field reads ${{H_{\text{surf}} = p_{y}\sigma_{z}' - \left(p_{z}+By \right)\sigma_{y}' + m\sigma_{x}',}}$ 
where we have made a Peierls substitution such that $p_{z} \to p_{z} + By$, and we have assumed that both $B$ and $m$ are positive. 
To facilitate the derivation, we perform a basis transformation through a $-2\pi/3$ radian spin rotation $U$ along the $[1,1,1]$ axis such that $U^{\dagger} (\sigma_{x}',\sigma_{y}',\sigma_{z}') U = (\sigma_{z}',\sigma_{x}',\sigma_{y}')$. 
The transformed Hamiltonian then reads
\begin{align}
    H_{\text{surf}} = p_{y}\sigma_{y}' - \left(p_{z}+By \right)\sigma_{x}' + m\sigma_{z}'. \label{eq:surf_H_chiral_HOTI_transformed}
\end{align}
Fourier transforming Eq.~(\ref{eq:surf_H_chiral_HOTI_transformed}) to replace {{$p_{z}$}} by the wavenumber $k_{z}$, and defining a $k_{z}$-dependent ladder operator
\begin{align}
    & {{a^{\dagger}_{k_{z}} = \frac{1}{\sqrt{2B}}\left(\left(k_{z}+By \right)-ip_{y} \right),}}  \label{eq:U1_ladder}
\end{align}
where $[a_{k_{z}},a^{\dagger}_{k_{z}}]=1$, we can rewrite Eq.~(\ref{eq:surf_H_chiral_HOTI_transformed}) as
\begin{align}
    {{H_{\text{surf}}(k_{z}) = \begin{bmatrix}
    m & -\sqrt{2B} a_{k_{z}} \\
    -\sqrt{2B} a^{\dagger}_{k_{z}} & -m
    \end{bmatrix}.}}
    \label{eq:chiral_reexpress_H_surf}
\end{align}
We can solve for the eigenstates and energy eigenvalues of Eq.~(\ref{eq:chiral_reexpress_H_surf}) to find
\begin{widetext}
\begin{align}
    & {{\psi^{-}_{k_{z},n=0} = e^{ik_{z}z} \begin{bmatrix}
    0 \\ \ket{0,k_{z}}
    \end{bmatrix},\  E^{-}_{k_{z},n=0} = -m,}} \nonumber \\
    & {{\psi^{-}_{k_{z},n>0} = e^{ik_{z}z} \begin{bmatrix}
    \alpha_{-}(n) \ket{n-1,k_{z}}\\ \ket{n,k_{z}}
    \end{bmatrix},\  E^{-}_{k_{z},n>0} = - \sqrt{m^{2} + 2Bn},}}  \nonumber \\
    & {{\psi^{+}_{k_{z},n>0} = e^{ik_{z}z} \begin{bmatrix}
    \alpha_{+}(n) \ket{n-1,k_{z}}\\ \ket{n,k_{z}}
    \end{bmatrix},\  E^{+}_{k_{z},n>0} = + \sqrt{m^{2} + 2Bn},}} \nonumber \\
    & {{\text{where } \alpha_{\pm}(n) = \frac{-1}{\sqrt{2Bn}}\left(\pm\sqrt{m^{2}+2Bn}+m \right).}} \label{eq:U1_surface_EE_EV}
\end{align}
\end{widetext}
Here $n$ is a non-negative integer labelling the $U(1)$ Landau levels (LLs), and $\ket{n,k_{z}}$ is the $n^{\text{th}}$ simple harmonic oscillator (SHO) eigenstate localized along $y$ defined by the $a^{\dagger}_{k_{z}}$ in Eq.~(\ref{eq:U1_ladder}). 
Notice that the energies $E^{-}_{k_{z},n=0}$, $E^{-}_{k_{z},n>0}$ and $E^{+}_{k_{z},n>0}$ of these LLs shown in Eq.~(\ref{eq:U1_surface_EE_EV}) are all independent of {{$k_{z}$}}. 
As before, we now construct the low energy description of the edge-confined modes in the 2D modulated system from the above low energy surface theory in Eq.~(\ref{eq:surf_H_chiral_HOTI_transformed}). 
We identify $k_{z}$ in the surface theory as $\Delta \phi = \phi - \pi$, since we have flat bands as a function of $\phi$ in our 2D modulated system. 
We also identify $B$ with $2\pi q_{y}$ since in our examples of Fig.~\ref{fig:chiral_sliding_main_text} (a) and Fig.~\ref{fig:chiral_sliding_low_energy_main_text_temp} (b), we have $q_{x} = 0$ and the corresponding vector potential is $\vec{A} = (0,0,2\pi q_{y}y)$. 
When we project down to the 2D model, the surface electrons correspond to states confined in the left and right edges, as shown in Fig.~\ref{fig:chiral_sliding_main_text} (c) and Fig.~\ref{fig:chiral_sliding_low_energy_main_text_temp} (e)--(h). 
We again use $q_{y} = 0.02$ to demonstrate the low energy theory.

We now remark on the implications of our low-energy analysis. 
First, the spectrum in Eq.~(\ref{eq:U1_surface_EE_EV}) breaks particle-hole symmetry as there is a $-m$ energy eigenvalue but no $+m$ energy eigenvalue. 
This can be observed in Fig.~\ref{fig:chiral_sliding_low_energy_main_text_temp} (b), where there are no flat bands of edge-confined modes around $E \approx +0.2$, which corresponds to $E = +m$. 
We thus identify the flat bands in Fig.~\ref{fig:chiral_sliding_low_energy_main_text_temp} (b) marked by red, green and orange as $E^{+}_{k_{z},n=1}$, $E^{-}_{k_{z},n=0}$ and $E^{-}_{k_{z},n=1}$ in Eq.~(\ref{eq:U1_surface_EE_EV}).

Second, from Eq.~(\ref{eq:U1_surface_EE_EV}), the probability distributions for the states $\psi^{-}_{k_{z},n=0}$ and $\psi^{\pm}_{k_{z},n=1}$ are given by
\begin{align}
    & {{|\psi^{-}_{k_{z},n=0}|^2 \propto\left|\varphi_{0,B}(y+k_{z}/B) \right|^{2}}} \\
    & {{|\psi^{\pm}_{k_{z},n=1}|^2 \propto\left| \alpha_{\pm}(1) \right|^{2}\left|\varphi_{0,B}(y+k_{z}/B) \right|^{2} + \left|\varphi_{1,B}(y+k_{z}/B) \right|^{2}}}
\end{align}
up to a normalization factor, where $\varphi_{n,B}(y)$ is the $n^{\text{th}}$ eigenstate of an SHO localized along {{$y$}}. 
Notice that we have indicated the explicit $B$-dependence on $\varphi_{n,B}(y)$ since the cyclotron frequency and the localization of the wave function depend on the strength of magnetic field. 
This implies that $\psi^{-}_{k_{z},n=0}$ has a pure Gaussian distribution. Furthermore, we expect that $\psi^{-}_{k_{z},n=1}$ is more characteristic of an SHO first excited state than $\psi^{+}_{k_{z},n=1}$ since $\left| \alpha_{-}(1) \right|^{2} = \left( -\sqrt{m^{2} + 2B} + m \right)^{2}/(2B) < \left( \sqrt{m^{2} + 2B} + m \right)^{2}/(2B) = \left| \alpha_{+}(1) \right|^{2}$,
as we have assumed both $B$ and $m$ are positive. 
Figs.~\ref{fig:chiral_sliding_low_energy_main_text_temp} (e)--(g) show the 2D wave function probability distributions at $\phi = \pi$ for edge confined modes in different LLs in our lattice model, together with the insets showing the integrated wave function probability over all $x$-coordinates. 
While both Figs.~\ref{fig:chiral_sliding_low_energy_main_text_temp} (e) and (g) corresponds to $n = 1$ LL, the former is at the negative energy branch and the latter is at the positive energy branch.
Therefore Fig.~\ref{fig:chiral_sliding_low_energy_main_text_temp} (e) shows split peaks characteristic of the SHO first excited state, more so than Fig.~\ref{fig:chiral_sliding_low_energy_main_text_temp} (g). 
In contrast, Fig.~\ref{fig:chiral_sliding_low_energy_main_text_temp} (f), corresponding to the $n=0$ LL wave function, shows the Gaussian probability distribution characteristic of the SHO ground state. 
We see that the qualitative properties of the wave functions are all consistent with the low energy surface theory.

Third, the definition of the ladder operator in Eq.~(\ref{eq:U1_ladder}) implies that the center of the wave functions will be shifted by $-k_{z}/B$ from $y = 0$. 
Identifying $k_{z}$ in the low energy theory as $\Delta \phi = \phi - \pi$ and $B$ as $2\pi q_{y}$, we deduce that the distance $l$ that the edge-confined mode gets shifted along $y$ in the lattice model will be $l = -\Delta \phi/({2\pi q_{y}})$. 
Notice that the edge-confined mode in Fig.~\ref{fig:chiral_sliding_low_energy_main_text_temp} (h) at $\phi = 0.9\pi$ ($\Delta \phi = -0.1\pi$) is shifted by $\approx +2.5$ lattice constants along $y$ comparing with Fig.~\ref{fig:chiral_sliding_low_energy_main_text_temp} (f), which is at $\phi = \pi$ ($\Delta \phi = 0$). 
This is consistent with our prediction, as $l$ will  be $+2.5$ when $\Delta \phi = -0.1\pi$ and $q_{y} = 0.02$.

Fourth, although Eq.~(\ref{eq:U1_surface_EE_EV}) predicts non-degenerate energy levels for a single surface with a perpendicular $U(1)$ magnetic field, in Fig.~\ref{fig:chiral_sliding_main_text} (a) the flat band corresponding to the $E^{-}_{k_{z},n=0}$ level is highly degenerate. 
This is due to zone-folding effects, similar to what we observed for the corner mode dispersion in Fig.~\ref{fig:chiral_sliding_main_text} (a). 
As the gap-crossing modes are shifted outside $\phi = [0,2\pi)$, they get folded back to $\phi = [0,2\pi)$ together with the flat bands connected to them. 
Up to the degeneracy due to zone folding, the universal feature is that the edge-confined modes appearing in our 2D chiral DW system originate from the projection of surface electrons in a chiral HOTI with $U(1)$ Landau quantization.

Before moving on, let us remark on the robustness of our low-energy predictions to perturbations of the model. 
If we consider a more complicated modulated system with, for example, long-range and anisotropic hopping terms together with other on-site potentials, as long as the promoted 3D system still preserves inversion symmetry and the gap is not closed, the 3D system will still be in the same chiral HOTI phase. 
However, the low energy theories that we have constructed might be modified. 
For example, the low energy theory at the surfaces, which we model with Eq.~(\ref{eq:surf_H_chiral_HOTI_transformed}), may be modified as 
\begin{align}
    H_{\text{surf}}  =& \alpha_{y}p_{y}\sigma_{y}' - \alpha_{z}\left(p_{z}+By \right)\sigma_{x}' + m\sigma_{z}' + \Delta \sigma_{0}' \nonumber \\
    & + \mathcal{O}(p_{y}^{2},p_{z}^{2},p_{y}p_{z}).\label{eq:higher_order_terms}
\end{align}
Differences between $\alpha_{x}$ and $\alpha_{y}$ can lead to an anisotropic gapped Dirac cone. 
A nonzero $\Delta$ induces unequal masses in different subspace of $\vec{\sigma}'$ which can shift the entire energy spectrum. 
$\mathcal{O}(p_{y}^{2},p_{z}^{2},p_{y}p_{z})$ represents higher-order terms in the low energy theory which might cause nonlinearity in the band dispersion in Eq.~(\ref{eq:higher_order_terms}) without minimal coupling.
By the same reasoning, we might also have non-linear hinge mode energies with a quadratic momentum correction in Eq.~(\ref{eq:chiral_low_energy_hinge_H_1}).
All of these additional terms will change the energetic feature of the system, such as energy spectra, Fermi velocities, together with the detailed form of the wave functions, which will be inevitably different from Eq.~(\ref{eq:U1_surface_EE_EV}). 
Nevertheless, the following features are universal: (1) There will be electrons confined to the surface that undergo $U(1)$ Landau quantization, and therefore there will be states that are confined along some directions. 
Upon projecting down to the 2D modulated system, we will still obtain edge-confined modes. 
(2) There will be (non-)linear hinge mode dispersion that will be shifted along $k_{z}$ due to the minimal coupling. 
Therefore the statement that we can tune the range of $\phi$ where the gap-crossing corner modes appear by tuning the magnitude of the modulation wave vectors, will still hold.  
We use the low energy theories Eq.~(\ref{eq:chiral_low_energy_hinge_H_1}) and Eq.~(\ref{eq:surf_H_chiral_HOTI_transformed}) since these allow us to uncover the relation between the states in the promoted dimension and those in the original low dimensional modulated system in an analytically tractable way.

\subsection{\label{sec:chiral_sliding_confined_bulk_modes}Bulk states}

The above analysis on corner- and edge-confined modes shows that the corresponding higher dimensional description of our modulated system is a 3D chiral HOTI minimally coupled to a $U(1)$ gauge field. 
To complete our analysis, we will now focus on the bulk states.
As expected, the low energy description of the bulk-confined modes, shown in Fig.~\ref{fig:chiral_sliding_main_text} (d), will correspond to the low energy theory of bulk electrons in a 3D chiral HOTI minimally coupled to a $U(1)$ gauge field. 
We start with the Bloch Hamiltonian of the promoted 3D chiral HOTI (Eq.~(\ref{eq:lattice_model_chiral_sliding}) with $q_{x} = q_{y} = 0$) expanded around the $\Gamma$ point\cite{pozo2019quantization}, 
\begin{align}
    H_{\text{bulk}} = m_{\text{bulk}} \tau_{z}\sigma_{0} + \tau_{x}\vec{p} \cdot \vec{\sigma} + \tau_{0} \vec{M} \cdot \vec{\sigma}. \label{eq:3D_chiral_HOTO_bulk}
\end{align}
We have defined several parameters to make Eq.~(\ref{eq:3D_chiral_HOTO_bulk}) compact for later convenience, and introduced $\tau_{0} \vec{M} \cdot \vec{\sigma}$ where $\vec{M} = (M,M,M)$ corresponding to the ferromagnetic potential in Eq.~(\ref{eq:chiral_on_site}). 
We now couple this $H_{\text{bulk}}$ to $\vec{A} = By \hat{z}$, which is equivalent to Eq.~(\ref{A_U1}) with $q_{x} = 0$. 
This can be done via the minimal substitution $p_{z} \to p_{z} + By$. 
Fourier transforming along $x$ and $z$ to replace $p_{x}$ and $p_{z}$ by wavenumbers $k_{x}$ and $k_{z}$, and defining the $k_{z}$-dependent ladder operator as
\begin{align}
    a^{\dagger}_{k_{z}} = \frac{1}{\sqrt{2B}}\left( k_{z} + By - ip_{y} \right), \label{eq:3D_chiral_bulk_U1_ladder}
\end{align}
we can rewrite Eq.~(\ref{eq:3D_chiral_HOTO_bulk}) coupled to $\vec{A} = By \hat{z}$ in terms of $a_{k_{z}}$ and $a^{\dagger}_{k_{z}}$ as
\begin{widetext}
\begin{align}
    H_{\text{bulk}}(k_{x},k_{z}) = m_{\text{bulk}} \tau_{z}\sigma_{0} + \tau_{x} \begin{bmatrix}
    \sqrt{\frac{B}{2}} \left(a_{k_{z}}+a^{\dagger}_{k_{z}} \right) & k_{x} -\sqrt{\frac{B}{2}} \left(a_{k_{z}}-a^{\dagger}_{k_{z}} \right) \\
    k_{x} +\sqrt{\frac{B}{2}} \left(a_{k_{z}}-a^{\dagger}_{k_{z}} \right) & -\sqrt{\frac{B}{2}} \left(a_{k_{z}}+a^{\dagger}_{k_{z}} \right)
    \end{bmatrix} + \tau_{0} \vec{M} \cdot \vec{\sigma}. \label{eq:3D_bulk_LL_Hamiltonian}
\end{align}
\end{widetext}
We have numerically shown in SM\cite{SM} that the effective theory in Eq.~(\ref{eq:3D_bulk_LL_Hamiltonian}) captures several properties of the flat bulk bands in Fig.~\ref{fig:chiral_sliding_low_energy_main_text_temp} (b) with relatively small $q_{y} = 0.02$, such as energy asymmetry with respect to $E = 0$ and the confinement direction of the bulk states due to $U(1)$ Landau quantization.

From the above analysis on corner-, edge-, and bulk-confined modes, we conclude that we can characterize this topological 2D modulated system with chiral sliding modes in terms of a 3D chiral HOTI coupled to a $U(1)$ gauge field. 
In addition, such 2D modulated systems provide a platform to examine the properties of a 3D chiral HOTI, by sliding the DW order parameter $\phi$.

\section{\label{sec_helical_HOTI_sliding_modes}Helical Higher-Order Sliding Modes and $SU(2)$ Gauge Fields }

Next, we will generalize our formalism to time-reversal invariant spinful systems. 
In doing so, we will see that incommensurate modulations induce coupling to $SU(2)$ gauge fields in the dimensionally promoted models.
$SU(2)$ gauge fields can be used to represent spin-orbit coupling\cite{YiLi_SU2_Hofstadter}, which is ubiquitous in topological states of matter. 
For example, $SU(2)$ gauge fields in 3D and 4D generates $SU(2)$ LLs that give rise to 3D TIs and 4D QHEs\cite{YiLi_TI_SU2_LL,zhang2001four}. 
A non-Abelian $SU(2)$ Peierls phase in 2D and 3D lattices can also lead to 2D and 3D TIs\cite{SU2_gauge_in_2D_Goldman,YiLi_SU2_Hofstadter}. 
In addition, in response to a bulk $SU(2)$ gauge flux insertion, a 2D TI can bind various quasi-particle excitations such as spinons, holons and chargeons\cite{qispincharge}. 
In this section, we present a 2D modulated system that allows us to simulate a 3D helical HOTI coupled to an $SU(2)$ gauge field.

\begin{figure}[t]
\hspace{-0.5cm}
\includegraphics[width=\columnwidth]{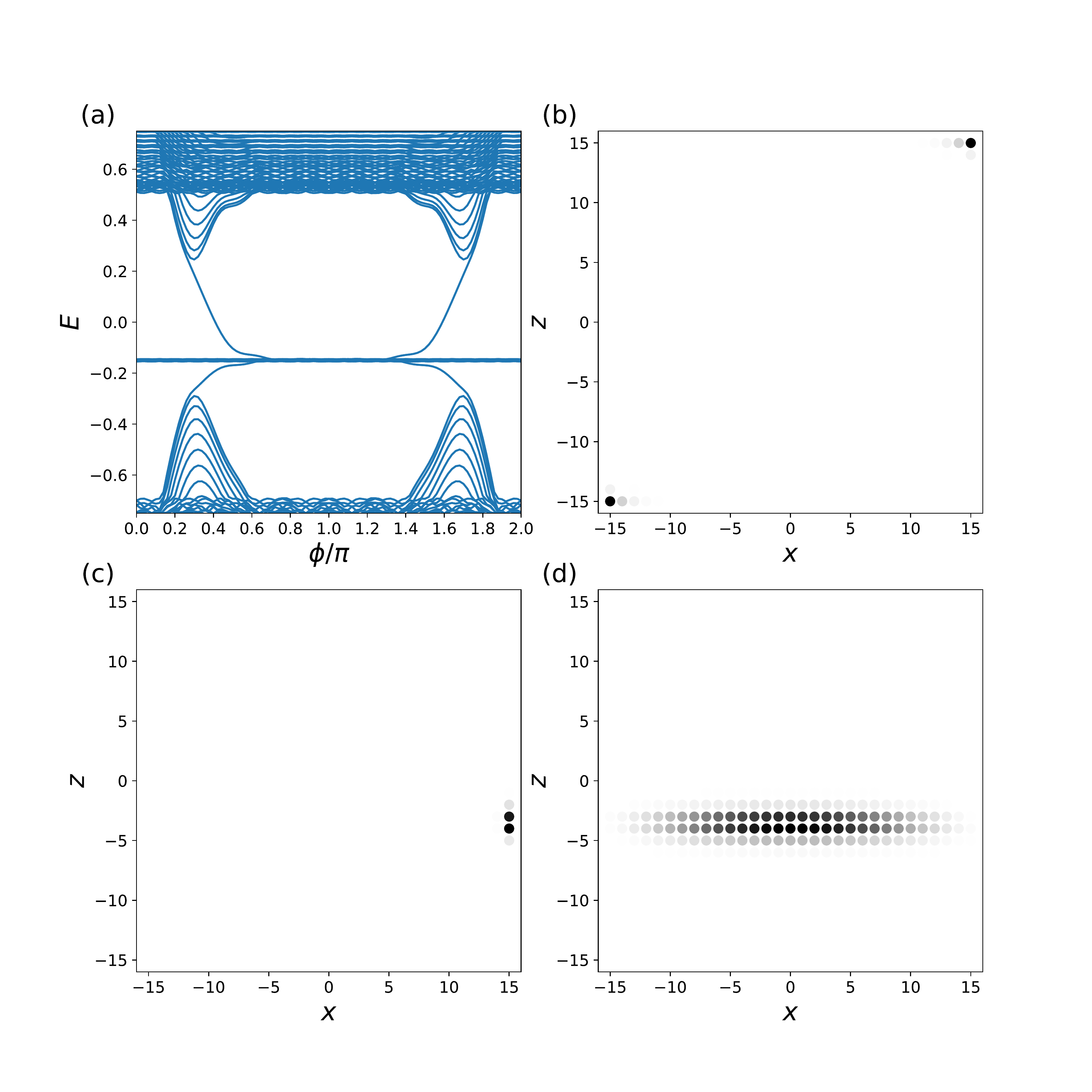}
\caption{(a) $\phi$-sliding spectrum of the 2D helical model in Eq.~(\ref{eq:H_helical_sliding}) with parameters given in the text. 
(b) Summation of the probability density of the doubly-degenerate corners modes at $\phi = 0.4\pi$ and $E=0.0146$. 
The two corner modes at the same $\phi$ are related to each other by the $\mathcal{I}\mathcal{T}$-symmetry, and hence they are localized at inversion-related corners and have opposite spins. 
(c) $\&$ (d) Probability distribution of edge and bulk modes at $\phi = 0.9\pi$ and $E = -0.1459$ and $E = 0.5227$, respectively. 
The darker (black) color in (b)--(d) implies higher probability density. 
In (b), (c) and (d), the $x$- and $z$-coordinate both range from $-15 ,\ldots, +15$.}
\label{fig:helical_sliding_main_text}
\end{figure}

\subsection{Dimensionally Promoted Helical Model}

We start by considering the following 2D Hamiltonian on a square lattice with one modulated on-site potential $[V(\vec{q},\vec{n},\phi)]$:
\begin{equation}
\begin{aligned}
H = {} & \sum_{\vec{n}} \psi^{\dagger}_{\vec{n}+\hat{x}} [H_{+\hat{x}}]\psi_{\vec{n}} + \psi^{\dagger}_{\vec{n}+\hat{z}} [H_{+\hat{z}}]\psi_{\vec{n}} + \text{h.c.}\\
&+ \sum_{\vec{n}} \psi^{\dagger}_{\vec{n}}\left( [H_{\text{on-site}}] + [V(\vec{q},\vec{n},\phi)] \right)\psi_{\vec{n}}, \label{eq:H_helical_sliding}
\end{aligned}
\end{equation}
where the unmodulated couplings are are
\begin{align}
    & [H_{+\hat{x}}] = \frac{v_{x}}{2}\tau_{z}\mu_{0}\sigma_{0} - \frac{u_{x}}{2i}\tau_{y}\mu_{y}\sigma_{0}, \\
    & [H_{+\hat{z}}] = \frac{v_{z}}{2}\tau_{z}\mu_{0}\sigma_{0} - \frac{u_{z}}{2i}\tau_{x}\mu_{0}\sigma_{0}, \\
    & [H_{\text{on-site}}] = m_{1}\tau_{z}\mu_{0}\sigma_{0}+m_{2}\tau_{z}\mu_{x}\sigma_{0} + m_{3}\tau_{z}\mu_{z}\sigma_{0} \nonumber \\ 
    & +m_{v_{1}}\tau_{0}\mu_{z}\sigma_{0} + m_{v_{2}}\tau_{0}\mu_{x}\sigma_{0}. \label{eq:helical_H_on_site}
\end{align}
The matrices $\vec{\tau}$, $\vec{\mu}$ and $\vec{\sigma}$, are Pauli matrices and denote the orbital, sub-lattice and spin degrees of
freedom, respectively. 
The hopping matrices $[H_{+\hat{x}}]$ and $[H_{+\hat{z}}]$, together with the on-site potential $[H_{\text{on-site}}]$ respect both inversion and TR symmetries. 
The inversion and TR operations are represented by $\tau_{z}$ and $i\tau_{z}\sigma_{y}\mathcal{K}$, respectively\cite{Wieder_spin_decoupled_helical_HOTI}. 
These hoppings give rise to low energy four-component Dirac fermions in each spin subspace, realizing a topological critical point. 
The modulated on-site energy is given by
\begin{eqnarray}\nonumber
[V(\vec{q},\vec{n},\phi)]&=& v_{y}\tau_{z}\mu_{0}\begin{bmatrix}
    \cos\theta^{+}_{\vec{q},\vec{n},\phi} & 0 \\ 0 & \cos\theta^{-}_{\vec{q},\vec{n},\phi}
    \end{bmatrix}\\
&& + v_{H} \tau_{y}\mu_{z}\begin{bmatrix}
    \sin\theta^{+}_{\vec{q},\vec{n},\phi} & 0 \\ 0 & \sin\theta^{-}_{\vec{q},\vec{n},\phi}
    \end{bmatrix},
    \label{eq:helical_sliding_modulation_H}
\end{eqnarray}
where $\theta^{\pm}_{\vec{q},\vec{n},\phi} = 2\pi \vec{q}\cdot \vec{n}\pm\phi$, $\vec{q} = (q_{x},q_{z})$ is the modulation wave vector in 2D, $\vec{n} \in \mathbb{Z}^{2}$ is the lattice position, and $\phi$ is the sliding phase.
The first term in Eq.~(\ref{eq:helical_sliding_modulation_H}) modulates the mass $m_{1}\tau_{z}\mu_{0}\sigma_{0}$ in Eq.~(\ref{eq:helical_H_on_site}), which may represent unequal on-site energy for $s$ and $p$ orbitals, with forward ($-\phi$) and backward ($+\phi$) sliding phase in each spin subspace\cite{SU2_gauge_in_2D_Goldman,SU2_gauge_2D_to_1D_Goldman}. 
The second term in Eq.~(\ref{eq:helical_sliding_modulation_H}) describes a modulation of the on-site energy which mixes $s$ and $p$ orbitals with unequal strength for different sublattices. 
Similarly, we have forward and backward sliding phases in different spin subspaces for the second term. 
Since the modulation in Eq.~(\ref{eq:helical_sliding_modulation_H}) has opposite phase offsets in each spin subspace, it may be induced from spin-orbit coupled spin ordering. 
This modulation is TR- and inversion-symmetric only when $\phi=0$, $\pi$. 
Note, however, that the product of inversion and TR symmetry, which we will denote $\mathcal{I}\mathcal{T}$-symmetry, is preserved for all values of $\phi$.
If we denote the third, synthetic dimension as $y$, this 2D model is equivalent to the inversion and TR symmetric 3D helical HOTI model of Ref.~\onlinecite{Wieder_spin_decoupled_helical_HOTI}, coupled to an $SU(2)$ gauge field given by
\begin{align}
    \vec{A} =  (0,2\pi (q_{x}x+q_{z}z)\sigma_{z},0). \label{SU2_A}
\end{align}
This matrix-valued $\vec{A}$ produces a constant $SU(2)$ magnetic field\cite{Estienne_2011} $\vec{B} = \vec{\nabla} \cross \vec{A} - i \vec{A} \cross \vec{A}$, determined from the field strength\cite{eguchi1980gravitation} $F_{\mu \nu} = \partial_{\mu} A_{\nu} - \partial_{\nu} A_{\mu} - i \left[ A_{\mu},A_{\nu}\right]$, and given by
\begin{align}
    \vec{B} = (-2\pi q_{z}\sigma_{z},0,2\pi q_{x} \sigma_{z} ). \label{SU2_B}
\end{align}
This constant $SU(2)$ field strength preserves both inversion and TR symmetry in 3D, up to a spin-dependent gauge transformation (see SM\cite{SM}). 
Notice that Eq.~(\ref{SU2_B}) implies that the $SU(2)$ magnetic field in this example can be interpreted as a $U(1)$ magnetic field with opposite sign for spin-up and spin-down electrons\cite{SU2_gauge_in_2D_Goldman,SU2_gauge_2D_to_1D_Goldman}. 
We then expect that, for a suitable choice of parameters such that the $SU(2)$ gauge field does not close the bulk gap in 3D, the insulating ground-state will be in the same non-trivial helical HOTI phase as the model with $\vec{q}=0$\cite{khalaf,hotis,Po2017}.
Therefore, in 3D, our promoted model will support an odd number of pairs of sample-encircling helical hinge modes respecting inversion and TR symmetries\cite{khalaf,Wieder_spin_decoupled_helical_HOTI}.
Upon projected back to 2D, the helical hinge modes in 3D become $\mathcal{I}\mathcal{T}$-related pairs of corner modes at the same $\phi$ in the 2D modulated system. 
In the SM\cite{SM}, we give the form of the 3D dimensionally-promoted model in position-space. 

\subsection{Calculation of the Spectrum}

Let us now numerically verify these conclusions. 
Fig.~\ref{fig:helical_sliding_main_text} (a) shows the $\phi$-sliding spectrum of Eq.~(\ref{eq:H_helical_sliding}) with parameters $m_{1} = -3$, $m_{2} = 0.3$, $m_{3} = 0.2$, $m_{v_{1}} = -0.4$, $m_{v_{2}} = 0.2$, $v_{x}=v_{z}=u_{x}=u_{z} = 1$, $v_{y} = 2$, $v_{H} = 1.2$\cite{Wieder_spin_decoupled_helical_HOTI}, and $\vec{q} = (0,q_{z})$ where $q_{z} = 0.11957$ \cite{shi2019charge}. 
The system size is $31 \times 31$. 
There are doubly-degenerate pairs of states which cross the gap as a function of $\phi$, where the degeneracy is protected by $\mathcal{IT}$-symmetry. 
We see from the wave functions that these are corner modes related by $\mathcal{IT}$-symmetry, as shown in Fig.~\ref{fig:helical_sliding_main_text} (b) for the branch with negative slope around $\phi \approx 0.4\pi$. 
In the other branch of doubly-degenerate gap-crossing states with positive slope, the corner modes are the inversion-symmetric counterpart (where recall that inversion symmetry leaves spin invariant) to those in Fig.~\ref{fig:helical_sliding_main_text} (b). 
Therefore, as $\phi$ slides from $0$ to $2\pi$, this model realizes a $\mathbb{Z}_{2}$ pump\cite{fu2006time,teo2010topological} as one of the pairs of corner states will merge into the occupied state subspace (with Fermi level $E_{F}=0$) while the other pair will flow out. 
In our specific examples, the two states in each  $\mathcal{I}\mathcal{T}$-related pair at the same $\phi$ are spin eigenstates and therefore in this case the $\mathbb{Z}_2$ pump is a spin pump; our conclusions, however, hold even when spin is not conserved.

As mentioned earlier, and in analogy with our chiral HOTI model, the corner modes here are equivalent to hinge modes along $y$ in 3D. 
The corresponding low energy theory for these corner modes is 
\begin{align}
    H_{\text{corner}}= v_{F} \left( \phi \sigma_{z}' + 2\pi \left( q_{x}x_{\text{corner}} + q_{z}z_{\text{corner}} \right) \sigma_{0}' \right), \label{eq:H_helical_hinge_1}
\end{align}
where $v_{F}$ is the group velocity of the hinge modes in 3D. 
We use the Pauli matrices $\vec{\sigma}'$ to denote the effective basis where in each subspace the states have opposite spin together with some orbital and sub-lattice textures. 
We have assumed without loss of generality that there is only one pair of helical hinge modes at the hinge along $y$ in the promoted 3D system. 
By denoting $\phi$ as $k_{y}$, which is the crystal momentum along $y$, we recognize Eq.~(\ref{eq:H_helical_hinge_1}) as the hinge mode dispersion $H(k_{y}) = v_{F}k_{y}\sigma_{z}'$ in 3D minimally coupled to an $SU(2)$ gauge field described by Eq.~(\ref{SU2_A}).
Similar to Sec.~\ref{sec_chiral_HOTI}, as we vary $\vec{q}$--which is equivalent to changing the strength and (spatial) direction of the $SU(2)$ gauge field in Eqs.~(\ref{SU2_A}) and (\ref{SU2_B})--the dispersion of the spin-polarized corner modes will shift along the $\phi$-axis. 
In the SM\cite{SM} we present a complete low energy theory analysis for the corner modes with the same structure as Sec.~\ref{sec_chiral_HOTI}.

In addition, we show in Figs.~\ref{fig:helical_sliding_main_text} (c) and (d) the probability density for the edge- and bulk-confined modes in the flat bands of Fig.~\ref{fig:helical_sliding_main_text} (a). 
Similar to the corner modes, these can be respectively understood in terms of 3D low energy surface and bulk theories minimally coupled to an $SU(2)$ gauge field, leading to an $SU(2)$ Landau quantization\cite{YiLi_TI_SU2_LL,YiLi_SU2_Hofstadter}. 
The relevant surface theory describes a time-reversed pair of Chern insulators. 
The relevant bulk theory is the $\vec{k}\cdot\vec{p}$ expansion around $\Gamma$ of the promoted 3D helical HOTI Hamiltonian\cite{Wieder_spin_decoupled_helical_HOTI}. 
We provide further details in the SM\cite{SM}.
Together with the corner mode analysis, we see that this topological 2D modulated system with helical sliding modes can be characterized by a 3D lattice model coupled to an $SU(2)$ gauge field. 
In addition, we have shown how 2D modulated systems can provide a platform to examine $SU(2)$ gauge physics in higher dimensions, by sliding the phase $\phi$ of the DW order parameter.

\section{\label{sec:Weyl_CDW}Weyl-CDWs and 4D topological modes}

As a final demonstration of our dimensional promotion formalism and its utility to investigating physics in more than 3D, we consider the mean-field state of a correlated inversion-symmetric 3D Weyl semimetal with CDW distortion (Weyl-CDW)\cite{dynamical_axion_insulator_BB,gooth2019evidence,shi2019charge,wang2013chiral,CDW_Weyl_Sehayek,CDW_in_Q1D_Cohn,Monopole_CDW_in_Weyl_Yi_Li,yu2020dynamical}. 
It has been shown that such a system can realize various topological phases. 
Depending on the phase $\phi$ of the CDW order parameter, the system can interpolate between quantum anomalous Hall (QAH) and {\it obstructed} QAH (oQAH) phase\cite{dynamical_axion_insulator_BB}. 
This is due to the $\pi$ mod $2\pi$ axion angle difference $\delta \theta_{\phi} = \theta(\phi = \pi) - \theta(\phi = 0)$ for the system with $\phi = 0$ and $\phi = \pi$, in the thermodynamic limit. 
Physically, this leads to a Hall conductance difference
\begin{align}
    \left|G_{xy}(\phi = \pi) - G_{xy}(\phi = 0) \right| = e^{2}/h \text{ mod } 2e^{2}/h \label{eq:main_G_xy_eqn}
\end{align}
for a semi-infinite slab [see also Eq.~(\ref{eq:agnostic}) below, as well as Refs.~\onlinecite{olsen2020gapless,2020_Axion_coupling_Vanderbilt}]. 
In this section, we analyze a minimal model of a 3D inversion-symmetric magnetic Weyl-CDW system, which admits a dimensional promotion to 4D with a $U(1)$ gauge field. 
We will explain the origin of the background QAH response and the interpolation between QAH and oQAH phases using the corresponding 4D theory. 
In the following, we will denote a sample infinite along $x$ and $y$ with finite thickness $L_z$ along the $z$ direction as an $xy$-slab. 
Similarly, we will use the term $y$-rod to denote a sample infinite along $y$ and finite along $x$ and $z$ with size $L_{x}\times L_{z}$.

\subsection{3D Weyl-CDW Model and Dimensional Promotion}

\begin{figure*}[t]
\hspace{-0.5cm}
\includegraphics[width=\linewidth]{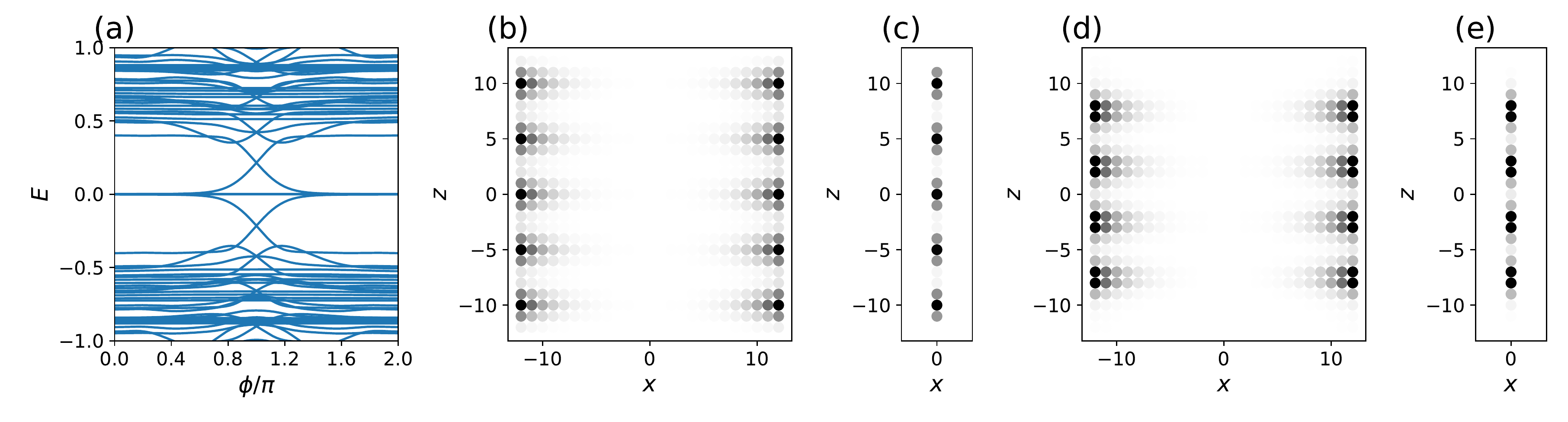}
\caption{(a) $\phi$-sliding spectrum of the Weyl-CDW model in a $y$-rod geometry at $k_{y} = 0$ with size $L_{x}\times L_{z} = 25 \times 25$, $t_{x} = -t_{y} = t_{z} = 1$, $m = 2$, $2|\Delta| = 0.75$ and $q = 1/5$. 
(b) The average probability distribution of the 10 zero modes at $\phi = 0$ in (a). 
These zero modes correspond to QAH surface states. 
(c) The average probability distribution of the 5 non-trivial states at $\vec{k} = \Gamma$ of the $xy$-slab at $\phi = 0$, which in total lead to $G_{xy}(\phi = 0) = -5 e^{2}/h$. 
(d) The average probability distribution of the 8 zero modes at $\phi = \pi$ in (a). 
These zero modes correspond to QAH surface states. 
(e) The average probability distribution of the 4 non-trivial states at $\vec{k} = \Gamma$ of the $xy$-slab at $\phi = \pi$, which in total lead to $G_{xy}(\phi = \pi) = -4 e^{2}/h$. 
The darker (black) color in (b)--(e) implies higher probability density. 
In (b) and (d), the $x$- and $z$-coordinate both range from $-12 ,\ldots, +12$. 
In (c) and (e), the $z$-coordinate ranges from $-12 ,\ldots, +12$.}
\label{fig:temporary_3d_weyl_cdw_fig}
\end{figure*}

To begin, we consider electrons on a 3D cubic-lattice with Hamiltonian $H = H_{0} + H_{CDW}(\phi) $. 
Here $H_0$ is a periodic tight-binding Hamiltonian given by
\begin{align}
    H_0&=\left(\sum_{\vec{n}}\left[-it_x\psi^\dag_{\vec{n}+\hat{x}}\sigma_x \psi_{\vec{n}}-it_y \psi^\dag_{\vec{n}+\hat{y}}\sigma_y \psi_{\vec{n}}+t_z\psi^\dag_{\vec{n}+\hat{z}}\sigma_z \psi_{\vec{n}}\right]\right. \nonumber \\
    &+\sum_{\vec{n}}\frac{m}{2}\left(\psi^\dag_{\vec{n}+\hat{x}}\sigma_z \psi_{\vec{n}} + \psi^\dag_{\vec{n}+\hat{y}}\sigma_z \psi_{\vec{n}}  -2 \psi^\dag_{\vec{n}}\sigma_z \psi_{\vec{n}} \right) \nonumber \\
    &\left.-\sum_{\vec{n}} t_z \left(\cos (\pi q)\right) \psi^\dag_{\vec{n}}\sigma_z c_{\vec{n}}\right) +\mathrm{h.c.}
\end{align}
in position space. 
The corresponding Bloch Hamiltonian is
\begin{align}
    &H_0(\vec{k})= -2[t_x\sin (k_x)\sigma_x +t_y\sin (k_y)\sigma_y]  \nonumber \\
    &  -m[2-\cos (k_x) - \cos (k_y)]\sigma_{z} +2t_z[\cos (k_z) -\cos (\pi q)]\sigma_{z} \label{eq:Bloch_H_0}  , 
\end{align}
with $m/2 \ge t_{x},-t_{y},t_{z} >0$. 
We take for the on-site modulation
\begin{align}
    H_{CDW}(\phi)&=2|\Delta|\sum_{\vec{n}}\cos(2\pi q n_{z}+\phi)\psi^\dag_{\vec{n}}\sigma_z \psi_{\vec{n}}.
\end{align}
Here $2|\Delta|$ is the strength of the CDW modulation, $2\pi q$ is the magnitude of the modulation wave vector $2\pi \vec{q}=(0,0,2\pi q)$ and $\phi$ is the phase of CDW order parameter. 
We again use $\vec{\sigma}$ to denote the Pauli matrices, which here index an orbital degree of freedom. 
The inversion and TR operation are represented by  $\sigma_{z}$ and $\mathcal{K}$, respectively (note that this is a model of spinless electrons).
The Hamiltonian $H_{0}$ then describes a TR-breaking, inversion-symmetric magnetic Weyl semimetal (WSM) with Weyl nodes at $\vec{k} = (0,0,\pm \pi q)$, see Eq.~(\ref{eq:Bloch_H_0})\cite{mccormick2017minimal}.
The perturbation $H_{CDW}(\phi)$ is the CDW modulation that couples these two Weyl nodes and opens a gap in the bulk spectrum~\cite{dynamical_axion_insulator_BB}. 
Note that in this simple model, we have chosen the modulation wavevector to be exactly equal to the Weyl node separation vector for simplicity of analysis. 
Even though the bulk is gapped, the surface of this 3D Weyl-CDW is gapless, due to the presence of QAH surface states. 
In Fig.~\ref{fig:temporary_3d_weyl_cdw_fig} (a) we show the $\phi$-sliding spectrum for a $y$-rod of $H_{0}+H_{CDW}(\phi)$ at $k_{y} = 0$ with size $L_{x} \times L_{z} = 25 \times 25$, $t_{x} = - t_{y} = t_{z} = 1$, $m = 2$, $2|\Delta|=0.75$ and $q = 1/5$. 
This corresponds to a commensurate Weyl-CDW system. 
The mid-gap zero modes in Fig.~\ref{fig:temporary_3d_weyl_cdw_fig} (a) correspond to the QAH surface states. 
In Fig.~\ref{fig:temporary_3d_weyl_cdw_fig} (b) we show the probability distribution of the 10 zero modes at $\phi = 0$. 
Together with Wilson loop and Berry curvature calculation in the SM\cite{SM}, we verify that the corresponding $xy$-slab with $L_{z} = 25$ carries a slab Hall conductance $ G_{xy}(\phi = 0)  = -5 e^{2}/h$. 
We can then identify the weak Chern number\cite{qi2008topological,kohmoto1992diophantine,halperin1987possible,fukanemele} of the 3D periodic system with $5=25/5$ unit cells (since $q= 1/5$) as $\nu_{z} = -1$.

As in Sec.~\ref{sec_chiral_HOTI} and~\ref{sec_helical_HOTI_sliding_modes}, we identify $\phi$ with the crystal momentum $k_{w}$ along a fourth, synthetic direction denoted by $w$. 
Using the dimensional promotion procedure in Sec.~\ref{sec_Dimension_promotion}, we can promote $H_{0} + H_{CDW}(\phi)$ to a 4D nodal line system coupled to a $U(1)$ gauge field. 
In the SM\cite{SM} we give the explicit form of the promoted model in 4D position space. 
The corresponding 4D nodal line system (with $q=0$) has a Bloch Hamiltonian
\begin{align}
    &H(\vec{k})= -2[t_x\sin (k_x)\sigma_x +t_y\sin (k_y)\sigma_y] + 2|\Delta|\cos{(k_{w})}\sigma_{z}  \nonumber \\
    & -m[2-\cos (k_x) - \cos (k_y)]\sigma_{z} +2t_z[\cos (k_z) -\cos (\pi q)]\sigma_{z}. 
\label{eq:ham0}
\end{align}
The spectrum of this Hamiltonian features nodal lines at $k_{x}=k_{y} = 0$ defined by the implicit equation
\begin{align}
    t_{z}\cos{(k_{z})}  +  |\Delta| \cos{(k_{w})} = t_{z} \cos{(\pi q)}.
\end{align}
According to Eq.~(\ref{eq:expression_A}), we then couple this Hamiltonian to a 4D $U(1)$ gauge field given by
\begin{align}
    \vec{A} = (0,0,0,2\pi q z), \label{eq:U1_4D}
\end{align}
since $2\pi \vec{q} = 2\pi q \hat{z}$ in this system. 
This $\vec{A}$ only produces non-zero field strength threading the $zw$ plane,
\begin{align}
    F_{zw} = -F_{wz}= \partial_{z}A_{w} - \partial_{w}A_{z}=2 \pi q, \label{eq:Fzw}
\end{align}
where all other components of $F_{\mu \nu} = \partial_{\mu} A_{\nu} - \partial_{\nu} A_{\mu}$ are zero. 
We are now in a position to reinterpret the existence of a background QAH response and QAH surface states when the bulk gap is opened due to the CDW. 
We will see how these features emerge from the low energy approximation for this 4D system minimally coupled to Eq.~(\ref{eq:U1_4D}).

\subsection{Low Energy Theory Analysis}

\begin{widetext}

We start from the 4D Bloch Hamiltonian in Eq.~(\ref{eq:ham0}). 
Expanding around $\vec{k} = \vec{0}$, we have
\begin{align}
    H(\vec{k}) \approx -2[t_x k_{x} \sigma_{x} +t_y  k_{y} \sigma_{y}] + 2t_{z}  \left( 1 - \frac{k_{z}^{2}}{2} - \cos (\pi q) \right)\sigma_{z} + 2|\Delta| \left(  1 - \frac{k_{w}^{2}}{2}\right) \sigma_{z}. \label{eq:ham0_low_energy}
\end{align}
The nodal line in this low energy theory is an ellipse in the $k_{z}$-$k_{w}$ plane with $k_{x}=k_{y}=0$, defined by
\begin{align}
    t_{z}k_{z}^{2} +  |\Delta|k_{w}^{2} =  2t_{z}\left[ 1 - \cos{ (\pi q)} \right] + 2|\Delta| > 0.
\end{align}
Replacing the 4D wave vector $\vec{k}=(k_{x},k_{y},k_{z},k_{w})$ by the 4D momentum operator $\vec{p}=(p_{x},p_{y},p_{z},p_{w})$ using the so-called  envelope function approximation\cite{Envelope_function_approximation_1,Envelope_function_approximation_2,Envelope_function_approximation_3,Envelope_function_approximation_4,Envelope_function_approximation_5,Qi_Zhang_RMP,hasan2010colloquium,bernevig2006quantum}, the Hamiltonian governing the low energy dynamics reads
\begin{align}
    H = -2[t_x p_{x} \sigma_{x} +t_y p_{y} \sigma_{y}] + 2t_{z}  \left( 1 - \frac{p_{z}^{2}}{2} - \cos (\pi q)\right) \sigma_{z} + 2|\Delta| \left(  1 - \frac{p_{w}^{2}}{2}\right) \sigma_{z}. \label{eq:ham0_low_energy_p}
\end{align}
Next, let us minimally couple Eq.~(\ref{eq:ham0_low_energy_p}) to a {{4D}} $U(1)$ gauge field $\vec{A} = (0,0,0,2\pi qz)$ via a Peierls substitution such that $p_{w} \to p_{w} + 2\pi q z$. 
Eq.~(\ref{eq:ham0_low_energy_p}) then becomes
\begin{align}
    H = -2[t_x p_{x} \sigma_{x} +t_y p_{y} \sigma_{y}] + 2 \left( t_{z}\left[ 1 - \cos{(\pi q)} \right] + |\Delta| \right) \sigma_{z}  - \left( t_{z}p_{z}^{2} +|\Delta| \left( p_{w} +  2 \pi q z \right)^{2} \right) \sigma_{z},  \label{eq:after_4D_U1_couple}
\end{align}
where we have assumed that the particle carries $-1$ charge. 
Fourier transforming along $x$, $y$ and $w$, we may replace $p_{x}$, $p_{y}$ and $p_{w}$ by the corresponding wavenumbers $k_{x}$, $k_{y}$, $k_{w}$, such that
\begin{align}
    H(k_{x},k_{y},k_{w}) = -2[t_x k_{x} \sigma_{x} +t_y k_{y} \sigma_{y}] +2 \left( t_{z}\left[ 1 - \cos{(\pi q)} \right] + |\Delta| \right) \sigma_{z}  - \left( t_{z}p_{z}^{2} + |\Delta| \left( k_{w} + 2\pi q z \right)^{2} \right) \sigma_{z}. \label{eq:FT1_4D_U1_LL}
\end{align}
\end{widetext}

Notice that the coefficient of $\sigma_z$ in the final term in the Hamiltonian,
\begin{align}
    t_{z}p_{z}^{2} + |\Delta| \left( k_{w} + 2 \pi q z \right)^{2}, \label{eq:4D_U1_SHO_H}
\end{align}
is an SHO Hamiltonian along $z$ which can be diagonalized as
\begin{align}
    4 \pi q \sqrt{t_{z}|\Delta|} \left( n + \frac{1}{2} \right).
\end{align}
Here $n$ is a non-negative integer and the eigenvalue of the number operator $ a^{\dagger}_{k_{w},q} a_{k_{w},q}$ with
\begin{align}
    a^{\dag}_{k_{w},q} = \frac{1}{\sqrt{ 4\pi q}} \left( \frac{t_{z}}{|\Delta|} \right)^{\frac{1}{4}}\left[ \left({\frac{|\Delta|}{t_{z}}}\right)^{\frac{1}{2}} \left(k_{w}+ 2\pi q z \right) - ip_{z} \right]. \label{eq:4D_U1_LL_ladder}
\end{align}
The quantum number $n$ is the 4D $U(1)$ LL index. 
By restricting to a subspace of the full Hilbert space with fixed $n$ and $k_{w}$, we see that the 4D low energy Hamiltonian Eq.~(\ref{eq:after_4D_U1_couple}) may be decomposed into a direct sum of 2D low energy Chern insulators (CIs) in $xy$-plane parameterized by $n$ and $k_{w}$.
The Hamiltonian for these 2D CIs is given by
\begin{align}
    H_{\text{2D CI}}(n,k_{w}) = -2[t_x p_{x} \sigma_{x} +t_y p_{y} \sigma_{y}] + 2 m \sigma_{z}, \label{eq:2D_CI_subspace}
\end{align}
where
\begin{align}
    m = t_{z}\left(1 - \cos{(\pi q)} \right)+ |\Delta| - 2\pi q \sqrt{t_{z}|\Delta|}\left( n + \frac{1}{2} \right). \label{eq:mass_term_emerge_CIs}
\end{align}
Since we have restricted to the subspace with fixed $n$ and $k_{w}$ in Eq.~(\ref{eq:2D_CI_subspace}), according to Eq.~(\ref{eq:4D_U1_SHO_H}) the wave function along $z$ and $w$ will be SHO eigenstates centered at $z = -k_{w} / (2\pi q)$ multiplied by a plane wave $e^{ik_{w} w}$. 
Notice that the $k_{w}$-dependence of Eq.~(\ref{eq:2D_CI_subspace}) is due to the integer $n$ in Eq.~(\ref{eq:mass_term_emerge_CIs}) which is an eigenvalue of the number operator $a_{k_{w},q}^{\dagger}a_{k_{w},q}$. 
Therefore, the eigenstates in the low energy approximation take the form of plane waves in $w$, and Chern insulator eigenstates as a function of $(x,y)$ localized at different constant-$z$ planes for different $k_{w}$.
This provides a four-dimensional interpretation of the layer construction of the Weyl-CDW presented in Refs.~\onlinecite{dynamical_axion_insulator_BB,CDW_Weyl_Sehayek}.

As in a 3D nodal ring system with a perpendicular magnetic field~\cite{Nodal_ring_perpendicular_B}, Eq.~(\ref{eq:2D_CI_subspace})  can yield a gapped 4D bulk spectrum provided that $m \ne 0$ $\forall n \ge 0$. 
This insulating ground state will then carry non-trivial topology inherited from the nodal line system, since in Eq.~(\ref{eq:2D_CI_subspace}) we found that the gapped 4D continuum theory is composed of low energy 2D CIs. 
We then expect that there will be CI layers in the $xy$-plane of the corresponding 4D lattice model (see SM\cite{SM}). 
The CI layers will also be separated along $z$ by $2\pi / (2\pi q) = 1/q$ for a fixed $k_{w}$, due to the $2\pi$ periodicity of $k_{w}$ in the lattice model. 
In our current example this separation is $5$ since $q  =1/5$. 
Notice that $k_{w}$ is now interpreted as the crystal momentum along the $4^{\text{th}}$ dimension. 
To connect these observation in 4D back to the physical 3D Weyl-CDW system with Hamiltonian $H_{0} + H_{CDW}(\phi)$, we notice that each 3D Weyl-CDW system with a fixed $\phi$ corresponds to the 4D theory with a fixed $k_{w}$. 
Focusing on the $xy$-slab with $\phi = 0$ and thickness $L_{z} = 25$, in which $ G_{xy}(\phi = 0)  = -5 e^{2}/h$, we show wavefunctions corresponding to the only $5$ layers of non-trivial CIs separated from each other by $5$ lattice constants along $z$ in Fig.~\ref{fig:temporary_3d_weyl_cdw_fig} (c). 
Each of these CI layers carries Chern number $C = -1$ and contributes one chiral edge mode along $y$ in the ${y}$-rod, shown in Fig.~\ref{fig:temporary_3d_weyl_cdw_fig} (b). 
In the SM\cite{SM} we provide technical details on identifying the non-trivial CI layers using hybrid Wannier function, Berry phases and Berry curvature calculations for an $xy$-slab. 
We can thus regard the CDW-induced gap opening and the existence of background QAH response as the results of $U(1)$ Landau quantization in the 4D nodal line system.

Next, we address the interpolation between the QAH phase at $\phi = k_w=0$ and the oQAH phase at $\phi = k_w=\pi$ using the 4D theory. 
Before we turn to the 4D low energy theory, we begin with the observation that in Fig.~\ref{fig:temporary_3d_weyl_cdw_fig} (a), the number of mid-gap zero modes corresponding to QAH surface states decreases by $2$ as $\phi$ slides from $0$ to $\pi$; one state is lowered into the valence band, while one state is elevated to the conduction band.
This is consistent with the change in Hall conductance Eq.~(\ref{eq:main_G_xy_eqn}), which is derived in the thermodynamic limit where the 2D slab thickness $L_{z} \to \infty$ with infinitesimal but non-zero $2|\Delta|$~\cite{dynamical_axion_insulator_BB}. 
The ambiguity modulo $2e^{2}/h$ in the change of Hall conductance is due to the axion angle $\theta$, which is only well-defined mod $2\pi$, as shown below in Eq.~(\ref{eq:agnostic}). 
Taking $L_{z} \to \infty$ with infinitesimal $2|\Delta|$ ensures that the only effect the CDW modulation has is to open the gap at the Weyl points without inverting bands at other high-symmetry points in the 3D Brillouin zone. 
We have also verified that our choice for the parameters in Fig.~\ref{fig:temporary_3d_weyl_cdw_fig} (a) is adiabatically connected to this condition by increasing $L_{z}$ and decreasing $2|\Delta|$. 
The slab Hall conductance $G_{xy}$ of the $xy$-slab contains both an extensive contribution from the bulk QAH phase through the weak Chern number $\nu_{z}$, and an intensive contribution from axion angle $\theta$, which collectively gives\cite{vanderbiltaxion,2020_Axion_coupling_Vanderbilt}
\begin{align}
    G_{xy} = \frac{e^2}{h}(\nu_{z} l_{z} + \theta / \pi), \label{eq:agnostic}
\end{align}
where $l_{z}$ is the number of unit cells in the slab. 
In our examples for $q = 1/5$, $l_{z}$ will be given by $L_{z}/5$. 
Recall also that as we slide $\phi$ from $0$ to $\pi$, the bulk gap of the 3D Weyl-CDW system never closes, hence the $\nu_{z}$ is unchanged during the process. 
Putting this all together, we see that Eq.~(\ref{eq:main_G_xy_eqn}) implies that there is a $\pi$ mod $2\pi$ change in the axion angle between $\phi = 0$ and $\phi = \pi$. 
To be more specific, in our current examples we have $G_{xy}(\phi = 0) = -5 e^{2}/h$ and $G_{xy}(\phi = \pi) = -4 e^{2}/h$. 
This quantized change of $G_{xy}$ or $\theta$ can be explained again using the 4D low energy theory, as we now show.

\begin{figure*}[t]
\hspace{-0.5cm}
\includegraphics[width=\linewidth]{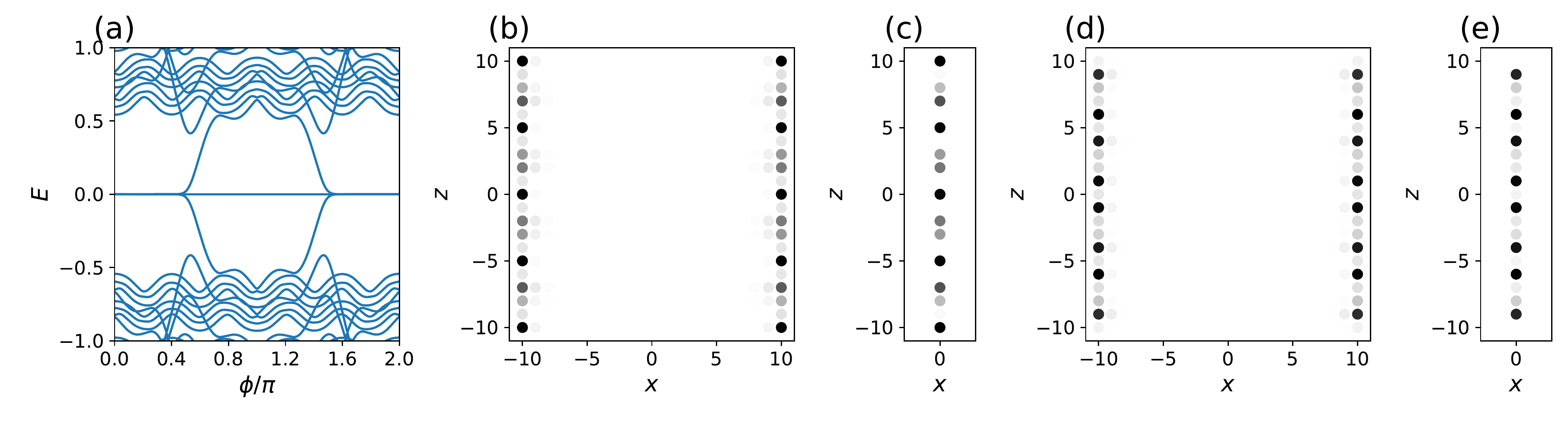}
\caption{(a) $\phi$-sliding spectrum of the Weyl-CDW model in a $y$-rod geometry at $k_{y} = 0$ with size $L_{x}\times L_{z} = 21 \times 21$, $t_{x} = -t_{y} = t_{z} = 1$, $m = 2$, $2|\Delta| = 2$ and $q = \tau /4$ where $\tau = (1+\sqrt{5})/2$. 
(b) The average probability distribution of the 18 zero modes at $\phi = 0$ in (a). 
These zero modes correspond to QAH surface states. 
(c) The average probability distribution of the 9 non-trivial states at $\vec{k} = \Gamma$ of the $xy$-slab at $\phi = 0$, which in total lead to $G_{xy}(\phi = 0) = -9 e^{2}/h$. 
(d) The average probability distribution of the 16 zero modes at $\phi = \pi$ in (a). 
These zero modes correspond to QAH surface states. 
(e) The average probability distribution of the 8 non-trivial states at $\vec{k} = \Gamma$ of the $xy$-slab at $\phi = \pi$, which in total lead to $G_{xy}(\phi = \pi) = -8 e^{2}/h$. 
The darker (black) color in (b)--(e) implies higher probability density. 
In (b) and (d), the $x$- and $z$-coordinate both range from $-10 ,\ldots, +10$. 
In (c) and (e), the $z$-coordinate ranges from $-10 ,\ldots, +10$.}
\label{fig:temporary_3d_weyl_cdw_fig_2}
\end{figure*}

Going back to the 4D low energy theory, Eq.~(\ref{eq:4D_U1_SHO_H}) predicts that if we shift $k_{w}$ to $k_{w} + \Delta k_{w}$, the corresponding CI layers described by the Hamiltonian in Eq.~(\ref{eq:2D_CI_subspace})--which are localized around $z = -k_{w}/(2\pi q)$--will be shifted in the $z$ direction by $\Delta z = -\Delta k_{w} / (2\pi q)$. 
Connecting this observation back to the physical 3D Weyl-CDW system, it implies that as we slide $\phi$ from $0$ to $\pi$, all the CI layers will be shifted by $\Delta z = -\pi / (2\pi q) = -1/(2q)$; for our choice of $q= 1/5$ this gives a shift of $\Delta z = -2.5$. 
We demonstrate this numerically in Figs.~\ref{fig:temporary_3d_weyl_cdw_fig} (d) and (e) which show the probability distribution of the $8$ QAH zero modes and the corresponding 4 non-trivial CI layers (with Chern number $C = -1$) for $\phi = \pi$. 
The physical interpretation of Eq.~(\ref{eq:main_G_xy_eqn}) is now clear: As we slide $\phi$ from $0$ to $\pi$, the non-trivial CI layers will be shifted by $\Delta z = -2.5$ unit cells, all in the same direction. 
Therefore, the bottom non-trivial CI layer at $\phi = 0$ and $z = -10$ depicted in Fig.~\ref{fig:temporary_3d_weyl_cdw_fig} (c) will be shifted outside the finite sample and hence will not appear when $\phi=\pi$. 
At $\phi = \pi$, there will be only $4$ non-trivial CI layers remaining. 
This leads to a change in the Hall conductance by $e^2/h$, as indicated by Eq.~(\ref{eq:main_G_xy_eqn}). 
Simultaneously, the number of QAH zero modes in the $y$-rod decreases by $2$ when we slide $\phi$ from $0$ to $\pi$. 
Physically, these two QAH zero modes are pushed toward the boundary of the system, due to the shifting of the bottom non-trivial CI layer. 
Therefore, their energies will be pushed toward the bulk continuum, leading to the inevitable appearance of gap-crossing bands as shown in Fig~\ref{fig:temporary_3d_weyl_cdw_fig} (a). 
Numerically, we have observed that in all of our examples (Figs.~\ref{fig:temporary_3d_weyl_cdw_fig} and~\ref{fig:temporary_3d_weyl_cdw_fig_2}), the zero modes in the band structure of the $y$-rod only appear at $k_{y} = 0$. 
Therefore, as far as the zero modes are concerned, we can focus on the energy spectrum of the $y$-rod at $k_{y}=0$, as in Figs.~\ref{fig:temporary_3d_weyl_cdw_fig} (a) and \ref{fig:temporary_3d_weyl_cdw_fig_2} (a). 
Analytically, this can be understood from the Hamiltonian of the low energy Chern insulator Eq.~(\ref{eq:2D_CI_subspace}) for each $n$ and $k_{w}$, which has zero energy edge modes only at $k_{y} = 0$\cite{hasan2010colloquium,Qi_Zhang_RMP,bernevigbook,Jackiw_Rebbi}.

To summarize, we have shown that the identity Eq.~(\ref{eq:main_G_xy_eqn}) can be regarded as a consequence of the $U(1)$ Landau quantization of a 4D nodal line system in which the localization centers along $z$ of the states are directly related to $k_{w}$.
We then identified $k_w$ with the sliding phase $\phi$ through our dimensional promotion formalism in Sec.~\ref{sec_Dimension_promotion}. 
The change in conductance as a function of $\phi$ can thus be regarded as a physical manifestation of the {\it Chern number polarization}, which can alternatively be computed in terms of $z$-localized hybrid Wannier centers~\cite{2020_Axion_coupling_Vanderbilt,zilberberg2018photonic,dynamical_axion_insulator_BB,yu2020dynamical,olsen2020gapless}.

Having demonstrated the utility of our dimensional promotion formalism for a 3D Weyl-CDW system coupled to a commensurate CDW with $q = 1/5$, we next explore the case of incommensurate modulations which are prevalent in nature\cite{gruner1988dynamics}. 
In particular, the experimentally intriguing Weyl-CDW (TaSe$_4$)$_2$I is incommensurate\cite{shi2019charge,tasei_original,heeger_tasei,tournier2013electronic,zhang2020first,shi2019charge}. 
We still consider $H_{0}+H_{CDW}(\phi)$ with $t_{x}=-t_{y}=t_{z}=1$, $m=2$, $2|\Delta|=2$. 
However, we now choose the modulation $q = \tau /4$ where $ \tau = (1+\sqrt{5})/2$ is the golden ratio. 
For an $xy$-slab we choose $L_{z} = 21$ and for $y$-rod we choose $L_{x} \times L_{z} = 21\times 21$. 
Since $q=\tau/4$ is an irrational number, the modulation $H_{CDW}(\phi)$ is incommensurate with $H_{0}$. 
Crucially though, we can use our dimensional promotion procedure regardless of whether or not the modulation is commensurate with the underlying lattice. 
The $U(1)$ gauge field to which the 4D nodal line system is coupled still takes the form in Eq.~(\ref{eq:U1_4D}).
The main difference is that now, the 4D system has an irrational flux $2\pi q = \pi\tau/2$ per plaquette in the $zw$ plane. 
We have verified that for the $xy$-slab we have $G_{xy}(\phi = 0) = -9 e^{2}/h$ and $G_{xy}(\phi = \pi) = -8 e^{2}/h$, consistent with Eq.~(\ref{eq:agnostic}). 
We also show in Fig.~\ref{fig:temporary_3d_weyl_cdw_fig_2} (a) the $\phi$-sliding spectrum of the $y$-rod at $k_{y}=0$. 
We see that there are $2$ fewer QAH zero modes at $\phi = \pi$ than at $\phi = 0$. 
The 18 and 16 QAH zero modes for $\phi = 0$ and $\phi = \pi$ are shown in Fig.~\ref{fig:temporary_3d_weyl_cdw_fig_2} (c) and (e), respectively. 
We again identify 9 and 8 non-trivial states in the $xy$-slab at $\vec{k} = \Gamma$ for $\phi = 0$ and $\phi = \pi$, and show their probability distributions in Fig.~\ref{fig:temporary_3d_weyl_cdw_fig_2} (b) and (d), respectively. 
In the SM\cite{SM} we present the details of the numerical methods for identifying non-trivial states in the $xy$-slab. 
The existence of non-zero QAH response and QAH zero modes can again be attributed to 4D $U(1)$ Landau quantization which gaps the 4D nodal line system, yielding a topologically non-trivial insulating ground state. 
In particular, we also have $\left| G_{xy}(\phi = \pi) - G_{xy}(\phi = 0) \right| = e^{2}/h$ mod $2e^{2}/h$. 
This can again be understood from the shifting of non-trivial CI layers. 
In this case, as $\phi$ slides from $0$ to $\pi$, all the non-trivial CI layers will be shifted downward by $\Delta z = -\pi / (2\pi q) = -2 / \tau \approx -1.236$ lattice constants. 
The non-trivial CI layer at the bottom ($z = -10$) of Fig.~\ref{fig:temporary_3d_weyl_cdw_fig_2} (b) will be shifted outside the finite size system and thus the absolute value of slab Hall conductance will be changed by $-1$. 
Consequently, the number of QAH zero modes in the $y$-rod at $k_{y}=0$ will be decreased by $2$. 
Therefore, together with the examples in Sec.~\ref{sec_chiral_HOTI} and Sec.~\ref{sec_helical_HOTI_sliding_modes}, we see that our dimensional promotion procedure provides a way to understand topological properties of systems with incommensurate modulations.

\subsection{Weyl-CDW and 4D Chern Number}

We can also understand the topological properties of the Weyl-CDW model from the perspective of 4D response theory. 
Combining the field strength in Eq.~(\ref{eq:Fzw}) with our analysis of the Hall conductance above allows us to formulate a $(4+1)$D field-theoretical description of the QAH response in a 3D Weyl-CDW system. 
The corresponding action is that of the $(4+1)$D Chern-Simon theory\cite{qi2008topological,zhang2001four}
\begin{align}
    S = \frac{C_{2}}{24 \pi^{2}} \sum_{\mu\nu\lambda\rho\sigma}\int d^{5}x \epsilon^{\mu \nu \lambda \rho \sigma} A_{\mu} \partial_{\nu} A_{\lambda} \partial_{\rho} A_{\sigma}, \label{eq:action}
\end{align}
where $C_{2}$ is the second Chern number, $A_{\mu}$ is the electromagnetic gauge potential and $\epsilon^{\mu \nu \lambda \rho \sigma}$ is the Levi-Civita symbol in $(4+1)$D.
The Greek indices here are taken to run over all $4+1$ dimensions. 
Eq.~(\ref{eq:action}) gives the electromagnetic response through
\begin{align}
    J^{\mu} = \frac{\delta S}{\delta A_{\mu}} =\frac{C_{2}}{32 \pi^{2}}  \sum_{\nu\lambda\rho\sigma}\epsilon^{\mu \nu \lambda \rho \sigma} F_{\nu \lambda} F_{\rho \sigma},
\end{align}
where $J^{\mu}$ is the current density along the $\mu$ direction. 
Since we have $F_{zw} = 2\pi q$, an electric field $E^{y}$ along $y$ (implying $F_{ty} = E_{y}$) 
will induce a Hall current density along $x$ through
\begin{align}
    J^{x} = q \frac{C_{2}}{2\pi}  E^{y}. \label{eq:Hall_current_x}
\end{align}
Integrating this along the $z$ direction, we find then that, with non-zero $C_{2}$, the Hall conductance $G_{xy}$ is proportional to $qL_z$. 
This is consistent with Eq.~(\ref{eq:agnostic}) and the the recent calculation\cite{dynamical_axion_insulator_BB} showing that the Hall conductance $G_{xy}$ of a 3D Weyl-CDW system is given by 
\begin{align}
    G_{xy} =  \left( |\vec{Q}|L_{z}+2\theta \right) \cdot e^{2} / (2\pi h), \label{eq:Eq_from_dynamical_axion_insulator_preprint}
\end{align}
where $L_{z}$ is the thickness of the $xy$-slab, $\vec{Q}$ is the CDW wave vector along $z$, which in our specific model system is $\vec{Q} = 2\pi q \hat{z}$, and $\theta$ is the bulk axion angle computed from the inversion-symmetric unit cell. 
As we take the thermodynamic limit $L_{z} \to \infty$, the axion angle contribution to $G_{xy}$ becomes negligible and thus $G_{xy}$ can also be regarded as proportional to the magnitude of CDW wave vector, which is consistent with Eq.~(\ref{eq:Hall_current_x}).
Therefore, the field strength in Eq.~(\ref{eq:Fzw}) indeed allows a sensible construction of higher dimensional continuum theory.

To see concretely that the 3D Weyl-CDW system indeed emulates a 4D system with non-zero $C_{2}$, we notice that for both examples in Fig.~\ref{fig:temporary_3d_weyl_cdw_fig} and Fig.~\ref{fig:temporary_3d_weyl_cdw_fig_2}, the system can be deformed into a limit where we have layers of decoupled Chern insulators localized along $z$. 
In the decoupled-layer limit, for the commensurate case, for example, $q = 1/5$, where we consider the single nontrivial band in each unit cell, this implies that $C_{2}$, which is defined through\cite{2D_QC_4D_QHE,zilberberg2018photonic,ozawa2016synthetic,4D_QHE_ultracold_atom,qi2008topological}
\begin{align}
    C_{2} = \frac{1}{4\pi^{2}} \int_{\mathbb{T}^{4}} d^{4}k  \left( \Omega_{xy}\Omega_{zw} + \Omega_{wx}\Omega_{zy} + \Omega_{zx} \Omega_{yw} \right)
\end{align}
becomes
\begin{align}
    C_{2} = \frac{1}{4\pi^{2}} \left(\int_{\mathbb{T}^{2}} dk_{x}dk_{y}  \Omega_{xy}\right)\left(\int_{\mathbb{T}^{2}} dk_{z}dk_{w}\Omega_{zw}\right)
\end{align}
in this limit, where $\Omega_{\mu \nu}$ is the Abelian Berry curvature in the $k_{\mu}$-$k_{\nu}$ plane. 
For both examples in Fig.~\ref{fig:temporary_3d_weyl_cdw_fig} and Fig.~\ref{fig:temporary_3d_weyl_cdw_fig_2}, we have identified the weak Chern number $\nu_{z} = -1$, implying that both systems have $\frac{1}{2 \pi} \int_{\mathbb{T}^{2}} dk_{x}dk_{y}  \Omega_{xy} = -1$. 
In fact, for 3D Weyl-CDW system it have been shown that there will always be background QAH response in the $xy$ plane\cite{dynamical_axion_insulator_BB}, implying that in the limit of decoupled Chern insulators we have $\frac{1}{2 \pi} \int_{\mathbb{T}^{2}} dk_{x}dk_{y}  \Omega_{xy} \ne 0$. 
Furthermore, as we shift the CDW sliding phase $\phi$, which is equivalent to shifting the momentum $k_{w}$, by $2\pi$, all the Chern insulating layers will be shifted by $\Delta z = -\Delta k_{w}/(2\pi q) = -2\pi /(2\pi q) = -1/q$, implying a non-trivial Thouless charge pump along $z$. 
Specifically, for Fig.~\ref{fig:temporary_3d_weyl_cdw_fig} with $q = 1/5$, all the Chern insulating layers will be shifted by $\Delta z = -5$, which is equal to the unit cell length along $z$, implying $\left| \frac{1}{2 \pi} \int_{\mathbb{T}^{2}} dk_{z}dk_{w}\Omega_{zw} \right|=1$. 
The fact that the 3D Weyl-CDW system can be viewed as layers of Chern insulators\cite{dynamical_axion_insulator_BB} and the expression $\Delta z = -1/q$ governing the charge pumping along $z$ as we vary $k_{w}$ by $2\pi$ collectively predict a non-zero $\frac{1}{2 \pi} \int_{\mathbb{T}^{2}} dk_{z}dk_{w}\Omega_{zw}$. 
Therefore, for a 3D Weyl-CDW system with QAH surface states\cite{dynamical_axion_insulator_BB}, the corresponding 4D theory is described by a (4+1)D Chern-Simon theory in Eq.~(\ref{eq:action}) with non-zero $C_{2}$. 
Furthermore, this result holds even as we deform away from the decoupled-layer limit, provided no energy gaps close. 
Thus the 3D Weyl-CDW system serves as a platform to study higher-dimensional topological field theories.

Let us conclude with two remarks. 
First, from the above analysis, we see that a 3D Weyl-CDW system with QAH surface states provides a platform to examine a 4D nodal line system gapped by a $U(1)$ gauge field and carries non-zero second Chern number $C_{2}$. 
Secondly, as opposed to Secs.~\ref{sec_chiral_HOTI} and~\ref{sec_helical_HOTI_sliding_modes} where we have in higher dimensions a gapped topological phase coupled to $U(1)$ or $SU(2)$ gauge fields, in the 4D model promoted from a 3D Weyl-CDW system it is precisely the coupling to a $U(1)$ gauge field that opens up a bulk gap, inducing emergent CI layers, QAH surface states and non-zero $C_{2}$.

\section{\label{sec:outlook}Outlook}

To conclude, we have shown in Secs.~\ref{sec_chiral_HOTI} and \ref{sec_helical_HOTI_sliding_modes} that higher-order topology in 3D can be probed in 2D DW systems. 
Furthermore, we showed in Sec.~\ref{sec:Weyl_CDW} how 3D Weyl-CDW systems with background QAH response can be used to study 4D topology. 
The next and natural step is to identify 3D systems with modulations coexisting with hinge or corner modes. 
This will be a platform for studying 4--or even higher--dimensional higher-order topology. 
Our dimensional promotion procedure in Sec.~\ref{sec_Dimension_promotion} can also be used together with the topological classification based on crystalline symmetries\cite{Po2017,khalaf2018higher,Chiu2016,NaturePaper} in the promoted dimensions, in order to explore topological crystalline phases in higher dimensions. 
With suitably chosen modulated systems, we may either study (1) how topological crystalline insulators diagnosed by symmetry-based indicators\cite{Po2017,comment,khalaf,ashvin-materials,po2020symmetry,watanabe2018structure} in the promoted dimensions respond to a background $U(1)$ or $SU(2)$ gauge fields, or (2) how topological semimetals\cite{armitage2018weyl} in the promoted dimensions can be gapped by background $U(1)$ or $SU(2)$ gauge fields. 
With the dimensional promotion procedure, we may also extend our studies of topological materials to those with space groups beyond 3D, known as superspace groups\cite{superspace1,superspace2,superspace3,superspace4,superspace5}. 
To extract the full information in higher dimension, a way to control the phase offset $\{\phi^{(i)} \}$ experimentally is needed, and currently applying electromagnetic fields to depin the (charge- or spin-)density waves is one practical approach\cite{gooth2019evidence,gruner1988dynamics}. 
In addition, since we can systematically compute the background continuous gauge field coupled to the dimensionally-promoted model, we can again use low dimensional modulated systems to study the low energy dynamics in higher dimensions, by minimally coupling the low energy theory to the known continuous gauge fields as in Sec.~\ref{sec:Weyl_CDW}. 
As our dimensional promotion procedure can be carried out for both commensurate and incommensurate modulations, this approach can be used to study topological properties of system with quasi-periodic potentials~\cite{Earliest_dimension_promotion_superspace,2D_QC_4D_QHE,zilberberg2018photonic,equivalence_Fibo_Harper_2012,Kraus_1D_QC_to_2D_QHE} where conventional band theory is not applicable. 
The general procedure will be to promote the dimension of these quasi-periodic systems and examine the response of possible topological phase in higher dimensions to a gauge field producing an irrational flux per plaquette. 
These techniques can be applied to analyze the DW phases in material systems of interest such as (TaSe$_4$)$_2$I\cite{gooth2019evidence,shi2019charge,zhang2020first,tournier2013electronic} and ZrTe$_5$\cite{tang2019three,qin2020theory,song2017instability,zhang2017transport}. 
This can also lead to interesting studies on the higher-dimensional Hofstadter butterfly, complementing the recent studies of Refs.~\onlinecite{Bernevig_fragile_LL,Herzog_Hof_topo}. 
\textcolor{black}{Another interesting direction is to introduce dynamics to the DW modulation. This can happen, for example, when the phase offsets $\{ \phi^{(i)}\}$ acquire non-adiabatic time-dependence and become $\{ \phi^{(i)}(t)\}$. Previous studies have focused on promoting the dimension of a periodically-driven system to a Floquet lattice, which under certain conditions can lead to topologically-protected quantized energy pump~\cite{time_periodic_1,time_periodic_2,Refael_Topo_freq_conversion_cavity_PRB}. We expect that richer phenomena in higher-dimensional space can be investigated when the DW modulations are not only periodic in real-space but also (1) periodic in time or (2) have general time-dependence.} Finally, we have shown in Sec.~\ref{sec_helical_HOTI_sliding_modes} the simplest case of how $SU(2)$ gauge field physics may be studied through a 2D modulated system. 
Recently, the spin-orbit-coupled Hofstadter models induced by non-Abelian $SU(2)$ gauge fields have also been studied both in 2D\cite{2D_SU2_butterfly_Joannopoulos} and 3D\cite{3D_SU2_butterfly_Joannopoulos}, where Dirac points with up to 16-fold degeneracy and various topological insulating states were found. 
We expect that 3D DW materials with different types of spin-orbit coupled modulations may enable simulation of various aspects of the physics of $SU(2)$ gauge fields in 4D or higher dimensions, including topological states and $SU(2)$ Hofstadter butterflies~\cite{YiLi_SU2_Hofstadter,2D_SU2_butterfly_Joannopoulos,3D_SU2_butterfly_Joannopoulos}. 
We hope that this work will lay the groundwork for the exciting future investigations mentioned above, and extend the search for exotic topological phases beyond 3D. 
In particular, there are many possible defects that one can imagine in a spin-orbit coupled density wave order parameter, each of which may correspond to a non-trivial response to $SU(2)$ gauge field defects in the higher-dimensional system.

\begin{acknowledgments}
The authors would like to thank Y. Li and B. Wieder for fruitful discussions. 
This work was supported by the Alfred P. Sloan Foundation, and by the National Science Foundation under grant DMR-1945058. 
Numerical computations made use of the Illinois Campus Cluster, a computing resource that is operated by the Illinois Campus Cluster Program (ICCP) in conjunction with the National Center for Supercomputing Applications (NCSA) and which is supported by funds from the University of Illinois at Urbana-Champaign. 
Numerical calculations in this work employed the open-source PythTB package\cite{PythTB}.
\end{acknowledgments}

\bibliography{refs}

\end{document}


\title{Supplemental Material for "Simulating Higher-Order Topological Insulators in Density Wave Insulators"}
\author{Kuan-Sen Lin}
\author{Barry Bradlyn}
\affiliation{Department of Physics and Institute for Condensed Matter Theory, University of Illinois at Urbana-Champaign, Urbana, IL, 61801-3080, USA}

\maketitle

Throughout the derivations in this Supplemental Material, we work in unit $\hbar = c= |e| = 1$, and the electron charge is taken to be $-|e| = -1$. 
Furthermore, the Einstein summation convention will not be used; whenever there is a summation over an index, we will write the summation explicitly.

\tableofcontents

\clearpage

\section{\label{sec:General_dimension_promotion}General dimensional promotion procedure}

In this section, we develop a general procedure to promote a $d$-dimensional ($d$D) modulated system to a $(d+N)$D lattice model coupled to a $U(1)$ gauge field. 
Generalizing Sec.~III in the main text, here we allow for both on-site energy and hopping modulations, general (potentially non-cubic) lattice structures in the higher-dimensional model, as well as arbitrary orbital positions. 
Sec.~\ref{subsec:formalism} is devoted to the development of the method. 
Specific examples are given in Sec.~\ref{subsec:examples_general_dim_promotion}.

\subsection{\label{subsec:formalism}Formalism}

\subsubsection{\label{subsubsec:__dimensional_promotion}Dimensional promotion}
Suppose we have a $d$D periodic electronic system with lattice vectors $\{\vec{a}_{1},\cdots,\vec{a}_{d} \}$, which form a general $d$D Bravais lattice. 
Note that the $\vec{a}_i$ need not be mutually orthogonal. 
We add $N$ mutually incommensurate on-site and hopping modulations with modulation wave vectors $\{\vec{q}^{(1)},\cdots,\vec{q}^{(N)}\}$, such that the Hamiltonian is given by
\begin{equation}
    H_{\text{low-dim}} = \sum_{\vec{n},\vec{m}} {\psi}^{\dagger}_{\vec{n}+\vec{m}} \left[H_{\vec{m}}\right] {\psi}_{\vec{n}} + {\sum_{\vec{n},\vec{l}}}  \sum_{i=1}^{N} {\psi}^{\dagger}_{\vec{n}+\vec{l}} \left[H_{\vec{l},\vec{n}}^{(i)} \right] {\psi}_{\vec{n}} + \sum_{\vec{n}} \sum_{i=1}^{N}{\psi}^{\dagger}_{\vec{n}} \left[ V^{(i)}_{\vec{n}} \right] {\psi}_{\vec{n}}. \label{eq:H_low_general}
\end{equation}
Here $\vec{n} = (n_{1},\cdots,n_{d})$, $\vec{m} = (m_{1},\cdots,m_{d})$ and $\vec{l} = (l_{1},\cdots,l_{d})\in \mathbb{Z}^{d}$ index lattice positions in reduced coordinates. 
For example, the vector $\vec{n}$ uniquely denotes the lattice site $\sum_{j=1}^d n_j\vec{a}_j$. 
The multi-component operator ${\psi}^{\dagger}_{\vec{n}}$ creates an electron at site $\vec{n}$ with a given set of spin and orbital degrees of freedom.
The matrix $\left[H_{\vec{m}}\right]$ is the unmodulated hopping matrix connecting lattice position $\vec{n}$ to $\vec{n}+\vec{m}$, with matrix indices encoding the spin and orbital dependence of the hopping. 
Similarly, $\left[H_{\vec{l},\vec{n}}^{(i)} \right]$ is the $i^{\text{th}}$ ($i = 1 ,\ldots, N$) modulated hopping matrix connecting lattice position $\vec{n}$ to $\vec{n}$+$\vec{l}$, and $\left[ V^{(i)}_{\vec{n}} \right]$ is the $i^{\text{th}}$ ($i = 1 ,\ldots, N$) modulated on-site energy at lattice position $\vec{n}$. 

Each $\vec{q}^{(i)}$ is associated with a set of modulated couplings $\left[V^{(i)}_{\vec{n}} \right]$ and $\left[ H^{(i)}_{\vec{l},\vec{n}} \right]$. 
Note it is possible that for a given $\vec{q}^{(i)}$, $\left[ H^{(i)}_{\vec{l},\vec{n}} \right]$ is only non-zero for some $\vec{l}$.
Here we sum over all $\vec{l}$, and set $\left[ H^{(i)}_{\vec{l},\vec{n}} \right] =0$ if for a given $\vec{q}^{(i)}$ there is no modulated hopping matrix with the given $\vec{l}$. 
In most practical cases of interest, $\vec{l}$ will be restricted to a nearest-neighbor hopping. 
Note that in the situations considered in the main text we took $\left[ H^{(i)}_{\vec{l},\vec{n}} \right]=0$ for all $\vec{l}$. 
Analogous considerations let us analyze systems with modulated hoppings but no modulated on-site potentials by taking $[V_{\vec{n}}^{(i)}]=0$.

We require $H_{\text{low-dim}}$ to be Hermitian, such that
\begin{align}
    & \left[ H_{-\vec{m}} \right]^{\dagger} = \left[H_{\vec{m}} \right],  \label{eq:Hermitian_1}  \\
    & \left[ H^{(i)}_{-\vec{l},\vec{n}+\vec{l}} \right]^{\dagger} = \left[ H^{(i)}_{\vec{l},\vec{n}} \right], \label{eq:Hermitian_2} \\
    &  \left[ V^{(i)}_{\vec{n}} \right]^{\dagger} = \left[ V^{(i)}_{\vec{n}} \right].\label{eq:Hermitian_3} 
\end{align}
Eq.~(\ref{eq:Hermitian_2}) is another statement that the hopping process from $\vec{n}$ to $\vec{n}+\vec{l}$ is the conjugate of the hopping process from $\vec{n}+\vec{l}$ to  $\vec{n}$. 
We assume that the modulations are periodic with functional dependence
\begin{align}
    & \left[ H_{\vec{l},\vec{n}}^{(i)} \right] = \left[ g^{(i)}_{\vec{l}} \left( 2\pi \vec{q}^{(i)} \cdot \left( n_{1}\vec{a}_{1}+\cdots+n_{d}\vec{a}_{d} \right) + \phi^{(i)} \right)\right], \label{eq:H_form_1} \\
    & \left[ V^{(i)}_{\vec{n}} \right] = \left[f^{(i)}  \left( 2\pi \vec{q}^{(i)} \cdot  \left( n_{1}\vec{a}_{1}+\cdots+n_{d}\vec{a}_{d} \right) + \phi^{(i)} \right)\right], \label{eq:H_form_2}
\end{align}
such that
\begin{align}
    & \left[ g^{(i)}_{\vec{l}} (x)\right] = \left[ g^{(i)}_{\vec{l}} (x + 2\pi)\right], \\
    & \left[f^{(i)} (x)\right] = \left[f^{(i)} (x+2\pi)\right].
\end{align}
Therefore, we can expand $\left[ H_{\vec{l},\vec{n}}^{(i)} \right]$ and $\left[ V^{(i)}_{\vec{n}} \right]$ in terms of Fourier components,
\begin{align}
    & \left[ H_{\vec{l},\vec{n}}^{(i)} \right] = \sum_{p } \left[ H_{\vec{l},p}^{(i)} \right] e^{i p \left( 2\pi \vec{q}^{(i)} \cdot \left(\sum_{j=1}^{d}n_{j}\vec{a}_{j} \right) + \phi^{(i)} \right) }, \label{eq:general_FT_1} \\
    &  \left[ V^{(i)}_{\vec{n}} \right] = \sum_{p } \left[ V^{(i)}_{p} \right] e^{i p \left( 2\pi \vec{q}^{(i)} \cdot \left(\sum_{j=1}^{d}n_{j}\vec{a}_{j} \right) + \phi^{(i)} \right) }, \label{eq:general_FT_2}
\end{align}
where $p \in \mathbb{Z}$. The matrices $\left[ H_{\vec{l},p}^{(i)} \right]$ and $\left[ V^{(i)}_{p} \right]$ are the $p^{\text{th}}$ (matrix-valued) Fourier components of $\left[ H_{\vec{l},\vec{n}}^{(i)} \right]$ and $\left[ V^{(i)}_{\vec{n}} \right]$, respectively. 
Eqs.~(\ref{eq:Hermitian_2}) and (\ref{eq:Hermitian_3}) also imply
\begin{align}
    & \left[ H^{(i)}_{-\vec{l},-p} \right]^{\dagger} e^{ip 2 \pi \vec{q}^{(i)} \cdot \sum_{j=1}^{d}l_{j}\vec{a}_{j}} = \left[ H^{(i)}_{\vec{l},p} \right], \label{eq:Hermitian_4}\\
    & \left[ V^{(i)}_{-p} \right]^{\dagger} = \left[ V^{(i)}_{p} \right]. \label{eq:Hermitian_5}
\end{align}

Although Eq.~(\ref{eq:Hermitian_4}) involves several different indices, it will be crucial in forming our dimensionally-promoted Hamiltonian. 
As such, it is illuminating to prove it as follows. 
Starting from Eq.~(\ref{eq:general_FT_1}), we can obtain
\begin{align}
    \left[ H_{-\vec{l},\vec{n}}^{(i)} \right] = \sum_{p } \left[ H_{-\vec{l},p}^{(i)} \right] e^{i p \left( 2\pi \vec{q}^{(i)} \cdot \left(\sum_{j=1}^{d}n_{j}\vec{a}_{j} \right) + \phi^{(i)} \right) }
\end{align}
by changing $\vec{l} \to -\vec{l}$. 
We next shift $\vec{n} \to \vec{n} + \vec{l}$ such that
\begin{align}
    \left[ H_{-\vec{l},\vec{n}+\vec{l}}^{(i)} \right] = \sum_{p } \left[ H_{-\vec{l},p}^{(i)} \right] e^{i p \left( 2\pi \vec{q}^{(i)} \cdot \left(\sum_{j=1}^{d}n_{j}\vec{a}_{j} \right) + \phi^{(i)} \right) } e^{ip 2\pi \vec{q}^{(i)} \cdot \sum_{j=1}^{d}l_{j}\vec{a}_{j}}. \label{eq:general_before_Hermitian_1}
\end{align}
Taking Hermitian conjugate of Eq.~(\ref{eq:general_before_Hermitian_1}) and setting $p \to -p$, we have
\begin{align}
    \left[ H_{-\vec{l},\vec{n}+\vec{l}}^{(i)} \right]^{\dagger} = \sum_{p } \left[ H_{-\vec{l},-p}^{(i)} \right]^{\dagger} e^{i p \left( 2\pi \vec{q}^{(i)} \cdot \left(\sum_{j=1}^{d}n_{j}\vec{a}_{j} \right) + \phi^{(i)} \right) } e^{ip 2\pi \vec{q}^{(i)} \cdot \sum_{j=1}^{d}l_{j}\vec{a}_{j}}. \label{eq:pf_1}
\end{align}
In order for Eq.~(\ref{eq:pf_1}) being consistent with the Hermiticity constraint Eq.~(\ref{eq:Hermitian_2}), we deduce that Eq.~(\ref{eq:Hermitian_4}) must hold.

To continue with our development of the dimensional promotion procedure, we plug Eq.~(\ref{eq:general_FT_1}) and Eq.~(\ref{eq:general_FT_2}) into Eq.~(\ref{eq:H_low_general}) such that Hamiltonian of the $d$D modulated system can be written as
\begin{align}
    H_{\text{low-dim}} & = \sum_{\vec{n},\vec{m}} {\psi}^{\dagger}_{\vec{n}+\vec{m}} \left[ H_{\vec{m}}\right] {\psi}_{\vec{n}}  \nonumber \\
    & + {\sum_{\vec{n},\vec{l}}}  \sum_{i=1}^{N} \sum_{p_{i}} {\psi}^{\dagger}_{\vec{n}+\vec{l}} \left[H_{\vec{l},p_{i}}^{(i)}\right] e^{i  p_{i} \left( 2\pi \vec{q}^{(i)} \cdot  \left(\sum_{j=1}^{d}n_{j}\vec{a}_{j} \right) + \phi^{(i)} \right) } {\psi}_{\vec{n}} \nonumber \\
    & + \sum_{\vec{n}} \sum_{i=1}^{N} \sum_{p_{i}}{\psi}^{\dagger}_{\vec{n}} \left[ V^{(i)}_{p_{i}} \right] e^{i  p_{i} \left( 2\pi \vec{q}^{(i)} \cdot  \left(\sum_{j=1}^{d}n_{j}\vec{a}_{j} \right) + \phi^{(i)} \right) } {\psi}_{\vec{n}}. \label{eq:H_low_general_2}
\end{align}
Notice that for each $i$ (which indexes the modulation wave vectors) we sum over all $p_{i} \in \mathbb{Z}$. 
To promote Eq.~(\ref{eq:H_low_general_2}) to a $(d+N)$D space, let us introduce a set of additional lattice vectors $\{\vec{c}_{1},\cdots,\vec{c}_{N} \}$ such that, together with $\{\vec{a}_{1},\cdots,\vec{a}_{d} \}$, we have a linearly independent basis spanning a $(d+N)$D lattice. 
Notice that this $(d+N)$D lattice is not necessarily orthorhombic, since we do not require the lattice vectors $\{\vec{a}_{1},\cdots,\vec{a}_{d} \}$ and $\{\vec{c}_{1},\cdots,\vec{c}_{N} \}$ be pairwise orthogonal. 
Next, we introduce the corresponding reciprocal lattice vectors $\{\vec{g}_{1},\cdots,\vec{g}_{d}\}$ and $\{\vec{G}_{1},\cdots,\vec{G}_{N}\}$ in the $(d+N)$D reciprocal space such that
\begin{align}
    & \vec{g}_{i} \cdot \vec{a}_{j} = 2 \pi \delta_{ij},\ i = 1 ,\ldots, d,\ {\text{and }}j = 1 ,\ldots, d, \label{eq:reciprocal_1} \\
    & \vec{g}_{i} \cdot \vec{c}_{j} = 0,\ i = 1 ,\ldots, d,\ {\text{and }}j = 1 ,\ldots, N, \label{eq:reciprocal_2} \\
    & \vec{G}_{i} \cdot \vec{a}_{j} = 0,\ i = 1 ,\ldots, N,\ {\text{and }}j = 1 ,\ldots, d, \label{eq:reciprocal_3} \\
    & \vec{G}_{i} \cdot \vec{c}_{j} = 2\pi \delta_{ij},\ i = 1 ,\ldots, N,\ {\text{and }}j = 1 ,\ldots, N. \label{eq:reciprocal_4}
\end{align}
We can now identify $\phi^{(i)} \in [0,2\pi)$ with $\vec{k} \cdot \vec{c}_{i}$, which is $2\pi$ times the coefficient of $\vec{k}$ along $\vec{G}_{i}$ (see Eq.~(\ref{eq:reciprocal_4})); the periodicity of $\phi^{(i)}$ is reflected in the periodicity of the $(d+N)$D Brillouin zone. 
We then promote the $d$D model to a $(d+N)$D space by summing over $\{ \vec{k}\cdot \vec{c}_{1},\cdots,\vec{k}\cdot \vec{c}_{N} \} \in \mathbb{T}^{N}$, where $\mathbb{T}^N$ denotes the $N$-torus represented as a parallelepiped with boundary spanned by $\{\vec{G}_{1},\cdots,\vec{G}_{N} \}$, and opposite edges identified. 
We further label the original creation and annihilation operators by additional symbols $\{ \vec{k}\cdot \vec{c}_{1},\cdots,\vec{k}\cdot \vec{c}_{N} \} \in \mathbb{T}^{N}$, such that
\begin{align}
    H_{\text{high-dim}} =&\sum_{\vec{k}\cdot\vec{c}_{1},\cdots,\vec{k}\cdot\vec{c}_{N}} \sum_{\vec{n},\vec{m}}  {\psi}^{\dagger}_{\vec{n}+\vec{m},\vec{k}\cdot \vec{c}_{1},\cdots,\vec{k}\cdot \vec{c}_{N}} \left[H_{\vec{m}}\right] \psi_{\vec{n},\vec{k}\cdot \vec{c}_{1},\cdots,\vec{k}\cdot \vec{c}_{N}}   \nonumber \\
    & + \sum_{\vec{k}\cdot\vec{c}_{1},\cdots,\vec{k}\cdot\vec{c}_{N}} {\sum_{\vec{n},\vec{l}}}  \sum_{i=1}^{N} \sum_{p_{i}}  {\psi}^{\dagger}_{\vec{n}+\vec{l},\vec{k}\cdot \vec{c}_{1},\cdots,\vec{k}\cdot \vec{c}_{N}}  \left[H_{\vec{l},p_{i}}^{(i)}\right] e^{ip_{i}\vec{k}\cdot\vec{c}_{i}} e^{i  p_{i}  2\pi \vec{q}^{(i)} \cdot \left(\sum_{j=1}^{d}n_{j}\vec{a}_{j} \right)   } {\psi}_{\vec{n},\vec{k}\cdot \vec{c}_{1},\cdots,\vec{k}\cdot \vec{c}_{N}} \nonumber \\
    & +  \sum_{\vec{k}\cdot\vec{c}_{1},\cdots,\vec{k}\cdot\vec{c}_{N}}\sum_{\vec{n}}  \sum_{i=1}^{N} \sum_{p_{i}}  {\psi}^{\dagger}_{\vec{n},\vec{k}\cdot \vec{c}_{1},\cdots,\vec{k}\cdot \vec{c}_{N}} \left[V^{(i)}_{p_{i}}\right] e^{ip_{i}\vec{k}\cdot\vec{c}_{i}} e^{i2\pi p_{i} \vec{q}^{(i)}\cdot \left(\sum_{j=1}^{d}n_{j}\vec{a}_{j} \right)}{\psi}_{\vec{n},\vec{k}\cdot \vec{c}_{1},\cdots,\vec{k}\cdot \vec{c}_{N}}, \label{eq:non_ortho_3}
\end{align}
where the summation of each $\vec{k} \cdot \vec{c}_{i}$ is from $0$ to $2\pi$. 
Notice that the summation over $\{\vec{k}\cdot \vec{c}_{1},\cdots,\vec{k} \cdot \vec{c}_{N} \}$ is outside of $\sum_{i=1}^{N} \sum_{p_{i}} $ over all integer $p_{i}$ for each modulation $i = 1 ,\ldots, N$. 
This means that we are summing over $\{\vec{k}\cdot \vec{c}_{1},\cdots, \vec{k}\cdot\vec{c}_{N} \}$ for each term in Eq.~(\ref{eq:H_low_general_2}). 
As mentioned in the main text, the physical motivation to sum over $\{\vec{k}\cdot \vec{c}_{1},\cdots,\vec{k}\cdot\vec{c}_{N} \}$ is because a single set of $\{\vec{k}\cdot \vec{c}_{1},\cdots,\vec{k}\cdot\vec{c}_{N} \}$ does not contain the full information of the lattice model in the promoted $(d+N)$D space. 
Only when we consider all values of $\{\vec{k}\cdot \vec{c}_{1},\cdots,\vec{k}\cdot\vec{c}_{N} \}$, can we obtain the exact form of the promoted lattice model, shown in Eq.~(\ref{eq:non_ortho_4}) below. 
We remind the readers that repeated indices are not implicitly summed.

Note that ${\psi}^{\dagger}_{\vec{n},\vec{k}\cdot \vec{c}_{1},\cdots,\vec{k}\cdot \vec{c}_{N}}$ is the Fourier transform of $\psi^{\dagger}_{\vec{n},\vec{\nu}}$ through
\begin{align}
    {\psi}^{\dagger}_{\vec{n},\vec{k}\cdot \vec{c}_{1},\cdots,\vec{k}\cdot \vec{c}_{N}} = \frac{1}{\sqrt{L}} \sum_{\vec{\nu}} e^{i \vec{k}\cdot\left( \nu_{1}\vec{c}_{1}+ \cdots + \nu_{N}\vec{c}_{N} \right)} \psi^{\dagger}_{\vec{n},\vec{\nu}}, \label{eq:FT_general}
\end{align}
where $\vec{\nu} = (\nu_{1},\cdots,\nu_{N}) \in \mathbb{Z}^{N}$ and $L$ is the size of the system in the additional dimensions (taken to infinity at the end of the calculation). We can thus take an inverse Fourier transform to find
\begin{align}
    H_{\text{high-dim}} =& \sum_{\vec{n},\vec{m},\vec{\nu}}  {\psi}^{\dagger}_{\vec{n}+\vec{m},\vec{\nu}} \left[ H_{\vec{m}}\right] \psi_{\vec{n},\vec{\nu}}   \nonumber \\
    & + {\sum_{\vec{n},\vec{l},\vec{\nu}}}  \sum_{i=1}^{N} \sum_{p_{i}}  {\psi}^{\dagger}_{\vec{n}+\vec{l},\vec{\nu}-p_{i}\hat{\nu}_{i}}  \left[H_{\vec{l},p_{i}}^{(i)}\right] e^{i  2\pi p_{i}   \vec{q}^{(i)} \cdot \left(\sum_{j=1}^{d}n_{j}\vec{a}_{j} \right)   } {\psi}_{\vec{n},\vec{\nu}} \nonumber \\
    & + \sum_{\vec{n},\vec{\nu}}\sum_{i=1}^{N} \sum_{p_{i}} {\psi}^{\dagger}_{\vec{n},\vec{\nu}-p_{i}\hat{\nu}_{i}} \left[V^{(i)}_{p_{i}}\right]  e^{i2\pi p_{i} \vec{q}^{(i)}\cdot \left(\sum_{j=1}^{d}n_{j}\vec{a}_{j} \right)}{\psi}_{\vec{n},\vec{\nu}}, \label{eq:non_ortho_4}
\end{align}
where ${\psi}^{\dagger}_{\vec{n},\vec{\nu}}$ denotes the electron creation operator for an electron at the lattice position $\sum_{j=1}^{d}n_{j}\vec{a}_{j} + \sum_{j=1}^{N}\nu_{j}\vec{c}_{j}$, and ${\psi}^{\dagger}_{\vec{n}+\vec{l},\vec{\nu} - p_{i}\hat{\nu}_{i}}$ denotes the electron creation operator for an electron at the lattice position $\sum_{j=1}^{d}(n_{j} + l_{j})\vec{a}_{j} + \sum_{j=1}^{N}\nu_{j}\vec{c}_{j} - p_{i} \vec{c}_{i}$. 
Notice that a $d$D modulated system with phase offsets $\{\phi^{(1)},\cdots,\phi^{(N)}\}$ is described by the Bloch Hamiltonian (see Eq.~(\ref{eq:non_ortho_3})) of the promoted $(d+N)$D lattice model with fixed crystal momenta $\{\vec{k}\cdot \vec{c}_{1},\cdots,\vec{k}\cdot \vec{c}_{N} \}$ via our identification of $\vec{k} \cdot \vec{c}_{i}$ with $\phi^{(i)}$. 
For later convenience, we can also rewrite Eq.~(\ref{eq:non_ortho_4}) as
\begin{align}
    H_{\text{high-dim}} =& \sum_{\vec{n},\vec{m},\vec{\nu}}  {\psi}^{\dagger}_{\vec{n}+\vec{m},\vec{\nu}} \left[H_{\vec{m}}\right] \psi_{\vec{n},\vec{\nu}}    \label{eq:non_ortho_5_1} \\
    & + {\sum_{\vec{n},\vec{l},\vec{\nu}}} \sum_{i=1}^{N} \sum_{p_{i}}  {\psi}^{\dagger}_{\vec{n}+\vec{l},\vec{\nu}-p_{i}\hat{\nu}_{i}}  \left[H_{\vec{l},p_{i}}^{(i)}\right]e^{-i \pi p_{i} \vec{q}^{(i)}\cdot \left( \sum_{j=1}^{d} l_{j}\vec{a}_{j} \right) }e^{i 2\pi p_{i}   \vec{q}^{(i)} \cdot \left(\sum_{j=1}^{d}\left( n_{j} + \frac{1}{2}l_{j} \right)\vec{a}_{j} \right)   } {\psi}_{\vec{n},\vec{\nu}} \label{eq:non_ortho_5_2} \\
    & + \sum_{\vec{n},\vec{\nu}}\sum_{i=1}^{N} \sum_{p_{i}} {\psi}^{\dagger}_{\vec{n},\vec{\nu}-p_{i}\hat{\nu}_{i}} \left[V^{(i)}_{p_{i}}\right]  e^{i2\pi p_{i} \vec{q}^{(i)}\cdot \left(\sum_{j=1}^{d}n_{j}\vec{a}_{j} \right)}{\psi}_{\vec{n},\vec{\nu}}, \label{eq:non_ortho_5_3}
\end{align}
where we have used
\begin{align}
    \sum_{j=1}^{d}n_{j}\vec{a}_{j} = \sum_{j=1}^{d}\left( n_{j} + \frac{1}{2}l_{j} \right)\vec{a}_{j} - \sum_{j=1}^{d} \frac{1}{2}l_{j} \vec{a}_{j}.
\end{align}
This will prove useful when we go to identify the gauge fields appearing in the hopping matrix elements. 
We remind the readers again that the product $p_{i}\vec{q}^{(i)}$ in Eqs.~(\ref{eq:non_ortho_5_1}--\ref{eq:non_ortho_5_3}) does not imply a summation over $i$, rather it denotes the product of integer $p_{i}$ and the $i^{th}$ modulation wave vector $\vec{q}^{(i)}$. 
Eqs.~(\ref{eq:non_ortho_5_1}--\ref{eq:non_ortho_5_3}) can then be interpreted as a $(d+N)$D lattice model with Hamiltonian
\begin{align}
    H_{\text{high-dim}} =& \sum_{\vec{n},\vec{m},\vec{\nu}}  {\psi}^{\dagger}_{\vec{n}+\vec{m},\vec{\nu}} \left[H_{\vec{m}}\right] \psi_{\vec{n},\vec{\nu}}    \label{eq:non_ortho_6_1} \\
    & + {\sum_{\vec{n},\vec{l},\vec{\nu}}} \sum_{i=1}^{N} \sum_{p_{i}}  {\psi}^{\dagger}_{\vec{n}+\vec{l},\vec{\nu}-p_{i}\hat{\nu}_{i}}  \left[H_{\vec{l},p_{i}}^{(i)}\right]e^{-i \pi p_{i} \vec{q}^{(i)}\cdot \left( \sum_{j=1}^{d} l_{j}\vec{a}_{j} \right) }{\psi}_{\vec{n},\vec{\nu}} \label{eq:non_ortho_6_2} \\
    & + \sum_{\vec{n},\vec{\nu}}\sum_{i=1}^{N} \sum_{p_{i}} {\psi}^{\dagger}_{\vec{n},\vec{\nu}-p_{i}\hat{\nu}_{i}} \left[V^{(i)}_{p_{i}}\right] {\psi}_{\vec{n},\vec{\nu}}, \label{eq:non_ortho_6_3}
\end{align}
coupled through a Peierls substitution\cite{Peierls_substitution} to a $U(1)$ gauge field 
\begin{align}
    \vec{A} = \frac{1}{2\pi} \sum_{i=1}^{N} \sum_{j=1}^{d} \vec{G}_{i} \left( \vec{q}^{(i)}\cdot \left[ \left( \vec{r} \cdot \vec{g}_{j} \right) \vec{a}_{j} \right] \right), \label{eq:non_ortho_A}
\end{align}
where $\vec{r} \in \mathbb{R}^{d+N}$. 
Notice that Eqs.~(\ref{eq:non_ortho_6_1}--\ref{eq:non_ortho_6_3}) represent the $(d+N)$D model {\it without} $U(1)$ gauge fields.
Although there are complex phases $e^{-i \pi p_{i} \vec{q}^{(i)}\cdot \left( \sum_{j=1}^{d} l_{j}\vec{a}_{j} \right) }$ in Eq.~(\ref{eq:non_ortho_6_2}), they do not depend on the reduced coordinates $(\vec{n},\vec{\nu}) \in \mathbb{Z}^{d+N}$ in the dimensionally-promoted lattice and so may be regarded as inherent phase factors in the $(d+N)$D model without $U(1)$ gauge field. 
To validate the identification Eq.~(\ref{eq:non_ortho_A}), we compute the various Peierls phase factors in the next section. 

\subsubsection{\label{sec:computing_Peierls_phase_1}Computation of Peierls phases}

Let us consider the terms in Eq.~(\ref{eq:non_ortho_6_3}) where the electrons hop from 
\begin{equation}
\vec{r}_{i} = \sum_{j=1}^{d}n_{j}\vec{a}_{j} + \sum_{j=1}^{N}\nu_{j}\vec{c}_{j}
\end{equation}
to 
\begin{equation}
 \vec{r}_{f} =\sum_{j=1}^{d}n_{j}\vec{a}_{j} + \sum_{j=1}^{N}\nu_{j}\vec{c}_{j} - p_{i} \vec{c}_{i}. 
\end{equation}
The Peierls phase can be computed from a straight line integral from $\vec{r}_{i}$ to $\vec{r}_{f} \in$ $\mathbb{R}^{d+N}$ through
\begin{align}
    \exp{-i \int_{\vec{r}_{i}}^{\vec{r}_{f}} d\vec{r} \cdot \vec{A}},
\end{align}
where we have worked in unit $\hbar = c= |e| = 1$ and the electron has charge $-|e| = -1$. 
The straight line connecting $\vec{r}_{i}$ to $\vec{r}_{f}$ is
\begin{align}
    \sum_{j=1}^{d}n_{j}\vec{a}_{j} + \sum_{j=1}^{N}\nu_{j}\vec{c}_{j} - p_{i} \vec{c}_{i}t,
\end{align}
where $ t\in [0,1]$ and the corresponding infinitesimal displacement vector is $-p_{i}\vec{c}_{i}dt$. 
The line integral can then be computed as follows:
\begin{align}
    \exp{-i \int_{\vec{r}_{i}}^{\vec{r}_{f}} d\vec{r} \cdot \vec{A}} & =  \exp{i \int_{0}^{1}dt  p_{i} \vec{c}_{i} \cdot \frac{1}{2\pi} \sum_{j=1}^{N} \sum_{k=1}^{d} \vec{G}_{j} \left( \vec{q}^{(j)}\cdot \left[ \left( \left( \sum_{l=1}^{d}n_{l}\vec{a}_{l} + \sum_{l=1}^{N}\nu_{l}\vec{c}_{l} - p_{i} \vec{c}_{i}t \right) \cdot \vec{g}_{k} \right) \vec{a}_{k} \right] \right) } \label{eq:Peierls_1_1} \\
    & = \exp{i \int_{0}^{1}dt  p_{i}  \sum_{j=1}^{N} \sum_{k=1}^{d} \delta_{ij} \left( \vec{q}^{(j)}\cdot \left[ \left( \left( \sum_{l=1}^{d}n_{l}\vec{a}_{l}  \right) \cdot \vec{g}_{k} \right) \vec{a}_{k} \right] \right) } \label{eq:Peierls_1_2} \\
    & = \exp{i 2\pi \int_{0}^{1}dt  p_{i}   \sum_{k=1}^{d} \left( \vec{q}^{(i)}\cdot \left[ \sum_{l=1}^{d}n_{l}\delta_{lk}    \vec{a}_{k} \right] \right) } \label{eq:Peierls_1_3} \\
    & = \exp{i 2\pi \int_{0}^{1}dt  p_{i}   \left( \vec{q}^{(i)}\cdot \sum_{k=1}^{d}  n_{k}  \vec{a}_{k}  \right) } \label{eq:Peierls_1_4} \\
    & = \exp{i 2\pi  p_{i}    \vec{q}^{(i)}\cdot \sum_{k=1}^{d}  n_{k}  \vec{a}_{k}  }. \label{eq:Peierls_1_5}
\end{align}
We used Eq.~(\ref{eq:reciprocal_2}) and Eq.~(\ref{eq:reciprocal_4}) in going from Eq.~(\ref{eq:Peierls_1_1}) to Eq.~(\ref{eq:Peierls_1_2}), Eq.~(\ref{eq:reciprocal_1}) in going from Eq.~(\ref{eq:Peierls_1_2}) to Eq.~(\ref{eq:Peierls_1_3}) and then finally do an integral $\int_{0}^{1}dt=1$ to obtain Eq.~(\ref{eq:Peierls_1_5}). 
Eq.~(\ref{eq:Peierls_1_5}) is exactly the additional phase factor in Eq.~(\ref{eq:non_ortho_5_3}) compared to Eq.~(\ref{eq:non_ortho_6_3}).

Next, consider Eq.~(\ref{eq:non_ortho_6_2}) where the electrons hop from 
\begin{equation}
\vec{r}_{i} = \sum_{j=1}^{d}n_{j}\vec{a}_{j} + \sum_{j=1}^{N}\nu_{j}\vec{c}_{j}
\end{equation}
to 
\begin{equation}
\vec{r}_{f} =\sum_{j=1}^{d}n_{j}\vec{a}_{j} + \sum_{j=1}^{N}\nu_{j}\vec{c}_{j} + \sum_{j=1}^{d}l_{j}\vec{a}_{j} - p_{i} \vec{c}_{i}.
\end{equation} 
The straight line connecting $\vec{r}_{i}$ to $\vec{r}_{f}$ is
\begin{align}
    \sum_{j=1}^{d}n_{j}\vec{a}_{j} + \sum_{j=1}^{N}\nu_{j}\vec{c}_{j} + \left(  \sum_{j=1}^{d}l_{j}\vec{a}_{j} - p_{i} \vec{c}_{i} \right) t,
\end{align}
where $ t= [0,1]$ and the corresponding infinitesimal displacement vector is $\left(  \sum_{j=1}^{d}l_{j}\vec{a}_{j} - p_{i} \vec{c}_{i} \right) dt$. 
The line integral can then be computed as follows:
\begin{align}
    & \exp{-i \int_{\vec{r}_{i}}^{\vec{r}_{f}} d\vec{r} \cdot \vec{A}} \\
    & =  \exp{-i \int_{0}^{1}dt  \left(  \sum_{r=1}^{d}l_{r}\vec{a}_{r} - p_{i} \vec{c}_{i} \right) \cdot \frac{1}{2\pi} \sum_{j,k} \vec{G}_{j} \left( \vec{q}^{(j)}\cdot \left[ \left( \left( \sum_{r=1}^{d}n_{r}\vec{a}_{r} + \sum_{r=1}^{N}\nu_{r}\vec{c}_{r} + \left(  \sum_{r=1}^{d}l_{r}\vec{a}_{r} - p_{i} \vec{c}_{i} \right) t \right) \cdot \vec{g}_{k} \right) \vec{a}_{k} \right] \right) } \label{eq:Peierls_2_1} \\
    & =  \exp{i \int_{0}^{1}dt  p_{i} \vec{c}_{i} \cdot \frac{1}{2\pi} \sum_{j,k} \vec{G}_{j} \left( \vec{q}^{(j)}\cdot \left[ \left( \left( \sum_{r=1}^{d}\left( n_{r} + l_{r}t \right)\vec{a}_{r}   \right) \cdot \vec{g}_{k} \right) \vec{a}_{k} \right] \right) } \label{eq:Peierls_2_2} \\
    & =  \exp{i 2\pi \int_{0}^{1}dt  p_{i}   \sum_{j,k} \delta_{ij} \left( \vec{q}^{(j)}\cdot \left[  \sum_{r=1}^{d}\left( n_{r} + l_{r}t \right)\delta_{rk}  \vec{a}_{k} \right] \right) } \label{eq:Peierls_2_3} \\
    & =  \exp{i 2\pi \int_{0}^{1}dt  p_{i}   \sum_{k=1}^{d}  \left( \vec{q}^{(i)}\cdot \left[  \left( n_{k} + l_{k}t   \right) \vec{a}_{k} \right] \right) } \label{eq:Peierls_2_4} \\
    & =  \exp{i 2\pi  p_{i}  \vec{q}^{(i)}\cdot \left( \sum_{k=1}^{d}   \left( n_{k} + \frac{1}{2}l_{k}   \right) \vec{a}_{k} \right) }. \label{eq:Peierls_2_5}
\end{align}
We used Eq.~(\ref{eq:reciprocal_2}) and Eq.~(\ref{eq:reciprocal_3}) in going from Eq.~(\ref{eq:Peierls_2_1}) to Eq.~(\ref{eq:Peierls_2_2}), Eq.~(\ref{eq:reciprocal_1}) and Eq.~(\ref{eq:reciprocal_4}) in going from Eq.~(\ref{eq:Peierls_2_2}) to Eq.~(\ref{eq:Peierls_2_3}) and then finally do integrals $\int_{0}^{1}dt=1$ and $\int_{0}^{1}tdt=\frac{1}{2}$ to obtain Eq.~(\ref{eq:Peierls_2_5}). 
Eq.~(\ref{eq:Peierls_2_5}) is exactly the additional phase factor in Eq.~(\ref{eq:non_ortho_5_2}) compared to Eq.~(\ref{eq:non_ortho_6_2}). 
Crucially, we see that our redefinition Eq.~(\ref{eq:non_ortho_A}) accounts for the factor of $1/2$ arising in the line integral of the vector potential.

Finally, consider Eq.~(\ref{eq:non_ortho_6_1}) where the electrons hop from 
\begin{equation}
\vec{r}_{i} = \sum_{j=1}^{d}n_{j}\vec{a}_{j} + \sum_{j=1}^{N}\nu_{j}\vec{c}_{j}
\end{equation}
to
\begin{equation} 
 \vec{r}_{f} =\sum_{j=1}^{d}n_{j}\vec{a}_{j} + \sum_{j=1}^{N}\nu_{j}\vec{c}_{j} + \sum_{j=1}^{d}m_{j}\vec{a}_{j}.
\end{equation}
The straight line connecting $\vec{r}_{i}$ to $\vec{r}_{f}$ is
\begin{align}
    \sum_{j=1}^{d}n_{j}\vec{a}_{j} + \sum_{j=1}^{N}\nu_{j}\vec{c}_{j} + \left(  \sum_{j=1}^{d}m_{j}\vec{a}_{j} \right) t,
\end{align}
where $ t= [0, 1]$ and the corresponding infinitesimal displacement vector is $\left(  \sum_{j=1}^{d}m_{j}\vec{a}_{j} \right) dt$. 
The line integral can then be computed as follows:
\begin{align}
    & \exp{-i \int_{\vec{r}_{i}}^{\vec{r}_{f}} d\vec{r} \cdot \vec{A}} \\
    & =  \exp{-i \int_{0}^{1}dt  \left(  \sum_{r=1}^{d}m_{r}\vec{a}_{r} \right) \cdot \frac{1}{2\pi} \sum_{j,k} \vec{G}_{j} \left( \vec{q}^{(j)}\cdot \left[ \left( \left( \sum_{r=1}^{d}n_{r}\vec{a}_{r} + \sum_{r=1}^{N}\nu_{r}\vec{c}_{r} + \left(  \sum_{r=1}^{d}m_{r}\vec{a}_{r} \right) t \right) \cdot \vec{g}_{k} \right) \vec{a}_{k} \right] \right) } \label{eq:Peierls_3_1} \\
    & = 1. \label{eq:Peierls_3_2}
\end{align}
We used Eq.~(\ref{eq:reciprocal_3}) in going from Eq.~(\ref{eq:Peierls_3_1}) to Eq.~(\ref{eq:Peierls_3_2}). 
This means that no additional phase factors arise if we compare Eq.~(\ref{eq:non_ortho_5_1}) and Eq.~(\ref{eq:non_ortho_6_1}). 
Thus, we see that our dimensionally-promoted Hamiltonian is consistent with Eqs.~(\ref{eq:non_ortho_6_1}--\ref{eq:non_ortho_6_3}) coupled via a Peierls substitution to the vector potential Eq.~(\ref{eq:non_ortho_A}).

\subsubsection{\label{sec:remarks_consistency_check_and_summary}Remarks, consistency checks and summary}

Let us briefly recap what we have developed. 
We began with a $d$D system with modulated hoppings and on-site energies, together with lattice vectors $\{\vec{a}_{1},\cdots,\vec{a}_{d}\}$. 
Upon dimensional promotion, we can choose {\it any} linearly independent additional lattice vectors $\{\vec{c}_{1},\cdots,\vec{c}_{N}\}$ which together with $\{\vec{a}_{1},\cdots,\vec{a}_{d}\}$ span the promoted $(d+N)$D space. 
The corresponding $U(1)$ gauge field is given in Eq.~(\ref{eq:non_ortho_A}), which has linear dependence on $\vec{r}$, the position vector in the promoted $(d+N)$D space. 
This implies the field strength $F_{\mu \nu} = \partial_{\mu} A_{\nu} - \partial_{\nu} A_{\mu}$ is constant in space. 
If we choose $\{\vec{a}_{1},\cdots,\vec{a}_{d},\vec{c}_{1},\cdots,\vec{c}_{N} \}$ to be orthogonal unit vectors in the $(d+N)$D Cartesian coordinates, and if the model only has modulated on-site energies, the general dimensional promotion procedure present here will reduce to the one present in Sec.~III of the main text.

In this more general construction in terms of non-orthogonal lattice vectors with modulated hopping terms, the mapping from the $d$D modulated system to the promoted $(d+N)$D lattice coupled to a $U(1)$ gauge field also requires no additional parameters. 
The Hamiltonians before and after the dimensional promotion, together with the convention for the Fourier series expansions are summarized in Table~\ref{tab:model_summary}. 
The hopping matrices in $(d+N)$D space, and the corresponding Peierls phases are summarized in Table~\ref{tab:hopping_1}. 
To use Table~\ref{tab:hopping_1} we multiply the hopping matrix entry and the Peierls phases to obtain the Hamiltonian matrix elements (with background $U(1)$ gauge fields) in the promoted $(d+N)$D space in Eqs.~(\ref{eq:non_ortho_5_1}--\ref{eq:non_ortho_5_3}). 
We remind the readers that there are phase factors $e^{-i p_{i} \pi \vec{q}^{(i)}\cdot \left( \sum_{j=1}^{d}l_{j}\vec{a}_{j} \right)}$ in the hopping matrix from $(\vec{n},\vec{\nu})$ to $(\vec{n}+\vec{l},\vec{\nu}-p_{i}\hat{\nu}_{i})$, as shown in Table~\ref{tab:hopping_1}. 
This means that these phase factors are included in the definition of the hopping matrices in the $(d+N)$D model, in addition to the Peierls phase factor. 
In addition, we reemphasize that the phases $\phi^{(i)}$ correspond to $\vec{k}\cdot \vec{c}_{i}$ where $\vec{k}$ is the crystal momentum $\in \mathbb{T}^{N}$ in the dimensionally-promoted Brillouin zone.
The modulation wave vectors $\vec{q}^{(i)}$ enter the definition of the $(d+N)$D $U(1)$ gauge fields in Eq.~(\ref{eq:non_ortho_A}).

\begin{table}[h]
\centering
\begin{tabular}{|c|c|}
\hline
Original $d$D modulated system & Eq.~(\ref{eq:H_low_general}) \\
\hline
Fourier expansion convention of on-site and hopping modulations & Eqs.~(\ref{eq:general_FT_1}--\ref{eq:general_FT_2})\\
\hline
Promoted $(d+N)$D system with $U(1)$ gauge fields & Eqs.~(\ref{eq:non_ortho_5_1}--\ref{eq:non_ortho_5_3}) \\
\hline
Promoted $(d+N)$D system without $U(1)$ gauge fields & Eqs.~(\ref{eq:non_ortho_6_1}--\ref{eq:non_ortho_6_3}) \\
\hline
\end{tabular}
\caption{Relevant equations in the general dimensional promotion formalism.}
\label{tab:model_summary}
\end{table}

\begin{table}[h]
\centering
\begin{tabular}{|c|c|c|}
\hline
Hopping from $(\vec{n},\vec{\nu})$ to & Hopping matrices & Peierls phases  \\
\hline
\hline
$(\vec{n}+\vec{m},\vec{\nu})$ & $\left[H_{\vec{m}}\right]$ & $1$ \\
\hline
$(\vec{n},\vec{\nu}-p_{i}\hat{\nu}_{i})$ & $ \left[V^{(i)}_{p_{i}}\right] $ & $e^{i2\pi p_{i} \vec{q}^{(i)}\cdot \left(\sum_{j=1}^{d}n_{j}\vec{a}_{j} \right)}$ \\
\hline
$(\vec{n}+\vec{l},\vec{\nu}-p_{i}\hat{\nu}_{i})$ & $\left[H_{\vec{l},p_{i}}^{(i)}\right]e^{-i \pi p_{i} \vec{q}^{(i)}\cdot \left( \sum_{j=1}^{d} l_{j}\vec{a}_{j} \right) }$ & $e^{i 2\pi p_{i}   \vec{q}^{(i)} \cdot \left(\sum_{j=1}^{d}\left( n_{j} + \frac{1}{2}l_{j} \right)\vec{a}_{j} \right)   }$ \\
\hline
\end{tabular}
\caption{Hopping terms in the promoted $(d+N)$D model and the corresponding Peierls phases, expressed in terms of parameters from the $d$D modulated system in Eq.~(\ref{eq:H_low_general}). 
Notice that $p_{i}\hat{\nu}_{i}$ does not imply a summation over $i$.}
\label{tab:hopping_1}
\end{table}

Before moving on, some additional remarks are in order. 
First, the vector potential in Eq.~(\ref{eq:non_ortho_A}) satisfies
\begin{align}
    \vec{A}\left( \vec{r} \right) = \vec{A}\left( \vec{r} + \sum_{j=1}^{N} \nu_{j} \vec{c}_{j} \right)
\end{align}
for any $\{\nu_{1},\cdots,\nu_{N} \} \in \mathbb{Z}^{N}$, which is a direct consequence of Eq.~(\ref{eq:reciprocal_2}), and corresponds to a generalized Landau gauge condition. 
This allows us to recover our low dimensional modulated system, by  Fourier transforming the higher-dimensional model Eq.~(\ref{eq:non_ortho_4}) along $\vec{k}$ in the subspace spanned by $\{\vec{G}_{1},\cdots,\vec{G}_{N} \} \in \mathbb{T}^{N}$ and regarding $\vec{k} \cdot \vec{c}_{i}$ as $\phi^{(i)}$. 
Any fixed value of $\{ \phi^{(1)},\cdots,\phi^{(N)} \}$ then describes a lower-dimensional modulated system with fixed phase offsets. 
In other words, we can use a $d$D modulated system with controllable phase offsets $\{ \phi^{(1)},\cdots,\phi^{(N)} \}$ to map out, by varying the values of $\phi^{(i)}$, the whole spectrum of the promoted $(d+N)$D model coupled to a $U(1)$ gauge field. 
Second, the constraints Eq.~(\ref{eq:Hermitian_1}), Eq.~(\ref{eq:Hermitian_4}) and Eq.~(\ref{eq:Hermitian_5}) ensures that $H_{\text{high-dim}}$ in Eq.~(\ref{eq:non_ortho_4}) with Peierls phases, as well as Eqs.~(\ref{eq:non_ortho_6_1}--\ref{eq:non_ortho_6_3}) without Peierls phases are Hermitian. 
We will make use of this in the subsequent examples and only list non-redundant matrix elements of the Hamiltonian. 
Third, we have assumed so far that  for a given modulation $\vec{q}^{(i)}$, the phase offsets of the on-site and hopping modulation are the same (see Eqs.~(\ref{eq:H_form_1}) and (\ref{eq:H_form_2})). 
However, it is possible for a system to develop incoherence between the modulation of the on-site energy and hopping terms. 
In these cases, we can relax our requirement that all $\vec{q}^{(i)}$ are mutually incommensurate, and regard the modulated hopping terms and on-site energy as being described by two distinct but numerically equal modulation wave vectors. 
This will increase the number of additional dimensions necessary to represent the system.
We demonstrate how this situation can be handled using the 1D Rice-Mele chain later in Sec.~\ref{sec:1D_RM_chain_incoherence}. 
Fourth, in Refs.~\onlinecite{LL_fragile_Lian_Biao,Herzog_Hof_topo}, it is shown that in general the line integral of the Peierls phase should be taken on a piecewise linear path along which the Wannier functions have greatest overlap, as opposed to the linear path we used here. 
In our present situation, this means we are implicitly assuming that the total Hilbert space (occupied plus unoccupied states) for our dimensionally promoted model can be described by topologically trivial Wannier functions. 
The topological properties of the Wannier functions in the promoted $(d+N)$D space are an interesting direction for future investigation, including investigating whether topologically non-trivial $(d+N)$D Wannier functions can imply any physical properties in the low dimensional modulated system. 
Lastly, we emphasize that our interpretation of the promoted models in $(d+N)$D, given in Eq.~(\ref{eq:non_ortho_4}) is not unique. 
Our interpretation is fixed by the requirement that the $U(1)$ gauge fields (Eq.~(\ref{eq:non_ortho_A})) have a spatially constant field strength $F_{\mu\nu}$ in $(d+N)$D. 
Therefore, this allows us to easily construct the continuum theory describing low energy dynamics in $(d+N)$D, as we know the underlying $U(1)$ gauge field, written as a function of $\vec{r} \in \mathbb{R}^{d+N}$. 
This completes our description on how to compute the $U(1)$ gauge field in a $(d+N)$D lattice with non-orthogonal lattice vectors, and the corresponding lattice model in $(d+N)$D space.

{
\subsubsection{Generalization to systems with arbitrary orbital positions in the original $d$D space}

In Sec.~\ref{subsubsec:__dimensional_promotion}--\ref{sec:remarks_consistency_check_and_summary}, we have implicitly chosen to promote the dimension of a $d$D modulated lattice model whose orbitals are located exactly at the $d$D lattice points $\sum_{j=1}^{d}n_{j}\vec{a}_{j}$ constructed from $\{ \vec{a}_{1},\ldots,\vec{a}_{d}\}$. 
This choice makes the emergence of the $U(1)$ gauge field and the identification of Peierls phases transparent. 
However, in many models of practical interest, not all of the orbitals are located at the $d$D lattice points. 
As the computation of Peierls phases depends on the actual orbital positions, we now examine the effect of orbital positions in our dimensional promotion method.

We consider a $d$D modulated system whose $\alpha^{\text{th}}$ orbital is located at the position $\vec{r}_{\alpha}$ measured from the origin of the unit cell. 
We will assume that $\vec{r}_{\alpha}$ can be written as a linear combination of the $d$D lattice vectors $\{\vec{a}_{1},\ldots,\vec{a}_{d}\}$, namely
\begin{align}
    \vec{r}_{\alpha} = \sum_{j=1}^{d} x^{j}_{\alpha} \vec{a}_{j}, \label{eq:r_alpha_expansion}
\end{align}
where $x^{j}_{\alpha} \in [0,1)$ denotes the fractional component of $\vec{r}_{\alpha}$ along $\vec{a}_{j}$ (this excludes certain quasi-$d$-dimensional models). 
Therefore, the actual position of the $\alpha^{\text{th}}$ orbital in the unit cell labelled by $\vec{n} \in \mathbb{Z}^{d}$ is $\sum_{j=1}^{d}n_{j}\vec{a}_{j} + \vec{r}_{\alpha}$. 
When all the orbitals are located at the lattice points $\sum_{j=1}^{d}n_{j}\vec{a}_{j}$, we will have $\vec{r}_{\alpha} = 0$ for all $\alpha$. 
The generic Hamiltonian for our $d$D system is still given by Eq.~(\ref{eq:H_low_general}). 
For later convenience, here we rewrite Eq.~(\ref{eq:H_low_general}) with an explicit summation over the orbital components as
\begin{equation}
    H_{\text{low-dim}} = \sum_{\alpha,\beta}\sum_{\vec{n},\vec{m}} {\psi}^{\dagger}_{\vec{n}+\vec{m},\alpha} \left[H_{\vec{m}}\right]_{\alpha,\beta} {\psi}_{\vec{n},\beta} + \sum_{\alpha,\beta}{\sum_{\vec{n},\vec{l}}}  \sum_{i=1}^{N} {\psi}^{\dagger}_{\vec{n}+\vec{l},\alpha} \left[H_{\vec{l},\vec{n}}^{(i)} \right]_{\alpha,\beta} {\psi}_{\vec{n},\beta} + \sum_{\alpha,\beta}\sum_{\vec{n}} \sum_{i=1}^{N}{\psi}^{\dagger}_{\vec{n},\alpha} \left[ V^{(i)}_{\vec{n}} \right]_{\alpha,\beta} {\psi}_{\vec{n},\beta}, \label{eq:H_low_general_alpha}
\end{equation}
where ${\psi}^{\dagger}_{\vec{n},\alpha}$ and ${\psi}_{\vec{n},\alpha}$ are the creation and annihilation operator of the $\alpha^{\text{th}}$ orbital at the unit cell labelled by $\vec{n} \in \mathbb{Z}^{d}$. 
As before, $\left[H_{\vec{m}}\right]_{\alpha,\beta}$, $\left[H_{\vec{l},\vec{n}}^{(i)} \right]_{\alpha,\beta}$ and $\left[ V^{(i)}_{\vec{n}} \right]_{\alpha,\beta}$ are the $(\alpha,\beta)$ entries for the unmodulated Hamiltonian $\left[H_{\vec{m}}\right]$, the hopping modulations $\left[H_{\vec{l},\vec{n}}^{(i)} \right]$ and the on-site modulations $\left[ V^{(i)}_{\vec{n}} \right]$, respectively. 
The dimensional promotion procedure is identical to that in Sec.~\ref{subsubsec:__dimensional_promotion} and thus we obtain the same $(d+N)$D model given by Eq.~(\ref{eq:non_ortho_4}). 
Again, for later convenience we rewrite Eq.~(\ref{eq:non_ortho_4}) with an explicit summation over the orbital components as
\begin{align}
    H_{\text{high-dim}} =& \sum_{\alpha,\beta}\sum_{\vec{n},\vec{m},\vec{\nu}}  {\psi}^{\dagger}_{\vec{n}+\vec{m},\vec{\nu},\alpha} \left[ H_{\vec{m}}\right]_{\alpha,\beta} \psi_{\vec{n},\vec{\nu},\beta}   \label{eq:H_high_dim_normal_hopping_alpha} \\
    & + \sum_{\alpha,\beta}{\sum_{\vec{n},\vec{l},\vec{\nu}}}  \sum_{i=1}^{N} \sum_{p_{i}}  {\psi}^{\dagger}_{\vec{n}+\vec{l},\vec{\nu}-p_{i}\hat{\nu}_{i},\alpha}  \left[H_{\vec{l},p_{i}}^{(i)}\right]_{\alpha,\beta} e^{i  2\pi p_{i}   \vec{q}^{(i)} \cdot \left(\sum_{j=1}^{d}n_{j}\vec{a}_{j} \right)   } {\psi}_{\vec{n},\vec{\nu},\beta} \label{eq:H_high_dim_hopping_modulation_alpha} \\
    & + \sum_{\alpha,\beta}\sum_{\vec{n},\vec{\nu}}\sum_{i=1}^{N} \sum_{p_{i}} {\psi}^{\dagger}_{\vec{n},\vec{\nu}-p_{i}\hat{\nu}_{i},\alpha} \left[V^{(i)}_{p_{i}}\right]_{\alpha,\beta}  e^{i2\pi p_{i} \vec{q}^{(i)}\cdot \left(\sum_{j=1}^{d}n_{j}\vec{a}_{j} \right)}{\psi}_{\vec{n},\vec{\nu},\beta}. \label{eq:H_high_dim_onsite_modulation_alpha}
\end{align}
We then take the actual position of the $\alpha^{\text{th}}$ orbital in the unit cell labelled by $(\vec{n},\vec{\nu}) \in (\mathbb{Z}^{d},\mathbb{Z}^{N})$ to be $\sum_{j=1}^{d}n_{j}\vec{a}_{j} + \sum_{j=1}^{N}\nu_{j}\vec{c}_{j} + \vec{r}_{\alpha}$ where we have assumed that there are no components of $\vec{r}_\alpha$ along $\vec{c}_{j}$ and Eq.~(\ref{eq:r_alpha_expansion}) still holds. 
However, due to the generic position $\vec{r}_{\alpha}$ of the orbitals, the separation of phase factors in Eqs.~(\ref{eq:H_high_dim_normal_hopping_alpha}--\ref{eq:H_high_dim_onsite_modulation_alpha}) between periodic hopping and the Peierls phase due to the gauge field in Eq.~(\ref{eq:non_ortho_A}) needs to be modified compared with those in Sec.~\ref{sec:computing_Peierls_phase_1}.

We now compute the various Peierls phases associated with terms in Eqs.~(\ref{eq:H_high_dim_normal_hopping_alpha}--\ref{eq:H_high_dim_onsite_modulation_alpha}) due to the gauge field in Eq.~(\ref{eq:non_ortho_A}). First, we consider the terms in Eq.~(\ref{eq:H_high_dim_onsite_modulation_alpha}) where the electrons hop from 
\begin{equation}
\vec{r}_{i} = \sum_{j=1}^{d}n_{j}\vec{a}_{j} + \sum_{j=1}^{N}\nu_{j}\vec{c}_{j} + \vec{r}_{\beta}
\end{equation}
to 
\begin{equation}
 \vec{r}_{f} =\sum_{j=1}^{d}n_{j}\vec{a}_{j} + \sum_{j=1}^{N}\nu_{j}\vec{c}_{j} - p_{i} \vec{c}_{i} + \vec{r}_{\alpha}. 
\end{equation}
The straight line connecting $\vec{r}_{i}$ to $\vec{r}_{f}$ is
\begin{align}
    \sum_{j=1}^{d}n_{j}\vec{a}_{j} + \sum_{j=1}^{N}\nu_{j}\vec{c}_{j} + \vec{r}_{\beta} + \left(- p_{i} \vec{c}_{i} +  \vec{r}_{\alpha} - \vec{r}_{\beta} \right)t,
\end{align}
where $ t\in [0,1]$ and the corresponding infinitesimal displacement vector is $\left(- p_{i} \vec{c}_{i} +  \vec{r}_{\alpha} - \vec{r}_{\beta} \right)dt$. The corresponding Peierls phase with a line integral can then be computed as follows:
\begin{align}
    & \exp{-i \int_{\vec{r}_{i}}^{\vec{r}_{f}} d\vec{r} \cdot \vec{A}} \\
    & =  \exp{-i \int_{0}^{1}dt  \left(- p_{i} \vec{c}_{i} +  \vec{r}_{\alpha} - \vec{r}_{\beta} \right) \cdot \frac{1}{2\pi} \sum_{j=1}^{N} \sum_{k=1}^{d} \vec{G}_{j} \left( \vec{q}^{(j)}\cdot \left[ \left( \begin{bmatrix}
    \sum_{l=1}^{d}n_{l}\vec{a}_{l} + \sum_{l=1}^{N}\nu_{l}\vec{c}_{l} + \vec{r}_{\beta}  \\
    +\left(- p_{i} \vec{c}_{i} +  \vec{r}_{\alpha} - \vec{r}_{\beta} \right)t
    \end{bmatrix}
    \cdot \vec{g}_{k} \right) \vec{a}_{k} \right] \right) } \label{eq:Peierls_1_1_alpha}\\
    & =  \exp{+i \int_{0}^{1}dt   p_{i} \vec{c}_{i}   \cdot \frac{1}{2\pi} \sum_{j=1}^{N} \sum_{k=1}^{d} \vec{G}_{j} \left( \vec{q}^{(j)}\cdot \left[ \left(
    \left(
    \sum_{l=1}^{d}n_{l}\vec{a}_{l} + \vec{r}_{\beta} + \left(  \vec{r}_{\alpha} - \vec{r}_{\beta} \right)t
    \right)
    \cdot \vec{g}_{k} \right) \vec{a}_{k} \right] \right) } \label{eq:Peierls_1_2_alpha} \\
    & =  \exp{+i \int_{0}^{1}dt   p_{i} \vec{c}_{i}   \cdot \frac{1}{2\pi} \sum_{j=1}^{N} \sum_{k=1}^{d} \vec{G}_{j} \left( \vec{q}^{(j)}\cdot \left[ \left(
    \left(
    \sum_{l=1}^{d}\left( n_{l} + x_{\beta}^{l} + t x_{\alpha}^{l} - t x_{\beta}^{l} \right)\vec{a}_{l} 
    \right)
    \cdot \vec{g}_{k} \right) \vec{a}_{k} \right] \right) } \label{eq:Peierls_1_3_alpha}\\
    & =  \exp{+i2\pi \int_{0}^{1}dt   p_{i}  \sum_{j=1}^{N} \sum_{k=1}^{d} \delta_{ij} \left( \vec{q}^{(j)}\cdot \left[ 
    \sum_{l=1}^{d}\left(n_{l} + x_{\beta}^{l}+tx^{l}_{\alpha} - t x^{l}_{\beta}\right)\delta_{lk} 
   \vec{a}_{k} \right] \right) } \label{eq:Peierls_1_4_alpha} \\
   & =  \exp{+i2\pi \int_{0}^{1}dt   p_{i}  \left( \vec{q}^{(i)}\cdot \left[ 
    \sum_{l=1}^{d}\left(n_{l} + x_{\beta}^{l}+tx^{l}_{\alpha} - t x^{l}_{\beta}\right)
     \vec{a}_{l} \right] \right) } \label{eq:Peierls_1_5_alpha} \\
    & =  \exp{+i2\pi   p_{i}  \vec{q}^{(i)}\cdot \left[ 
    \sum_{l=1}^{d}\left(n_{l} + x_{\beta}^{l}+\frac{1}{2}x^{l}_{\alpha} - \frac{1}{2} x^{l}_{\beta}\right)
     \vec{a}_{l} \right]  }\label{eq:Peierls_1_6_alpha} \\
    & =  \exp{+i2\pi   p_{i}  \vec{q}^{(i)}\cdot \left[ 
    \sum_{l=1}^{d}\left(n_{l}  + \frac{x^{l}_{\alpha} + x^{l}_{\beta}}{2} \right)
     \vec{a}_{l} \right]  } \label{eq:Peierls_1_7_alpha} .
\end{align}
To go from Eq.~(\ref{eq:Peierls_1_1_alpha}) to Eq.~(\ref{eq:Peierls_1_2_alpha}) we have used Eq.~(\ref{eq:reciprocal_2}) and the fact that $(\vec{r}_{\alpha} - \vec{r}_{\beta}) \cdot \vec{G}_{j} = 0$ because of Eqs.~(\ref{eq:reciprocal_3}) and (\ref{eq:r_alpha_expansion}). 
We then use Eq.~(\ref{eq:r_alpha_expansion}) to go from Eq.~(\ref{eq:Peierls_1_2_alpha}) to Eq.~(\ref{eq:Peierls_1_3_alpha}). 
To go from Eq.~(\ref{eq:Peierls_1_3_alpha}) to Eq.~(\ref{eq:Peierls_1_4_alpha}), we have used Eqs.~(\ref{eq:reciprocal_1}) and (\ref{eq:reciprocal_4}). 
Performing the summation over $j$ and $k$ we obtain Eq.~(\ref{eq:Peierls_1_5_alpha}). 
Using $\int_{0}^{1}dt  = 1$ and $\int_{0}^{1}tdt  = \frac{1}{2}$ we arrive at Eqs.~(\ref{eq:Peierls_1_6_alpha}--\ref{eq:Peierls_1_7_alpha}).

Next, let us consider the terms in Eq.~(\ref{eq:H_high_dim_hopping_modulation_alpha}) where the electrons hop from 
\begin{equation}
\vec{r}_{i} = \sum_{j=1}^{d}n_{j}\vec{a}_{j} + \sum_{j=1}^{N}\nu_{j}\vec{c}_{j} + \vec{r}_{\beta}
\end{equation}
to 
\begin{equation}
\vec{r}_{f} =\sum_{j=1}^{d}n_{j}\vec{a}_{j} + \sum_{j=1}^{N}\nu_{j}\vec{c}_{j} + \sum_{j=1}^{d}l_{j}\vec{a}_{j} - p_{i} \vec{c}_{i} + \vec{r}_{\alpha}.
\end{equation}
The straight line connecting $\vec{r}_{i}$ to $\vec{r}_{f}$ is
\begin{align}
    \sum_{j=1}^{d}n_{j}\vec{a}_{j} + \sum_{j=1}^{N}\nu_{j}\vec{c}_{j} + \vec{r}_{\beta} + \left( \sum_{j=1}^{d}l_{j}\vec{a}_{j}- p_{i} \vec{c}_{i} +  \vec{r}_{\alpha} - \vec{r}_{\beta} \right)t,
\end{align}
where $ t\in [0,1]$ and the corresponding infinitesimal displacement vector is $\left( \sum_{j=1}^{d}l_{j}\vec{a}_{j}- p_{i} \vec{c}_{i} +  \vec{r}_{\alpha} - \vec{r}_{\beta} \right)dt$. 
The corresponding Peierls phase with a line integral can then be computed as follows:
\begin{align}
    & \exp{-i \int_{\vec{r}_{i}}^{\vec{r}_{f}} d\vec{r} \cdot \vec{A}} \\
    & =  \exp{-i \int_{0}^{1}dt  \left( \sum_{r=1}^{d}l_{r}\vec{a}_{r}- p_{i} \vec{c}_{i} +  \vec{r}_{\alpha} - \vec{r}_{\beta} \right) \cdot \frac{1}{2\pi} \sum_{j=1}^{N} \sum_{k=1}^{d} \vec{G}_{j} \left( \vec{q}^{(j)}\cdot \left[ \left( \begin{bmatrix}
    \sum_{r=1}^{d}n_{r}\vec{a}_{r} + \sum_{r=1}^{N}\nu_{r}\vec{c}_{r} + \vec{r}_{\beta}  \\
    + \left( \sum_{r=1}^{d}l_{r}\vec{a}_{r}- p_{i} \vec{c}_{i} +  \vec{r}_{\alpha} - \vec{r}_{\beta} \right)t
    \end{bmatrix}
    \cdot \vec{g}_{k} \right) \vec{a}_{k} \right] \right) } \label{eq:Peierls_2_1_alpha}\\
    & =  \exp{+i \int_{0}^{1}dt    p_{i} \vec{c}_{i} \cdot \frac{1}{2\pi} \sum_{j=1}^{N} \sum_{k=1}^{d} \vec{G}_{j} \left( \vec{q}^{(j)}\cdot \left[ \left( \begin{bmatrix}
    \sum_{r=1}^{d}n_{r}\vec{a}_{r} + \vec{r}_{\beta}  \\
    + \left( \sum_{r=1}^{d}l_{r}\vec{a}_{r} +  \vec{r}_{\alpha} - \vec{r}_{\beta} \right)t
    \end{bmatrix}
    \cdot \vec{g}_{k} \right) \vec{a}_{k} \right] \right) } \label{eq:Peierls_2_2_alpha}\\
    & =  \exp{+i \int_{0}^{1}dt    p_{i} \vec{c}_{i} \cdot \frac{1}{2\pi} \sum_{j=1}^{N} \sum_{k=1}^{d} \vec{G}_{j} \left( \vec{q}^{(j)}\cdot \left[ \left( \left( \sum_{r=1}^{d}\left( n_{r} + tl_{r} + x^{r}_{\beta} + t x^{r}_{\alpha} - t x^{r}_{\beta} \right)\vec{a}_{r}  \right)
    \cdot \vec{g}_{k} \right) \vec{a}_{k} \right] \right) } \label{eq:Peierls_2_3_alpha}\\
    & =  \exp{+i2\pi \int_{0}^{1}dt    p_{i}  \sum_{j=1}^{N} \sum_{k=1}^{d} \delta_{ij} \left( \vec{q}^{(j)}\cdot \left[  \sum_{r=1}^{d}\left( n_{r} + tl_{r} + x^{r}_{\beta} + t x^{r}_{\alpha} - t x^{r}_{\beta} \right)\delta_{rk}   \vec{a}_{k} \right] \right) } \label{eq:Peierls_2_4_alpha}\\
    & =  \exp{+i2\pi \int_{0}^{1}dt    p_{i}    \vec{q}^{(i)}\cdot \left[  \sum_{r=1}^{d}\left( n_{r} + tl_{r} + x^{r}_{\beta} + t x^{r}_{\alpha} - t x^{r}_{\beta} \right) \vec{a}_{r} \right] } \label{eq:Peierls_2_5_alpha}\\
    & =  \exp{+i2\pi  p_{i}    \vec{q}^{(i)}\cdot \left[  \sum_{r=1}^{d}\left( n_{r} + \frac{1}{2}l_{r} + x^{r}_{\beta} + \frac{1}{2} x^{r}_{\alpha} - \frac{1}{2} x^{r}_{\beta} \right) \vec{a}_{r} \right] } \label{eq:Peierls_2_6_alpha}\\
    & =  \exp{+i2\pi  p_{i}    \vec{q}^{(i)}\cdot \left[  \sum_{r=1}^{d}\left( n_{r} + \frac{l_{r} +x^{r}_{\alpha} + x^{r}_{\beta} }{2}  \right) \vec{a}_{r} \right] }. \label{eq:Peierls_2_7_alpha}
\end{align}
To go from Eq.~(\ref{eq:Peierls_2_1_alpha}) to Eq.~(\ref{eq:Peierls_2_2_alpha}) we have used Eqs.~(\ref{eq:reciprocal_2}--\ref{eq:reciprocal_3}), and again the fact that $(\vec{r}_{\alpha} - \vec{r}_{\beta}) \cdot \vec{G}_{j} = 0$ because of Eqs.~(\ref{eq:reciprocal_3}) and (\ref{eq:r_alpha_expansion}).
 We then use Eq.~(\ref{eq:r_alpha_expansion}) to go from Eq.~(\ref{eq:Peierls_2_2_alpha}) to Eq.~(\ref{eq:Peierls_2_3_alpha}). 
 To go from Eq.~(\ref{eq:Peierls_2_3_alpha}) to Eq.~(\ref{eq:Peierls_2_4_alpha}), we have used Eqs.~(\ref{eq:reciprocal_1}) and (\ref{eq:reciprocal_4}). 
 Performing the summation over $j$ and $k$ we obtain Eq.~(\ref{eq:Peierls_2_5_alpha}). 
 Using $\int_{0}^{1}dt  = 1$ and $\int_{0}^{1}tdt  = \frac{1}{2}$ we arrive at Eqs.~(\ref{eq:Peierls_2_6_alpha}--\ref{eq:Peierls_2_7_alpha}).

Finally, let us consider the terms in Eq.~(\ref{eq:H_high_dim_normal_hopping_alpha}) where the electrons hop from 
\begin{equation}
\vec{r}_{i} = \sum_{j=1}^{d}n_{j}\vec{a}_{j} + \sum_{j=1}^{N}\nu_{j}\vec{c}_{j} + \vec{r}_{\beta}
\end{equation}
to 
\begin{equation}
\vec{r}_{f} =\sum_{j=1}^{d}n_{j}\vec{a}_{j} + \sum_{j=1}^{N}\nu_{j}\vec{c}_{j} + \sum_{j=1}^{d}m_{j}\vec{a}_{j} + \vec{r}_{\alpha}.
\end{equation}
The straight line connecting $\vec{r}_{i}$ to $\vec{r}_{f}$ is
\begin{align}
    \sum_{j=1}^{d}n_{j}\vec{a}_{j} + \sum_{j=1}^{N}\nu_{j}\vec{c}_{j} + \vec{r}_{\beta} + \left( \sum_{j=1}^{d}m_{j}\vec{a}_{j}+  \vec{r}_{\alpha} - \vec{r}_{\beta} \right)t,
\end{align}
where $ t\in [0,1]$ and the corresponding infinitesimal displacement vector is $\left( \sum_{j=1}^{d}m_{j}\vec{a}_{j}+  \vec{r}_{\alpha} - \vec{r}_{\beta} \right)dt$. 
The corresponding Peierls phase with a line integral can then be computed as follows:
\begin{align}
    & \exp{-i \int_{\vec{r}_{i}}^{\vec{r}_{f}} d\vec{r} \cdot \vec{A}} \\
    & =  \exp{-i \int_{0}^{1}dt  \left( \sum_{r=1}^{d}m_{r}\vec{a}_{r}+  \vec{r}_{\alpha} - \vec{r}_{\beta} \right) \cdot \frac{1}{2\pi} \sum_{j=1}^{N} \sum_{k=1}^{d} \vec{G}_{j} \left( \vec{q}^{(j)}\cdot \left[ \left( \begin{bmatrix}
    \sum_{r=1}^{d}n_{r}\vec{a}_{r} + \sum_{r=1}^{N}\nu_{r}\vec{c}_{r} + \vec{r}_{\beta}  \\
    + \left( \sum_{r=1}^{d}m_{r}\vec{a}_{r}+  \vec{r}_{\alpha} - \vec{r}_{\beta} \right)t
    \end{bmatrix}
    \cdot \vec{g}_{k} \right) \vec{a}_{k} \right] \right) } \label{eq:Peierls_3_1_alpha}\\ 
    & = 1. \label{eq:Peierls_3_2_alpha}
\end{align}
To go from Eq.~(\ref{eq:Peierls_3_1_alpha}) to Eq.~(\ref{eq:Peierls_3_2_alpha}) we have used Eq.~(\ref{eq:reciprocal_3}) and again the fact that $(\vec{r}_{\alpha} - \vec{r}_{\beta}) \cdot \vec{G}_{j} = 0$ because of Eqs.~(\ref{eq:reciprocal_3}) and (\ref{eq:r_alpha_expansion}).

With this knowledge of the Peierls phases in Eqs.~(\ref{eq:Peierls_1_7_alpha}), (\ref{eq:Peierls_2_7_alpha}), and (\ref{eq:Peierls_3_2_alpha}), we can rewrite $H_{\text{high-dim}}$ in Eqs.~(\ref{eq:H_high_dim_normal_hopping_alpha}--\ref{eq:H_high_dim_onsite_modulation_alpha}) as 
\begin{align}
    H_{\text{high-dim}} =& \sum_{\alpha,\beta}\sum_{\vec{n},\vec{m},\vec{\nu}}  {\psi}^{\dagger}_{\vec{n}+\vec{m},\vec{\nu},\alpha} \left[ H_{\vec{m}}\right]_{\alpha,\beta} \psi_{\vec{n},\vec{\nu},\beta}   \label{eq:H_high_dim_normal_hopping_alpha_split} \\
    & + \sum_{\alpha,\beta}{\sum_{\vec{n},\vec{l},\vec{\nu}}}  \sum_{i=1}^{N} \sum_{p_{i}}  {\psi}^{\dagger}_{\vec{n}+\vec{l},\vec{\nu}-p_{i}\hat{\nu}_{i},\alpha}  \left[H_{\vec{l},p_{i}}^{(i)}\right]_{\alpha,\beta} e^{-i\pi  p_{i}    \vec{q}^{(i)}\cdot  \left(\sum_{j=1}^{d}\left(  l_{j} +x^{j}_{\alpha} + x^{j}_{\beta}  \right) \vec{a}_{j} \right) } e^{i2\pi  p_{i}    \vec{q}^{(i)}\cdot \left( \sum_{j=1}^{d}\left( n_{j} + \frac{l_{j} +x^{j}_{\alpha} + x^{j}_{\beta} }{2}  \right) \vec{a}_{j} \right) }  {\psi}_{\vec{n},\vec{\nu},\beta} \label{eq:H_high_dim_hopping_modulation_alpha_split} \\
    & + \sum_{\alpha,\beta}\sum_{\vec{n},\vec{\nu}}\sum_{i=1}^{N} \sum_{p_{i}} {\psi}^{\dagger}_{\vec{n},\vec{\nu}-p_{i}\hat{\nu}_{i},\alpha} \left[V^{(i)}_{p_{i}}\right]_{\alpha,\beta}  e^{-i\pi   p_{i}  \vec{q}^{(i)}\cdot \left(
    \sum_{j=1}^{d}\left( x^{j}_{\alpha} + x^{j}_{\beta} \right)
     \vec{a}_{j} \right)  } 
    e^{i2\pi   p_{i}  \vec{q}^{(i)}\cdot \left( 
    \sum_{j=1}^{d}\left(n_{j}  + \frac{x^{j}_{\alpha} + x^{j}_{\beta}}{2} \right)
     \vec{a}_{j} \right)  } 
    {\psi}_{\vec{n},\vec{\nu},\beta}. \label{eq:H_high_dim_onsite_modulation_alpha_split}
\end{align}
Eqs.~(\ref{eq:H_high_dim_normal_hopping_alpha_split}--\ref{eq:H_high_dim_onsite_modulation_alpha_split}) can then be interpreted as a $(d+N)$D lattice model with Hamiltonian
\begin{align}
    H_{\text{high-dim}} =& \sum_{\alpha,\beta}\sum_{\vec{n},\vec{m},\vec{\nu}}  {\psi}^{\dagger}_{\vec{n}+\vec{m},\vec{\nu},\alpha} \left[ H_{\vec{m}}\right]_{\alpha,\beta} \psi_{\vec{n},\vec{\nu},\beta}   \label{eq:H_high_dim_normal_hopping_alpha_no_gauge} \\
    & + \sum_{\alpha,\beta}{\sum_{\vec{n},\vec{l},\vec{\nu}}}  \sum_{i=1}^{N} \sum_{p_{i}}  {\psi}^{\dagger}_{\vec{n}+\vec{l},\vec{\nu}-p_{i}\hat{\nu}_{i},\alpha}  \left[H_{\vec{l},p_{i}}^{(i)}\right]_{\alpha,\beta} e^{-i\pi  p_{i}    \vec{q}^{(i)}\cdot  \left(\sum_{j=1}^{d}\left(  l_{j} +x^{j}_{\alpha} + x^{j}_{\beta}  \right) \vec{a}_{j} \right) }  {\psi}_{\vec{n},\vec{\nu},\beta} \label{eq:H_high_dim_hopping_modulation_alpha_no_gauge} \\
    & + \sum_{\alpha,\beta}\sum_{\vec{n},\vec{\nu}}\sum_{i=1}^{N} \sum_{p_{i}} {\psi}^{\dagger}_{\vec{n},\vec{\nu}-p_{i}\hat{\nu}_{i},\alpha} \left[V^{(i)}_{p_{i}}\right]_{\alpha,\beta}  e^{-i\pi   p_{i}  \vec{q}^{(i)}\cdot \left(
    \sum_{j=1}^{d}\left( x^{j}_{\alpha} + x^{j}_{\beta} \right)
     \vec{a}_{j} \right)  } 
    {\psi}_{\vec{n},\vec{\nu},\beta}, \label{eq:H_high_dim_onsite_modulation_alpha_no_gauge}
\end{align}
which is periodic with lattice vectors $\{\vec{a}_{1},\ldots,\vec{a}_{d},\vec{c}_{1},\ldots,\vec{c}_{N} \}$, coupled through a Peierls substitution\cite{Peierls_substitution} to a $U(1)$ gauge field given in Eq.~(\ref{eq:non_ortho_A}). 
Notice that we have regarded
\begin{align}
    e^{i2\pi  p_{i}    \vec{q}^{(i)}\cdot \left( \sum_{j=1}^{d}\left( n_{j} + \frac{l_{j} +x^{j}_{\alpha} + x^{j}_{\beta} }{2}  \right) \vec{a}_{j} \right) } 
\end{align}
and
\begin{align}
    e^{i2\pi   p_{i}  \vec{q}^{(i)}\cdot \left( 
    \sum_{j=1}^{d}\left(n_{j}  + \frac{x^{j}_{\alpha} + x^{j}_{\beta}}{2} \right)
     \vec{a}_{j} \right)  } 
\end{align}
in Eqs.~(\ref{eq:H_high_dim_hopping_modulation_alpha_split}) and (\ref{eq:H_high_dim_onsite_modulation_alpha_split}) as the Peierls phases (see Eqs.~(\ref{eq:Peierls_2_7_alpha}) and (\ref{eq:Peierls_1_7_alpha})) due to the $U(1)$ gauge field in Eq.~(\ref{eq:non_ortho_A}). 
Notice that there are no Peierls phases induced from the gauge field in Eq.~(\ref{eq:H_high_dim_normal_hopping_alpha_split}). 

We have thus generalized our dimensional promotion method to $d$D modulated lattice models whose orbitals are not located at the $d$D lattice points $\sum_{j=1}^{d}n_{j}\vec{a}_{j}$. 
Similar to Tables~\ref{tab:model_summary} and \ref{tab:hopping_1}, we have summarized the Hamiltonian before and after the dimensional promotion in Table~\ref{tab:model_summary_general_orbital_positions}, and the hopping matrix elements together with the Peierls phases in Table~\ref{tab:hopping_1_general_orbital_positions}. 
A notable feature is that now both the hopping matrix elements of the $(d+N)$D lattice model {\it without} the gauge field and the Peierls phases due to the gauge field in Eq.~(\ref{eq:non_ortho_A}) encode information of the orbital positions. 
When all the orbitals are located right at the lattice points we have $\vec{r}_{\alpha} = 0$ for all $\alpha$, namely $x^{j}_{\alpha}=0$ for all $j$ and $\alpha$ in Eq.~(\ref{eq:r_alpha_expansion}). 
In such cases, Table~\ref{tab:hopping_1_general_orbital_positions} effectively reduces to Table~\ref{tab:hopping_1}.


\begin{table}[h]
\centering
\begin{tabular}{|c|c|}
\hline
Original $d$D modulated system & Eq.~(\ref{eq:H_low_general_alpha}) \\
\hline
Promoted $(d+N)$D system with $U(1)$ gauge fields & Eqs.~(\ref{eq:H_high_dim_normal_hopping_alpha_split}--\ref{eq:H_high_dim_onsite_modulation_alpha_split}) \\
\hline
Promoted $(d+N)$D system without $U(1)$ gauge fields & Eqs.~(\ref{eq:H_high_dim_normal_hopping_alpha_no_gauge}--\ref{eq:H_high_dim_onsite_modulation_alpha_no_gauge}) \\
\hline
\end{tabular}
\caption{Relevant equations in the general dimensional promotion formalism with arbitrary orbital positions.}
\label{tab:model_summary_general_orbital_positions}
\end{table}

\begin{table}[h]
\centering
\begin{tabular}{|c|c|c|}
\hline
Hopping from $\beta^{\text{th}}$ orbital in unit cell $(\vec{n},\vec{\nu})$ to & Hopping matrix elements & Peierls phases  \\
\hline
\hline
$\alpha^{\text{th}}$ orbital in unit cell $(\vec{n}+\vec{m},\vec{\nu})$ & $\left[H_{\vec{m}}\right]_{\alpha,\beta}$ & $1$ \\
\hline
$\alpha^{\text{th}}$ orbital in unit cell $(\vec{n},\vec{\nu}-p_{i}\hat{\nu}_{i})$ & $ \left[V^{(i)}_{p_{i}}\right]_{\alpha,\beta}  e^{-i\pi   p_{i}  \vec{q}^{(i)}\cdot \left(     \sum_{j=1}^{d}\left( x^{j}_{\alpha} + x^{j}_{\beta} \right)      \vec{a}_{j} \right)  }  $ & $e^{i2\pi   p_{i}  \vec{q}^{(i)}\cdot \left(      \sum_{j=1}^{d}\left(n_{j}  + \frac{x^{j}_{\alpha} + x^{j}_{\beta}}{2} \right)      \vec{a}_{j} \right)  } $ \\
\hline
$\alpha^{\text{th}}$ orbital in unit cell $(\vec{n}+\vec{l},\vec{\nu}-p_{i}\hat{\nu}_{i})$ & $\left[H_{\vec{l},p_{i}}^{(i)}\right]_{\alpha,\beta} e^{-i\pi  p_{i}    \vec{q}^{(i)}\cdot  \left(\sum_{j=1}^{d}\left(  l_{j} +x^{j}_{\alpha} + x^{j}_{\beta}  \right) \vec{a}_{j} \right) } $ & $e^{i2\pi  p_{i}    \vec{q}^{(i)}\cdot \left( \sum_{j=1}^{d}\left( n_{j} + \frac{l_{j} +x^{j}_{\alpha} + x^{j}_{\beta} }{2}  \right) \vec{a}_{j} \right) } $ \\
\hline
\end{tabular}
\caption{Hopping terms in the promoted $(d+N)$D model with arbitrary orbital positions and the corresponding Peierls phases, expressed in terms of parameters from the $d$D modulated system in Eq.~(\ref{eq:H_low_general}). 
Notice that $p_{i}\hat{\nu}_{i}$ does not imply a summation over $i$.}
\label{tab:hopping_1_general_orbital_positions}
\end{table}

}

\subsection{\label{subsec:examples_general_dim_promotion}Examples}

To demonstrate our general construction, we will use Table~\ref{tab:model_summary} and Table~\ref{tab:hopping_1} to consider four examples: 
(1) promoting the 1D Rice-Mele chain to a 2D square lattice with $\pi$-flux, 
(2) promoting the 1D Rice-Mele chain with incoherent phase offsets in on-site and hopping modulations to a 3D cubic lattice coupled to a $U(1)$ gauge field, 
(3) promoting a 1D modulated system to a 2D hexagonal lattice with a perpendicular magnetic field, and 
(4) promoting a 2D modulated system with hexagonal lattice to a 3D hexagonal lattice coupled to a $U(1)$ gauge field.

\subsubsection{\label{sec:ex_1D_RM_chain} 1D Rice-Mele chain $\to$ 2D square lattice with $\pi$-flux}

The Hamiltonian of the 1D Rice-Mele\cite{RiceMele} chain oriented along the $x$-axis is given by
\begin{align}
    H_{\text{Rice-Mele}} = \sum_{n}  \left( t + \delta t (-1)^{n} \cos{\phi} \right)\psi^{\dagger}_{n+1}\psi_{n} + \text{h.c.} + \sum_{n} (-1)^{n+1} \Delta  \sin{\phi} \psi^{\dagger}_{n}\psi_{n},
    \label{eq:supp_Rice_Mele_1}
\end{align}
where $\psi^{\dagger}_{n}$ is the creation operator for an electron at position $n$, and $\text{h.c.}$ denotes the Hermitian conjugate of all terms before it. 
We can rewrite Eq.~(\ref{eq:supp_Rice_Mele_1}) as
\begin{align}
    H_{\text{Rice-Mele}} & = \sum_{n}  \left( t + \delta t \cos{(\pi n + \phi)} \right)\psi^{\dagger}_{n+1}\psi_{n} + \text{h.c.} - \sum_{n} \Delta  \sin{(\pi n + \phi)} \psi^{\dagger}_{n}\psi_{n} 
    \label{eq:supp_Rice_Mele_2} \\
    & = \sum_{n}  \left( t + \delta t \cdot \frac{e^{i\left( \pi n + \phi\right)} + e^{-i\left( \pi n + \phi\right)}}{2} \right)\psi^{\dagger}_{n+1}\psi_{n} + \text{h.c.} - \sum_{n} \Delta  \cdot \frac{e^{i\left( \pi n + \phi\right)} - e^{-i\left( \pi n + \phi\right)}}{2i} \psi^{\dagger}_{n}\psi_{n} .\label{eq:supp_Rice_Mele_3}
\end{align}
Thus we can see that 1D Rice-Mele chain is a 1D modulated system with modulation wave vector $q = 1/2$ ($2\pi q = \pi$). 
We now choose the second, synthetic lattice vector to be $\hat{y}$. 
The promote 2D system will then have
\begin{align}
    \vec{a}_{1} = (1,0),\ \vec{c}_{1} = (0,1),\ \vec{g}_{1} = 2\pi(1,0) \text{ and } \vec{G}_{1} = 2\pi (0,1),
\end{align}
in terms of the notation from Eqs.~(\ref{eq:reciprocal_1}--\ref{eq:reciprocal_4}). 
Notice that $\vec{a}_{1}$ and $\vec{c}_{1}$ describe a square lattice. 
Using the Fourier expansion convention in Table~\ref{tab:model_summary} and Table~\ref{tab:hopping_1} with $\vec{q} = (1/2,0)$ and Eq.~(\ref{eq:supp_Rice_Mele_3}), we obtain the hopping terms summarized in Table~\ref{tab:hopping_2}. 
Multiplying the entries for hopping matrices and Peierls phases, the Hamiltonian for the promoted 2D model is
\begin{align}
    H_{\text{2D}} = \sum_{n,m} \left( t\psi^{\dagger}_{n+1,m}\psi_{n,m}- \frac{i\delta t}{2}e^{i\pi \left(n+\frac{1}{2} \right)} \psi^{\dagger}_{n+1,m-1}\psi_{n,m}+ \frac{i\delta t}{2}e^{-i\pi \left(n+\frac{1}{2} \right)} \psi^{\dagger}_{n+1,m+1}\psi_{n,m}-  \frac{i\Delta}{2}e^{-i\pi n} \psi^{\dagger}_{n,m+1}\psi_{n,m} + \text{h.c.}  \right), \label{eq:1D_Rice_Mele_to_2D_2}
\end{align}
where $\psi^{\dagger}_{n,m}$ is the creation operator for an electron at position $(n,m)$, and the vector potential to which this 2D model is coupled is 
\begin{align}
    \vec{A} = \frac{1}{2\pi}  \vec{G}_{1} \left( \vec{q}\cdot \left[ \left( \vec{r} \cdot \vec{g}_{1}\right) \vec{a}_{1} \right] \right) = (0,\pi x).
\end{align}
This $\vec{A}$ produces a perpendicular magnetic field $\vec{B} = \pi \hat{z}$ in the promoted 2D system such that there is a $\pi$-flux per plaquette. 

\begin{table}[h]
\centering
\begin{tabular}{|c|c|c|}
\hline
Hopping from $(n,m)$ to & Hopping matrices & Peierls phases  \\
\hline
\hline
$(n+1,m)$ & $t$ & $1$ \\
\hline
$(n,m+1)$  & $ \Delta / (2i) = -i\Delta / 2$ & $e^{-i \pi n}$  \\
\hline 
$(n+1,m+1)$ & $(\delta t/ 2) \cdot e^{i \pi /2} = i\delta t/ 2 $  & $e^{-i \pi \left( n + \frac{1}{2} \right)}$ \\
\hline
$(n+1,m-1)$ & $(\delta t/ 2) \cdot e^{-i \pi /2} = -i\delta t/ 2 $ & $e^{i \pi \left( n + \frac{1}{2} \right)}$ \\
\hline
\end{tabular}
\caption{Hopping matrices and the corresponding Peierls phases for the promoted $2$D model from a 1D Rice-Mele chain.
The hopping matrices along the opposite directions of those listed here are omitted and can be obtained through Hermitian conjugation.}
\label{tab:hopping_2}
\end{table}

\subsubsection{\label{sec:1D_RM_chain_incoherence}1D Rice-Mele chain with phase offset incoherence $\to$ 3D cubic lattice coupled to a $U(1)$ gauge field}

Consider again a 1D Rice-Mele chain oriented along the $x$-axis as in Sec.~\ref{sec:ex_1D_RM_chain}. 
Now, however, we assume that the phase offsets in the hopping and on-site modulation can be different.
The $1$D Hamiltonian then reads
\begin{align}
    H_{\text{Rice-Mele}} = \sum_{n}  \left( t + \delta t (-1)^{n} \cos{\phi^{(1)}} \right)\psi^{\dagger}_{n+1}\psi_{n} + \text{h.c.} + \sum_{n} (-1)^{n+1} \Delta  \sin{\phi^{(2)}} \psi^{\dagger}_{n}\psi_{n},
    \label{eq:Rice_Mele_incoherence}
\end{align}
where $\psi^{\dagger}_{n}$ is the creation operator for an electron at position $n$, and $\text{h.c.}$ means the Hermitian conjugate of all terms before it. 
This situation arises when we are able to tune the phase offsets of the on-site and hopping modulations independently. 
We can write Eq.~(\ref{eq:Rice_Mele_incoherence}) as
\begin{align}
    H_{\text{Rice-Mele}} & = \sum_{n}  \left( t + \delta t \cos{(\pi n + \phi^{(1)})} \right)\psi^{\dagger}_{n+1}\psi_{n} + \text{h.c.} - \sum_{n} \Delta  \sin{(\pi n + \phi^{(2)})} \psi^{\dagger}_{n}\psi_{n} 
    \label{eq:Rice_Mele_incoherence_2} \\
    & = \sum_{n}  \left( t + \delta t \cdot \frac{e^{i\left( \pi n + \phi^{(1)}\right)} + e^{-i\left( \pi n + \phi^{(1)}\right)}}{2} \right)\psi^{\dagger}_{n+1}\psi_{n} + \text{h.c.} - \sum_{n} \Delta  \cdot \frac{e^{i\left( \pi n + \phi^{(2)}\right)} - e^{-i\left( \pi n + \phi^{(2)}\right)}}{2i} \psi^{\dagger}_{n}\psi_{n}. \label{eq:Rice_Mele_incoherence_3}
\end{align}
We now choose the second and third synthetic lattice vectors to be $\hat{y}$ and $\hat{z}$. 
The promote 3D system will then have
\begin{align}
    & \vec{a}_{1} = (1,0,0),\ \vec{c}_{1} = (0,1,0),\ \vec{c}_{2} = (0,0,1) \\
    & \vec{g}_{1} = 2\pi(1,0,0),\ \vec{G}_{1} = 2\pi(0,1,0),\ \vec{G}_{2} = 2\pi(0,0,1),
\end{align}
in terms of the notation from Eqs.~(\ref{eq:reciprocal_1}--\ref{eq:reciprocal_4}). 
Notice that $\{\vec{a}_{1},\vec{c}_{1},\vec{c}_{2} \}$ describe a 3D cubic lattice. 
Using the Fourier expansion convention in Table~\ref{tab:model_summary} and Table~\ref{tab:hopping_1} with $\vec{q}^{(1)}= \vec{q}^{(2)} = (1/2,0,0)$ and Eq.~(\ref{eq:Rice_Mele_incoherence_3}), we obtain the hopping terms in this case summarized in Table~\ref{tab:hopping_2_incoherence}. 
Multiplying the entries for hopping matrices and Peierls phases, the Hamiltonian for the promoted 3D model is
\begin{equation}
\hspace*{-0cm}
    H_{\text{3D}} = \sum_{n,m,l} \left( t\psi^{\dagger}_{n+1,m,l}\psi_{n,m,l}- \frac{i\delta t}{2}e^{i\pi \left(n+\frac{1}{2} \right)} \psi^{\dagger}_{n+1,m-1,l}\psi_{n,m,l}+ \frac{i\delta t}{2}e^{-i\pi \left(n+\frac{1}{2} \right)} \psi^{\dagger}_{n+1,m+1,l}\psi_{n,m,l}-  \frac{i\Delta}{2}e^{-i\pi n} \psi^{\dagger}_{n,m,l+1}\psi_{n,m,l} + \text{h.c.}  \right),
\end{equation}
where $\psi^{\dagger}_{n,m,l}$ is the creation operator for an electron at position $(n,m,l)$, and the vector potential to which this 3D model is coupled is 
\begin{align}
    \vec{A} = \frac{1}{2\pi} \left( \vec{G}_{1} \left(\vec{q}^{(1)}\cdot\left[ \left( \vec{r}\cdot \vec{g}_{1} \right)\vec{a}_{1} \right] \right) +  \vec{G}_{2} \left(\vec{q}^{(2)}\cdot\left[ \left( \vec{r}\cdot \vec{g}_{1} \right)\vec{a}_{1} \right] \right) \right) = (0,\pi x, \pi x).
\end{align}
This $\vec{A}$ produces a magnetic field $\vec{B} = (0,-\pi,\pi)$ in the promoted 3D system such that there is a $\pi$-flux threading through the plaquettes in $zx$- and $xy$-planes. 
In contrast to Sec.~\ref{sec:ex_1D_RM_chain} where the promoted 2D model is a Chern insulator (see Sec.~\ref{sec:Thouless_pump_1D_Rice_Mele}), the 1D Rice-Mele chain with phase offset incoherence promotes to a 3D gapless model. 
We can see this by noting that with $\phi^{(1)} = \pi /2$ and $\phi^{(2)} = 0$, Eq.~(\ref{eq:Rice_Mele_incoherence}) becomes
\begin{align}
    \sum_{n} \left( t  \psi^{\dagger}_{n+1}\psi_{n} + \text{h.c.}  \right),
\end{align}
which is the Hamiltonian for the 1D Rice-Mele chain at the gapless critical point.

\begin{table}[h]
\centering
\begin{tabular}{|c|c|c|}
\hline
Hopping from $(n,m,l)$ to & Hopping matrices & Peierls phases  \\
\hline
\hline
$(n+1,m,l)$ & $t$ & $1$ \\
\hline
$(n,m,l+1)$  & $ \Delta / (2i) = -i\Delta / 2$ & $e^{-i \pi n}$  \\
\hline 
$(n+1,m+1,l)$ & $(\delta t/ 2) \cdot e^{i \pi /2} = i\delta t/ 2 $  & $e^{-i \pi \left( n + \frac{1}{2} \right)}$ \\
\hline
$(n+1,m-1,l)$ & $(\delta t/ 2) \cdot e^{-i \pi /2} = -i\delta t/ 2 $ & $e^{i \pi \left( n + \frac{1}{2} \right)}$ \\
\hline
\end{tabular}
\caption{Hopping matrices and the corresponding Peierls phases for the promoted $3$D model for a $1$D Rice-Mele chain with an incoherence between the phase offsets of on-site and hopping modulations. 
The hopping matrices along the opposite directions of those listed here are omitted and can be obtained through Hermitian conjugation.}
\label{tab:hopping_2_incoherence}
\end{table}

\subsubsection{\label{sec:1D_to_2D_hexa}1D modulated system $\to$ 2D hexagonal lattice under a perpendicular magnetic field}

Let us now explore what happens when our dimensionally-promoted lattice vectors are non-orthogonal. 
Consider a $1$D modulated system with both on-site and nearest-neighbor hopping modulations, with Hamiltonian
\begin{align}
    H_{1D} = & \sum_{n_{1}} \left( \psi^{\dagger}_{n_{1}+1} [H_{\vec{a}_{1}}] \psi_{n_{1}} + \psi^{\dagger}_{n_{1}+1} e^{i \left( 2\pi \vec{q} \cdot \left( n_{1} + \frac{1}{2} \right)\vec{a}_{1} + \phi \right)} [H_{\vec{a}_{1}-\vec{a}_{2}}] \psi_{n_{1}} + \text{h.c.} \right)  \label{eq:1D_hexagonal_1} \\
    & + \sum_{n_{1}} \left(  \psi^{\dagger}_{n_{1}} [H_{0}] \psi_{n_{1}} + \psi^{\dagger}_{n_{1}} [V_{n_{1}}] \psi_{n_{1}} \right), \label{eq:1D_hexagonal_2}
\end{align}
where $\text{h.c.}$ denotes hermitian conjugation, and $\psi^{\dagger}_{n_{1}}$ is the creation operator for an electron at position $n_{1}\vec{a}_{1}$. 
Additionally, $[V_{n_{1}}]$ is a modulated on-site interaction which can be decomposed into
\begin{align}
    [V_{n_{1}}] = e^{-i \left( 2\pi \vec{q} \cdot n_{1}\vec{a}_{1} + \phi\right)} [H_{\vec{a}_{2}}] +  e^{i \left( 2\pi \vec{q} \cdot n_{1}\vec{a}_{1} + \phi \right)} [H_{\vec{a}_{2}}]^{\dagger}, \label{eq:1D_hexagonal_3}
\end{align}
and $\vec{a}_{1} = (1,0)$ and $\vec{q} = (q,0)$. 
We also have a modulated nearest-neighbor hopping term from site $n_{1}$ to $n_{1} + 1$ given by the second term
\begin{equation}
e^{i \left( 2\pi \vec{q} \cdot \left( n_{1} + \frac{1}{2} \right)\vec{a}_{1} + \phi \right)} [H_{\vec{a}_{1}-\vec{a}_{2}}]
\end{equation}
in Eq.~(\ref{eq:1D_hexagonal_1}). 
The $\vec{a}_{2}$ in Eq.~(\ref{eq:1D_hexagonal_1}--\ref{eq:1D_hexagonal_3}) will become useful when we promote the dimension: 
At this stage, $\vec{a}_{2}$ is an unspecified label, though we have anticipated that it will be identified with the synthetic direction in the promoted lattice. 

We promote the dimension of this 1D modulated system to 2D and choose the second, synthetic lattice vector as $\vec{a}_{2} = (1/2, \sqrt{3}/2)$. 
Thus, $\{\vec{a}_{1},\vec{a}_{2} \}$ forms a 2D hexagonal lattice. 
The reciprocal lattice vectors in this promoted 2D space are then
\begin{align}
    & \vec{g}_{1} = 2\pi \left( 1 , -\frac{1}{\sqrt{3}} \right), \\
    & \vec{g}_{2} = 2\pi \left( 0,\frac{2}{\sqrt{3}}\right),
\end{align}
where we bear in mind that, if compared with our previous notation from Eqs.~(\ref{eq:reciprocal_1}--\ref{eq:reciprocal_4}), we should identify $\vec{a}_{2} \leftrightarrow \vec{c}_{1}$ and $\vec{g}_{2} \leftrightarrow \vec{G}_{1}$. 
Using the Fourier expansion convention in Table~\ref{tab:model_summary} and Table~\ref{tab:hopping_1} with $\vec{q} = (q,0)$, and Eqs.~(\ref{eq:1D_hexagonal_1}--\ref{eq:1D_hexagonal_3}), we obtain the hopping terms in this case in Table~\ref{tab:hopping_3}. 
Multiplying the entries for hopping matrices and Peierls phases, the Hamiltonian for the promoted 2D model is
\begin{align}
    H_{2D}= &  \sum_{n_{1},n_{2}} \left( \psi^{\dagger}_{n_{1}+1,n_{2}} [H_{\vec{a}_{1}}] \psi_{n_{1},n_{2}} + \psi^{\dagger}_{n_{1},n_{2}+1} [H_{\vec{a}_{2}}]e^{-i 2\pi \vec{q} \cdot n_{1}\vec{a}_{1}} \psi_{n_{1},n_{2}} + \psi^{\dagger}_{n_{1}+1,n_{2}-1} [H_{\vec{a}_{1}-\vec{a}_{2}}] e^{i 2\pi \vec{q} \cdot \left( n_{1} + \frac{1}{2} \right)\vec{a}_{1}} \psi_{n_{1},n_{2}} + \text{h.c.} \right)  \\
    & + \sum_{n_{1},n_{2}} \psi^{\dagger}_{n_{1},n_{2}} [H_{0}] \psi_{n_{1},n_{2}},
\end{align}
where $\psi^{\dagger}_{n_{1},n_{2}}$ is the creation operator for an electron at position $n_{1}\vec{a}_{1}+ n_{2}\vec{a}_{2}$, and the vector potential to which this 2D model is coupled is 
\begin{align}
    \vec{A} = \frac{1}{2\pi}  \vec{g}_{2} \left( \vec{q}\cdot \left[ \left( \vec{r} \cdot \vec{g}_{1}\right) \vec{a}_{1} \right] \right) = \left( 0 , \frac{4\pi q}{\sqrt{3}} \left( x - \frac{y}{\sqrt{3}} \right) \right).
\end{align}
This $\vec{A}$ reproduces a perpendicular magnetic field $\vec{B} = \frac{4\pi q}{\sqrt{3}} \hat{z}$ in the promoted 2D system. 
We thus see that this 1D modulated system may be used to map out the Hofstadter spectrum of a hexagonal lattice with an irrational  magnetic flux.

\begin{table}[h]
\centering
\begin{tabular}{|c|c|c|}
\hline
Hopping from $n_{1}\vec{a}_{1} + n_{2}\vec{a}_{2}$ to & Hopping matrices & Peierls phase  \\
\hline
\hline
$n_{1}\vec{a}_{1} + n_{2}\vec{a}_{2}$ & $[H_{0}]$ & $1$ \\
\hline
$(n_{1}+1)\vec{a}_{1} + n_{2}\vec{a}_{2}$ & $[H_{\vec{a}_{1}}]$ & $1$ \\
\hline
$n_{1}\vec{a}_{1} + (n_{2}+1)\vec{a}_{2}$  & $[H_{\vec{a}_{2}}]$  & $e^{-i 2\pi \vec{q} \cdot n_{1}\vec{a}_{1}}$ \\
\hline
$(n_{1}+1)\vec{a}_{1} + (n_{2}-1)\vec{a}_{2}$ & $e^{i 2\pi \vec{q} \cdot \frac{1}{2}\vec{a}_{1}} [H_{\vec{a}_{1}-\vec{a}_{2}}]e^{-i \pi \vec{q}\cdot \vec{a}_{1}} = [H_{\vec{a}_{1}-\vec{a}_{2}}]$   & $e^{i 2\pi \vec{q}\cdot \left( n_{1} + \frac{1}{2} \right)\vec{a}_{1}}$\\
\hline
\end{tabular}
\caption{Hopping matrices and the corresponding Peierls phases for the promoted $2$D model for a 1D modulated chain described by Eqs.~(\ref{eq:1D_hexagonal_1}--\ref{eq:1D_hexagonal_3}). 
The hopping matrices along the opposite directions for those listed here are omitted and can be obtained through Hermitian conjugate.}
\label{tab:hopping_3}
\end{table}

\subsubsection{2D hexagonal lattice $\to$ 3D hexagonal lattice coupled to a $U(1)$ gauge field}

Consider a 2D modulated system with one on-site modulation, hexagonal lattice and Hamiltonian
\begin{align}
    H_{2D}= \sum_{\vec{n},\vec{m}} \psi^{\dagger}_{\vec{n}+\vec{m}} [H_{\vec{m}}] \psi_{\vec{n}} + \sum_{\vec{n}} \psi^{\dagger}_{\vec{n}} [V_{\vec{n}}] \psi_{\vec{n}}, \label{eq:2D_hexa_to_3D_hexa_1}
\end{align}
where $\vec{n} = (n_{1},n_{2})$, $\vec{m} = (m_{1},m_{2})$, $\vec{a}_{1} = \left(1/2,\sqrt{3}/2\right)$, $\vec{a}_{2} = \left(-1/2,\sqrt{3}/2\right)$, and $\psi^{\dagger}_{\vec{n}}$ is the creation operator for an electron at lattice position $n_{1} \vec{a}_{1} + n_{2} \vec{a}_{2}$.
The matrix $[H_{\vec{m}}]$ describes hopping terms from position $n_{1} \vec{a}_{1} + n_{2} \vec{a}_{2}$ to $(n_{1}+m_{1}) \vec{a}_{1} + (n_{2}+m_{2})\vec{a}_{2}$ which can include long range hopping terms. 
The on-site modulation can be expanded as 
\begin{align}
    [V_{\vec{n}}] = \sum_{p} [V_{p}] e^{i p \left( 2\pi \vec{q} \cdot \left(\sum_{j=1}^{2}n_{j}\vec{a}_{j} \right) + \phi \right) }, \label{eq:2D_hexa_to_3D_hexa_2}
\end{align}
where $p \in \mathbb{Z}$, $[V_{p}]$ is the $p^{\text{th}}$ Fourier coefficient of $[V_{\vec{n}}]$ and $\vec{q} = (q_{x},q_{y})$ is the modulation wave vector parallel to the 2D system. 
We next promote the dimension of this 2D system to 3D and choose the third, synthetic lattice vector $\vec{a}_{3}$ as $(0,0,1)$. 
The lattice vectors in the promoted 3D space are then
\begin{align}
    \vec{a}_{1} = \left( \frac{1}{2},\frac{\sqrt{3}}{2},0 \right),\  \vec{a}_{2} = \left( -\frac{1}{2},\frac{\sqrt{3}}{2},0 \right),\ \vec{a}_{3} = \left( 0,0,1 \right),
\end{align}
which describes a 3D hexagonal lattice. 
The corresponding reciprocal lattice vectors are
\begin{align}
    \vec{g}_{1} = 2\pi \left( 1, \frac{1}{\sqrt{3}},0 \right),\ \vec{g}_{2} = 2\pi \left( -1, \frac{1}{\sqrt{3}},0 \right),\ \vec{g}_{3} = 2\pi (0,0,1).
\end{align}
As in Sec.~\ref{sec:1D_to_2D_hexa}, we bear in mind that in comparing with Eqs.~(\ref{eq:reciprocal_1}--\ref{eq:reciprocal_4}), we should identify $\vec{a}_{3} \leftrightarrow \vec{c}_{1} $ and $\vec{g}_{3} \leftrightarrow \vec{G}_{1}$. 
Using the Fourier expansion convention in Table~\ref{tab:model_summary} and Table~\ref{tab:hopping_1} with $\vec{q} = (q_{x},q_{y})$, and Eqs.~(\ref{eq:2D_hexa_to_3D_hexa_1}--\ref{eq:2D_hexa_to_3D_hexa_2}), we obtain the hopping terms given in Table~\ref{tab:hopping_4}. 
Multiplying the entries for hopping matrices and Peierls phases, the Hamiltonian for the promoted 3D model is
\begin{align}
    H_{3D} = \sum_{\vec{n},\vec{m}} \psi^{\dagger}_{\vec{n} + \vec{m}} [H_{\vec{m}}] \psi_{\vec{n}} + \sum_{\vec{n},p} \psi^{\dagger}_{\vec{n} - (0,0,p)} [V_{p}] e^{i 2\pi p \vec{q} \cdot \left( \sum_{j=1}^{2}n_{j}\vec{a}_{j} \right)} \psi_{\vec{n}},
\end{align}
where $\vec{n} = (n_{1},n_{2},n_{3})$, $\vec{m }= (m_{1},m_{2},0)$, $\psi^{\dagger}_{\vec{n}}$ is the creation operator for an electron at lattice position $\sum_{j=1}^{3}n_{j}\vec{a}_{j}$, and the vector potential to which this 3D model is coupled is 
\begin{align}
    \vec{A} = \frac{1}{2\pi}  \vec{g}_{3} \left( \vec{q}\cdot \left[ \left( \vec{r} \cdot \vec{g}_{1}\right) \vec{a}_{1} + \left( \vec{r} \cdot \vec{g}_{2}\right) \vec{a}_{2} \right] \right) = \left( 0,0,2\pi\left(q_{x}x + q_{y}y \right) \right).
\end{align}
This $\vec{A}$ produces a magnetic field $\vec{B} = \left(2\pi q_{y},-2\pi q_{x},0 \right)$ in the promoted 3D space.

\begin{table}[h]
\centering
\begin{tabular}{|c|c|c|}
\hline
Hopping from $\sum_{j=1}^{3}n_{j}\vec{a}_{j}$ to & Hopping matrices & Peierls phase  \\
\hline
\hline
$\sum_{j=1}^{3}n_{j}\vec{a}_{j} + \sum_{j=1}^{2} m_{j}\vec{a}_{j}$  & $[H_{\vec{m}}]$ & $1$ \\
\hline
$\sum_{j=1}^{3}n_{j}\vec{a}_{j} - p \vec{a}_{3}$ & $[V_{p}]$ & $e^{i2\pi p \vec{q} \cdot \left(\sum_{j=1}^{2}n_{j}\vec{a}_{j} \right)}$ \\
\hline
\end{tabular}
\caption{Hopping matrices and the corresponding Peierls phases for the promoted $3$D model from a 2D hexagonal lattice. 
The hopping matrices along the opposite directions of those listed here are omitted and can be obtained through Hermitian conjugation.}
\label{tab:hopping_4}
\end{table}

\section{\label{sec:Thouless_pump_1D_Rice_Mele}Thouless pump and dimensional promotion for 1D Rice-Mele model}

\begin{figure}[h]
  \includegraphics[scale=0.4]{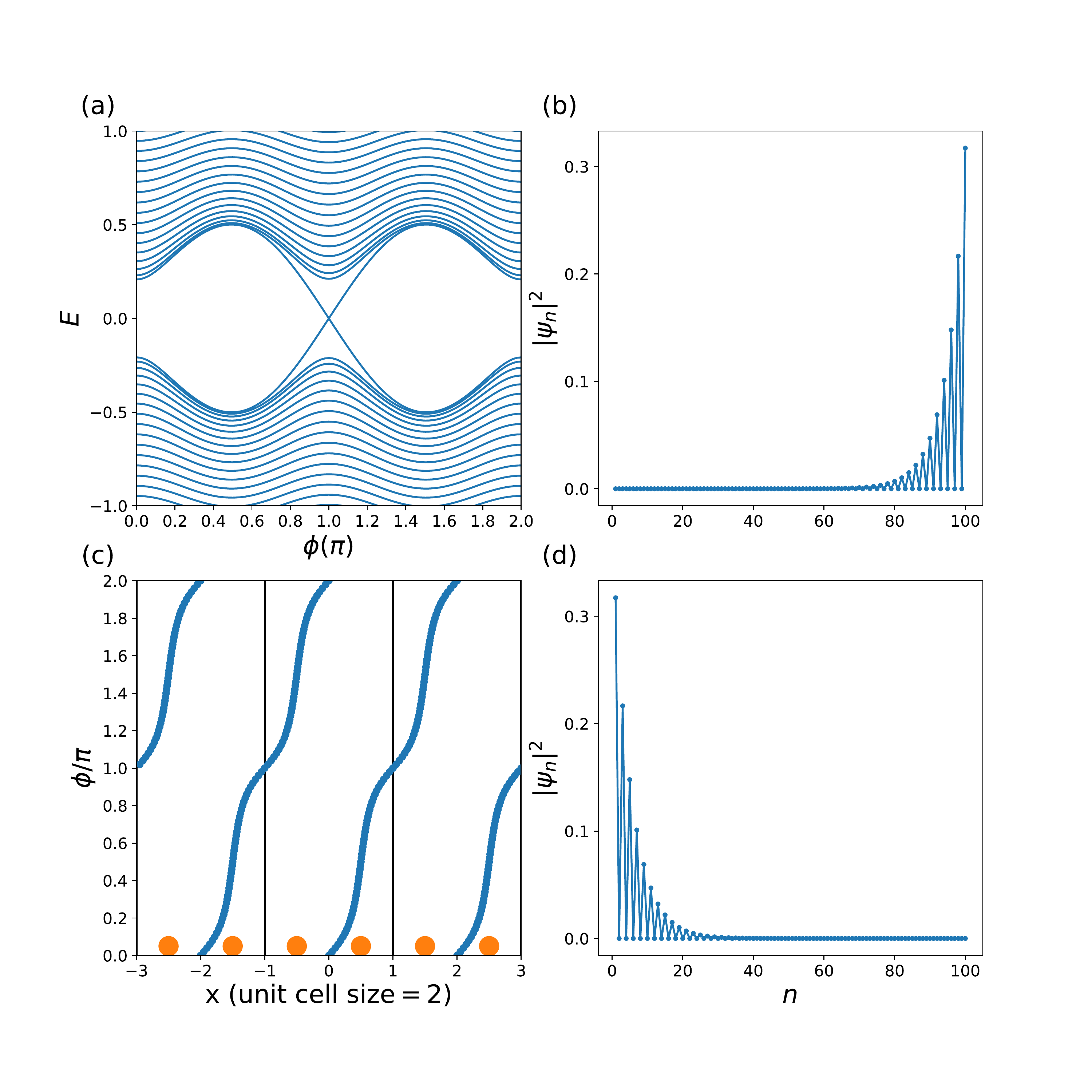}
  \caption{(a) $\phi$-sliding spectrum of the Rice-Mele model with $t = 1$, $\delta t=-0.1$ and $\Delta = 0.5$. 
  (c) The shifting of the valence band Wannier center (blue dots) as a function of $\phi$. 
  The vertical black lines denote the boundary of the unit cell, which has length $2$. 
  The orange dots denote the positions of the tight-binding basis orbitals. 
  The origin of the $x$-axis is placed at the middle of the bond with hopping $t - \delta t \cos{\phi}$ where $t = 1$ and $\delta t = -0.1$. 
  In other words, the unit cell is formed by the sites $n=1$ and $n=2$ in Eq.~(\ref{eq:supp_Rice_Mele}) and the origin is placed at the midpoint between these two sites. 
  The inversion centers at $\phi=0$ and $\pi$ lie at integer values of $x$, and as such the Wannier centers at these two $\phi$ are located either at the center of the unit cell ($\phi=0$) or the boundaries between unit cells ($\phi=\pi)$\cite{Aris2014}. 
  Also, this unit cell choice is commensurate with the finite size system in (a), (b) and (d), where we choose $n = 1 ,\ldots, 100$. 
  (b) $\&$ (d) Probability distribution of localized modes around the two ends at $\phi = 0.9 \pi$ and energy $E = -0.1545$ $\&$ $+0.1545$. 
  Notice that during the adiabatic pumping from $\phi =0$ to $2\pi$, it is the boundary state at the left [right] end, see (d) [(b)], that flows into [out of] the subspace of occupied states. 
  This is consistent with (c) where the Wannier center flows toward the right and reappear at the left boundary of the unit cell at $\phi = \pi$.}
  \label{Fig_Rice_Mele}
\end{figure}

Here we review how topological properties such as the Thouless pump\cite{Thouless_pump_original_paper} in the 1D Rice-Mele model\cite{RiceMele} (SSH chain\cite{su1979solitons}) can be attributed to a 2D lattice model coupled to a $U(1)$ gauge field, via our dimensional promotion procedure. 
Consider the 1D Rice-Mele model
\begin{align}
    H_{\text{Rice-Mele}} = \sum_{n}  \left( t + \delta t (-1)^{n} \cos{\phi} \right)c^{\dagger}_{n+1}c_{n} + \text{h.c.} + \sum_{n} (-1)^{n+1} \Delta  \sin{\phi} c^{\dagger}_{n}c_{n},
    \label{eq:supp_Rice_Mele}
\end{align}
where $c^{\dagger}_{n}$ is the creation operator for an electron at 1D sites $n \in \mathbb{Z}$, and $\text{h.c.}$ means the Hermitian conjugate of all terms before it. 
This Hamiltonian describes a $1$D chain with a twofold (Peierls) CDW distortion. 
Suppose we prepare the ground state (with Fermi level $E_{F}=0$) for $H_{\text{Rice-Mele}}$ with $t = 1$, $\delta t=-0.1$, $\Delta = 0.5$ at $\phi = 0$. 
Before dimensional promotion, let us first review the properties of this 1D Hamiltonian. 
As the phase $\phi$ adiabatically changes from $0$ to $2\pi$, the single valence band Wannier center\cite{Kohn59,Brouder2007,Marzari2012,ksv,fu2006time} in the bulk is pumped by one unit cell (two lattice sites), as shown in Fig.~\ref{Fig_Rice_Mele} (c). 
The bulk polarization then changes by $(-1)\times d = -d$. 
This bulk polarization change is quantized\cite{ksv,resta1994macroscopic} in units of the unit cell length $d=2$. 
This is the classic realization of the Thouless pump\cite{Thouless_pump_original_paper} in 1D. 
In Fig.~\ref{Fig_Rice_Mele} (a) we show the $\phi$-sliding spectrum (defined in Sec.~IV of the main text) for a finite chain with size $100$ and positions $n = 1 ,\ldots, 100$. 
We see that in addition to the gapped bulk states, there are two linearly dispersing modes that cross the bulk gap. 
This indicates that during the pumping process, localized charges at the right and left end (see Figs.~\ref{Fig_Rice_Mele} (b) and (d)) of the chain flow out of and into the occupied state subspace, respectively. 
This is the celebrated topological origin of the quantized polarization change in the bulk, and the existence of boundary states crossing the energy gap.

Let us now see how these properties emerge in our 2D dimensionally-promoted picture. 
Identifying $\phi$ as the crystal momentum $k_{y}$ along the second, synthetic dimension $y$, $H_{\text{Rice-Mele}}$ is equivalent to the Bloch Hamiltonian of a 2D model (see Sec.~\ref{sec:ex_1D_RM_chain} for the detailed derivation) 
\begin{align}
    H_{\text{2D}} = \sum_{n,m} \left( tc^{\dagger}_{n+1,m}c_{n,m}- \frac{i\delta t}{2}e^{i\pi \left(n+\frac{1}{2} \right)} c^{\dagger}_{n+1,m-1}c_{n,m}+ \frac{i\delta t}{2}e^{-i\pi \left(n+\frac{1}{2} \right)} c^{\dagger}_{n+1,m+1}c_{n,m}-  \frac{i\Delta}{2}e^{-in\pi} c^{\dagger}_{n,m+1}c_{n,m} \right)+ \text{h.c.}, \label{eq:1D_Rice_Mele_to_2D_1}
\end{align}
with a fixed $k_{y}$. Here $c^{\dagger}_{n,m}$ is the creation operator for an electron at 2D site $(n,m)$ on a square lattice. 
If, during the adiabatic process, $H_{\text{Rice-Mele}}$ is always gapped, the quantized polarization change is determined by the Chern number\cite{tknn,niu1984quantised,niu1985quantized,Aris2014,fu2006time,bernevigbook} of the occupied bands of $H_{\text{2D}}$. 
We can thus characterize the topological properties shown in Fig.~\ref{Fig_Rice_Mele} by a Chern number $C=1$. 
We notice that in addition to hopping terms from site $(n,m)$ to $(n\pm1,m)$ and $(n,m\pm1)$, $H_{\text{2D}}$ also has hopping terms going from site $(n,m)$ to $(n\pm1,m+1)$ and $(n\pm1,m- 1)$. 
This is in contrast to the standard textbook correspondence between the SSH chain and a 2D Chern insulator, where a re-embedding of orbitals within the enlarged unit cell is typically used to obtain a 2D model with only perpendicular hoppings along the $\hat{x}$ and $\hat{y}$ directions.
Note also that $H_{\text{2D}}$ is a 2D lattice model coupled to a $U(1)$ gauge field $\vec{A} = (0,\pi x)$ through the Peierls substitution\cite{Peierls_substitution} assuming the electron has charge $-1$. 
This $\vec{A}$ produces a uniform $U(1)$ magnetic field threading $\pi$-flux per plaquette, since the magnetic field $\vec{B} = \vec{\nabla} \cross \vec{A} = \pi \hat{z}$. 
Reinserting factors of $\hbar$ and $|e|$, this corresponds to half a flux quantum $\Phi_{0} = 2\pi \hbar / |e|$ per plaquette. 
The localized and mid-gap states in $H_{\text{Rice-Mele}}$ are then identified as the chiral edge modes due to the quantum Hall effect in $H_{\text{2D}}$.

\section{\label{sec:promoted_lattice_model} Promoted lattice models}

For completeness, in this section we give the promoted lattice models in position space corresponding to the 2D modulated system with helical sliding modes and the 3D Weyl semimetal with mean-field charge-density waves (CDWs) in Secs.~V and VI of the main text, respectively.

\subsection{\label{sec:promoted_helical_lattice_model}2D modulated system with helical sliding modes}

The promoted 3D lattice model\cite{Wieder_spin_decoupled_helical_HOTI} for the 2D modulated system with helical sliding modes in the main text is
\begin{align}
    H = \sum_{\vec{n}} & \left( \psi^{\dagger}_{\vec{n}+\hat{x}} [H_{+\hat{x}}]\psi_{\vec{n}} + \psi^{\dagger}_{\vec{n}-\hat{x}} [H_{+\hat{x}}]^{\dagger}\psi_{\vec{n}} +\psi^{\dagger}_{\vec{n}+\hat{z}} [H_{+\hat{z}}]\psi_{\vec{n}} + \psi^{\dagger}_{\vec{n}-\hat{z}} [H_{+\hat{z}}]^{\dagger}\psi_{\vec{n}}+ \psi^{\dagger}_{\vec{n}} [H_{\text{on-site}}] \psi_{\vec{n}} \nonumber \right. \\
    & \left. + \psi^{\dagger}_{\vec{n}+\hat{y}} [H_{+\hat{y}}] e^{-i\text{A}_{\vec{n}+\hat{y},\vec{n}}}\psi_{\vec{n}} + \psi^{\dagger}_{\vec{n}-\hat{y}} e^{i\text{A}_{\vec{n},\vec{n}-\hat{y}}}[H_{+\hat{y}}]^{\dagger} \psi_{\vec{n}}  \right) \label{eq:lattice_model_helical_sliding} 
\end{align}
with
\begin{align}
    & [H_{+\hat{x}}] = \frac{v_{x}}{2}\tau_{z}\mu_{0}\sigma_{0} - \frac{u_{x}}{2i}\tau_{y}\mu_{y}\sigma_{0},\\
    & [H_{+\hat{y}}] = \frac{v_{y}}{2}\tau_{z}\mu_{0}\sigma_{0} - \frac{v_{H}}{2 i} \tau_{y}\mu_{z}\sigma_{z},\\
    & [H_{+\hat{z}}] = \frac{v_{z}}{2}\tau_{z}\mu_{0}\sigma_{0} - \frac{u_{z}}{2i}\tau_{x}\mu_{0}\sigma_{0}, \\
    & [H_{\text{on-site}}] = m_{1}\tau_{z}\mu_{0}\sigma_{0}+m_{2}\tau_{z}\mu_{x}\sigma_{0} + m_{3}\tau_{z}\mu_{z}\sigma_{0}+m_{v_{1}}\tau_{0}\mu_{z}\sigma_{0} + m_{v_{2}}\tau_{0}\mu_{x}\sigma_{0}.
\end{align}
The $SU(2)$ lattice gauge field is given by
\begin{align}
    \text{A}_{\vec{n}+\hat{y},\vec{n}} =2\pi \left( q_{x}n_{x}+q_{z}n_{z} \right) \tau_{0}\mu_{0}\sigma_{z}. \label{eq:lattice_SU2_A}
\end{align}
The $SU(2)$ gauge field in the continuous coordinate representation is then (dropping the $\tau_{0}\mu_{0}$)
\begin{align}
    \vec{A} = (0,2\pi(q_{x}x+q_{z}z)\sigma_{z},0). \label{eq:A_SU2}
\end{align}
The corresponding $SU(2)$ magnetic field $\vec{B} = \vec{\nabla} \cross \vec{A} - i \vec{A} \cross \vec{A}$ is\cite{Estienne_2011}
\begin{align}
    \vec{B} = (-2\pi q_{z}\sigma_{z},0,2\pi q_{x}\sigma_{z}). \label{eq:B_SU2_1}
\end{align}
Eq.~(\ref{eq:lattice_model_helical_sliding}) describes a helical higher-order topological insulator (HOTI) coupled to a $SU(2)$ gauge field. 
If we Fourier transform Eq.~(\ref{eq:lattice_model_helical_sliding}) along $y$ and regard $k_{y}$ (wavenumber along $y$) as the sliding phase $\phi$, we can obtain the 2D modulated system in the main text.

\subsection{\label{sec:promoted_4D_lattice_model}3D Weyl semimetal with mean-field charge density waves}

The promoted 4D lattice model for the 3D Weyl semimetal with mean-field CDW order\cite{dynamical_axion_insulator_BB} in the main text is
\begin{align}
    H&=\left(\sum_{\vec{n}}\left[-it_x\psi^\dag_{\vec{n}+\hat{x}}\sigma_x \psi_{\vec{n}}-it_y \psi^\dag_{\vec{n}+\hat{y}}\sigma_y \psi_{\vec{n}}+t_z\psi^\dag_{\vec{n}+\hat{z}}\sigma_z \psi_{\vec{n}} + |\Delta| e^{-i2 \pi q n_{z}} \psi^\dag_{\vec{n}+\hat{w}} \sigma_{z} \psi_{\vec{n}} \right]\right. \nonumber \\
    & \left. +\sum_{\vec{n}}\frac{m}{2}\left(\psi^\dag_{\vec{n}+\hat{x}}\sigma_z \psi_{\vec{n}} + \psi^\dag_{\vec{n}+\hat{y}}\sigma_z \psi_{\vec{n}}  -2 \psi^\dag_{\vec{n}}\sigma_z \psi_{\vec{n}}  \right) -\sum_{\vec{n}} t_z \left(\cos (\pi q)\right) \psi^\dag_{\vec{n}}\sigma_z \psi_{\vec{n}}\right) +\mathrm{h.c.}, \label{eq:lattice_model_4D_nodal_line}
\end{align}
where the phase factors correspond to Peierls substitution of a 4D $U(1)$ gauge field
\begin{align}
    \vec{A} = (0,0,0,2\pi q z).
\end{align}
The only non-zero components of the $U(1)$ field strength $F_{\mu \nu} = \partial_{\mu} A_{\nu} - \partial_{\nu} A_{\mu}$ is
\begin{align}
    F_{zw} = -F_{wz} = \partial_{z}A_{w} - \partial_{w}A_{z}=2 \pi q, \label{eq:4D_U1_Fzw}
\end{align}
which threads through the $zw$ plane. 
Without coupling to a 4D $U(1)$ gauge field, the Bloch Hamiltonian of Eq.~(\ref{eq:lattice_model_4D_nodal_line}) with $q=0$ is
\begin{align}
    H(\vec{k})=& -2[t_x \sin (k_x) \sigma_x +t_y \sin (k_y) \sigma_y] +2t_z[\cos (k_z) -\cos (\pi q)] \sigma_{z} -m[2-\cos (k_x) - \cos (k_y)] \sigma_{z} + 2|\Delta|\cos{(k_{w})}\sigma_{z},
\label{eq:lattice_model_4D_nodal_line_bloch}
\end{align}
which has a nodal line in the $k_{z}$-$k_{w}$ plane (with $k_{x}=k_{y}=0$) defined by
\begin{align}
    t_{z}\cos(k_{z}) +  |\Delta| \cos(k_{w}) = t_{z}\cos{(\pi q)}. \label{eq:lattice_nodal_line}
\end{align}
Therefore, Eq.~(\ref{eq:lattice_model_4D_nodal_line}) describes a 4D nodal line system coupled to a 4D $U(1)$ gauge field where the corresponding field strength Eq.~(\ref{eq:4D_U1_Fzw}) threads through the area enclosed by the nodal line defined in Eq.~(\ref{eq:lattice_nodal_line}). 
If we Fourier transform Eq.~(\ref{eq:lattice_model_4D_nodal_line}) along $w$ and regard $k_{w}$ as the sliding phase $\phi$, we obtain the model for a 3D Weyl semimetal with mean-field CDW order given in the main text.

\section{\label{sec:inv_sym_gauge_tr}Inversion symmetry up to a gauge transformation}
 
In this section we will show that if a lattice has inversion symmetry, such as the 3D models promoted from our examples of 2D modulated systems with chiral and helical sliding modes, then upon coupling to a $U(1)$ or $SU(2)$ gauge field producing a constant magnetic field, the inversion symmetry is still preserved {\it up to a gauge transformation}. 
We will do this in details in the simplest case, which is a 2D square lattice coupled to a perpendicular magnetic field. 
We will briefly mention the generalization to the 3D cases, which corresponds to the dimensional promotion from 2D modulated systems.

\subsection{2D system with inversion symmetry}

Let us consider a 2D square lattice with only one degree of freedom within each unit cell labelled by $(n,m) \in \mathbb{Z}^{2}$
\begin{align}
    H = -t\sum_{n,m} \left( \psi^{\dagger}_{n+1,m}\psi_{n,m}  +\psi^{\dagger}_{n-1,m}\psi_{n,m}+ e^{-i2\pi bn}\psi^{\dagger}_{n,m+1}\psi_{n,m}+ e^{+i2\pi bn}\psi^{\dagger}_{n,m-1}\psi_{n,m} \right). \label{eq:inv_sym_gauge_tr_1}
\end{align}
Eq.~(\ref{eq:inv_sym_gauge_tr_1}) is coupled to a $U(1)$ gauge field $\vec{A} = 2\pi bx \hat{y}$ through Peierls substitution where we have assumed the particle carries charge $-1$. 
This $U(1)$ gauge field produces a constant perpendicular magnetic field $\vec{B} = 2\pi b \hat{z}$. 
If we turn off the $U(1)$ gauge field by setting $b = 0$, Eq.~(\ref{eq:inv_sym_gauge_tr_1}) will have inversion symmetry, where the inversion operators are defined by 
\begin{align}
    & \mathcal{I}_{n_{c},m_{c}} \psi^{\dagger}_{n,m} \mathcal{I}_{n_{c},m_{c}}^{-1} = \psi^{\dagger}_{2n_{c}-n,2m_{c}-m}, \label{eq:inv_sym_gauge_tr_2} \\
    & \mathcal{I}_{n_{c},m_{c}} \psi_{n,m} \mathcal{I}_{n_{c},m_{c}}^{-1} = \psi_{2n_{c}-n,2m_{c}-m}. \label{eq:inv_sym_gauge_tr_3}
\end{align}
Here $\mathcal{I}_{n_{c},m_{c}}$ is a unitary operator and $(n_{c},m_{c})$ is the inversion center (which can be any lattice point). We then have
\begin{align}
    \mathcal{I}_{n_{c},m_{c}} H \mathcal{I}_{n_{c},m_{c}}^{-1} = H\  \forall (n_{c},m_{c})
\end{align}
provided that the summation over $n$ and $m$ in Eq.~(\ref{eq:inv_sym_gauge_tr_1}) goes from $- \infty$ to $+\infty$. 
When $b \ne 0$, we must modify our unitary inversion operations in Eq.~(\ref{eq:inv_sym_gauge_tr_2}) and Eq.~(\ref{eq:inv_sym_gauge_tr_3}) to be
\begin{align}
    & \mathcal{I}_{n_{c},m_{c}} \psi^{\dagger}_{n,m} \mathcal{I}_{n_{c},m_{c}}^{-1} = \psi^{\dagger}_{2n_{c}-n,2m_{c}-m}e^{i4\pi b n_{c}(m-m_{c})}, \label{eq:inv_sym_gauge_tr_4} \\
    & \mathcal{I}_{n_{c},m_{c}} \psi_{n,m} \mathcal{I}_{n_{c},m_{c}}^{-1} = \psi_{2n_{c}-n,2m_{c}-m}e^{-i4\pi b n_{c}(m-m_{c})}, \label{eq:inv_sym_gauge_tr_5}
\end{align}
which acts as inversion through the center $(n_{c},m_{c})$ together with a gauge transformation.
With these modified inversion operations, the following three identities can be proved: 
\begin{align}
    & \mathcal{I}_{n_{c},m_{c}} H \mathcal{I}_{n_{c},m_{c}}^{-1} = H \ \forall (n_{c},m_{c}), \label{eq:inv_sym_gauge_tr_6} \\
    & \left( \mathcal{I}_{n_{c},m_{c}} \right)^{2} \psi^{\dagger}_{n,m} \left( \mathcal{I}_{n_{c},m_{c}}^{-1} \right)^{2} = \psi^{\dagger}_{n,m}, \label{eq:inv_sym_gauge_tr_7}\\
    & \left( \mathcal{I}_{n_{c},m_{c}} \right)^{2} \psi_{n,m} \left( \mathcal{I}_{n_{c},m_{c}}^{-1} \right)^{2} = \psi_{n,m}. \label{eq:inv_sym_gauge_tr_8}
\end{align}
We prove Eq.~(\ref{eq:inv_sym_gauge_tr_6}) as follows:
\begin{align}
    \mathcal{I}_{n_{c},m_{c}} H \mathcal{I}_{n_{c},m_{c}}^{-1} & = -t \sum_{n,m} \begin{bmatrix}
    \psi^{\dagger}_{2n_{c}-n-1,2m_{c}-m} e^{i 4\pi b n_{c}(m-m_{c})} \psi_{2n_{c}-n,2m_{c}-m} e^{-i 4\pi b n_{c}(m-m_{c})}  \\
    + \psi^{\dagger}_{2n_{c}-n+1,2m_{c}-m} e^{i 4\pi b n_{c}(m-m_{c})} \psi_{2n_{c}-n,2m_{c}-m} e^{-i 4\pi b n_{c}(m-m_{c})} \\
    + e^{-i2\pi b n } \psi^{\dagger}_{2n_{c}-n,2m_{c}-m-1}e^{i 4\pi b n_{c}(m+1-m_{c})} \psi_{2n_{c}-n,2m_{c}-m} e^{-i 4\pi b n_{c}(m-m_{c})}  \\
    + e^{+i2\pi b n } \psi^{\dagger}_{2n_{c}-n,2m_{c}-m+1}e^{i 4\pi b n_{c}(m-1-m_{c})} \psi_{2n_{c}-n,2m_{c}-m} e^{-i 4\pi b n_{c}(m-m_{c})}
    \end{bmatrix} \label{eq:pf_inv_1}\\
    &  = -t \sum_{n,m} \begin{bmatrix}
    \psi^{\dagger}_{2n_{c}-n-1,2m_{c}-m}   \psi_{2n_{c}-n,2m_{c}-m}   \\
    + \psi^{\dagger}_{2n_{c}-n+1,2m_{c}-m} \psi_{2n_{c}-n,2m_{c}-m}  \\
    + e^{-i2\pi b n } \psi^{\dagger}_{2n_{c}-n,2m_{c}-m-1}e^{i 4\pi b n_{c}} \psi_{2n_{c}-n,2m_{c}-m}   \\
    + e^{+i2\pi b n } \psi^{\dagger}_{2n_{c}-n,2m_{c}-m+1}e^{-i 4\pi b n_{c}} \psi_{2n_{c}-n,2m_{c}-m} 
    \end{bmatrix} \label{eq:pf_inv_2} \\
    &  = -t \sum_{n,m} \begin{bmatrix}
    \psi^{\dagger}_{n-1,m}   \psi_{n,m}   \\
    + \psi^{\dagger}_{n+1,m} \psi_{n,m}  \\
    + e^{-i2\pi b (-n+2n_{c}) } \psi^{\dagger}_{n,m-1}e^{i 4\pi b n_{c}} \psi_{n,m}   \\
    + e^{+i2\pi b (-n+2n_{c}) } \psi^{\dagger}_{n,m+1}e^{-i 4\pi b n_{c}} \psi_{n,m} 
    \end{bmatrix}  \label{eq:pf_inv_3} \\
    & = -t \sum_{n,m} \begin{bmatrix}
    \psi^{\dagger}_{n-1,m}   \psi_{n,m}   \\
    + \psi^{\dagger}_{n+1,m} \psi_{n,m}  \\
    + e^{+i2\pi bn } \psi^{\dagger}_{n,m-1}\psi_{n,m}   \\
    + e^{-i2\pi bn } \psi^{\dagger}_{n,m+1} \psi_{n,m} 
    \end{bmatrix}  \label{eq:pf_inv_4} \\
    & = H. \label{eq:pf_inv_5}
\end{align}
In Eq.~(\ref{eq:pf_inv_1}) we apply the inversion operation to each of the creation and annihilation operators according to Eq.~(\ref{eq:inv_sym_gauge_tr_4}) and Eq.~(\ref{eq:inv_sym_gauge_tr_5}). 
The first to fourth terms in Eq.~(\ref{eq:pf_inv_1}) correspond to the (transformed) first to fourth terms in Eq.~(\ref{eq:inv_sym_gauge_tr_1}). 
In Eq.~(\ref{eq:pf_inv_2}) we cancel out redundant exponential phase factors and reindex $n \to -n + 2n_{c}$ and $m\to -m+2m_{c}$ in Eq.~(\ref{eq:pf_inv_3}). 
Eq.~(\ref{eq:pf_inv_4}) and Eq.~(\ref{eq:pf_inv_5}) then show that the transformed Hamiltonian is the same. 
We can also prove Eq.~(\ref{eq:inv_sym_gauge_tr_7}) as follows:
\begin{align}
    \left( \mathcal{I}_{n_{c},m_{c}} \right)^{2} \psi^{\dagger}_{n,m} \left( \mathcal{I}_{n_{c},m_{c}}^{-1} \right)^{2} & = \mathcal{I}_{n_{c},m_{c}} \mathcal{I}_{n_{c},m_{c}} \psi^{\dagger}_{n,m} \mathcal{I}_{n_{c},m_{c}}^{-1}  \mathcal{I}_{n_{c},m_{c}}^{-1}  \\
    & =  \mathcal{I}_{n_{c},m_{c}} \psi^{\dagger}_{2n_{c}-n,2m_{c}-m}e^{i4\pi b n_{c}(m-m_{c})} \mathcal{I}_{n_{c},m_{c}}^{-1}\\
    & = \psi^{\dagger}_{2n_{c}-(2n_{c}-n),2m_{c}-(2m_{c}-m)} e^{i4\pi b n_{c}((2m_{c}-m)-m_{c})} e^{i4\pi b n_{c}(m-m_{c})}\\
    & = \psi^{\dagger}_{n,m} e^{i4\pi b n_{c}(m_{c}-m)} e^{i4\pi b n_{c}(m-m_{c})} \\
    & = \psi^{\dagger}_{n,m} .
\end{align}
We have used Eq.~(\ref{eq:inv_sym_gauge_tr_4}) twice above, first acting on $\psi^{\dagger}_{n,m}$ and then on $\psi^{\dagger}_{2n_{c}-n,2m_{c}-m}$. 
Eq.~(\ref{eq:inv_sym_gauge_tr_8}) can also be proved in a similar way as follows:
\begin{align}
    \left( \mathcal{I}_{n_{c},m_{c}} \right)^{2} \psi_{n,m} \left( \mathcal{I}_{n_{c},m_{c}}^{-1} \right)^{2} & = \mathcal{I}_{n_{c},m_{c}} \mathcal{I}_{n_{c},m_{c}} \psi_{n,m} \mathcal{I}_{n_{c},m_{c}}^{-1}  \mathcal{I}_{n_{c},m_{c}}^{-1}  \\
    & =  \mathcal{I}_{n_{c},m_{c}} \psi_{2n_{c}-n,2m_{c}-m}e^{-i4\pi b n_{c}(m-m_{c})} \mathcal{I}_{n_{c},m_{c}}^{-1}\\
    & = \psi_{2n_{c}-(2n_{c}-n),2m_{c}-(2m_{c}-m)} e^{-i4\pi b n_{c}((2m_{c}-m)-m_{c})} e^{-i4\pi b n_{c}(m-m_{c})}\\
    & = \psi_{n,m} e^{-i4\pi b n_{c}(m_{c}-m)} e^{-i4\pi b n_{c}(m-m_{c})} \\
    & = \psi_{n,m} .
\end{align}
Eq.~(\ref{eq:inv_sym_gauge_tr_6}) implies that if the $H$ in Eq.~(\ref{eq:inv_sym_gauge_tr_1}) has non-zero $b$ with the summation over $n$ and $m$ going from $- \infty$ to $+\infty$, then $H$ is invariant under the inversion operation $\mathcal{I}_{n_{c},m_{c}}$ defined by Eqs.~(\ref{eq:inv_sym_gauge_tr_4}) and (\ref{eq:inv_sym_gauge_tr_5}). 
Also, since $\psi^{\dagger}_{n,m}$ and $\psi_{n,m}$ denote the creation and annihilation operators for electronic states spanning the whole Hilbert space, Eq.~(\ref{eq:inv_sym_gauge_tr_7}) and Eq.~(\ref{eq:inv_sym_gauge_tr_8}) imply that $\left( \mathcal{I}_{n_{c},m_{c}} \right)^{2}$ is the identity operation. 

To complete the proof, we give a construction for the unitary operator $\mathcal{I}_{n_c,m_c}$.
The matrix representation of $\mathcal{I}_{n_{c},m_{c}}$ in Eq.~(\ref{eq:inv_sym_gauge_tr_5}) for sites at $(n,m)$ and $(2n_{c}-n,2m_{c}-m)$ is given by
\begin{align}
    \begin{bmatrix}
    0 & e^{-i4\pi b n_{c}(m-m_{c})} \\
    e^{-i4\pi b n_{c}(m_{c}-m)} & 0
    \end{bmatrix}, \label{eq:mat_I_nc_mc}
\end{align}
since
\begin{align}
    & \mathcal{I}_{n_{c},m_{c}} \psi_{n,m} \mathcal{I}_{n_{c},m_{c}}^{-1} = \psi_{2n_{c}-n,2m_{c}-m}e^{-i4\pi b n_{c}(m-m_{c})}, \\
    & \mathcal{I}_{n_{c},m_{c}} \psi_{2n_{c}-n,2m_{c}-m} \mathcal{I}_{n_{c},m_{c}}^{-1} = \psi_{n,m}e^{-i4\pi b n_{c}(m_{c}-m)}.
\end{align}
As we can see Eq.~(\ref{eq:mat_I_nc_mc}) is in fact a unitary matrix. 
Therefore, even though in this 2D lattice we have a constant perpendicular magnetic field which couples to the lattice through a Peierls substitution, there still exist unitary inversion operators with inversion centers at every lattice site which square to the identity and commute with the Hamiltonian.

\subsection{3D system with inversion symmetry}

We have shown that by defining unitary inversion operations up to a gauge transformation in Eq.~(\ref{eq:inv_sym_gauge_tr_4}) and Eq.~(\ref{eq:inv_sym_gauge_tr_5}), the inversion symmetry of Eq.~(\ref{eq:inv_sym_gauge_tr_1}) is preserved. 
Let us now discuss how this extends to our $3$D promoted systems. 
Since our $3$D model of a chiral HOTI from the main text is also coupled to a constant $U(1)$ gauge field given by $\vec{A} = (0,0,2\pi(q_{x}x+q_{y}y))$, we can construct the proper unitary inversion operators squaring to the identity, given by
\begin{align}
    & \mathcal{I}_{n_{x}^{c},n_{y}^{c},n_{z}^{c}} \psi^{\dagger}_{n_{x},n_{y},n_{z}} \mathcal{I}_{n_{x}^{c},n_{y}^{c},n_{z}^{c}}^{-1} = \psi^{\dagger}_{2n_{x}^{c}-n_{x},2n_{y}^{c}-n_{y},2n_{z}^{c}-n_{z}} [\mathcal{I}]^{-1} e^{i 4\pi \left( q_{x}n_{x}^{c} + q_{y}n_{y}^{c} \right) \left( n_{z} - n_{z}^{c} \right)}  , \\
    & \mathcal{I}_{n_{x}^{c},n_{y}^{c},n_{z}^{c}} \psi_{n_{x},n_{y},n_{z}} \mathcal{I}_{n_{x}^{c},n_{y}^{c},n_{z}^{c}}^{-1} = [\mathcal{I}] \psi_{2n_{x}^{c}-n_{x},2n_{y}^{c}-n_{y},2n_{z}^{c}-n_{z}} e^{-i 4\pi \left( q_{x}n_{x}^{c} + q_{y}n_{y}^{c} \right) \left( n_{z} - n_{z}^{c} \right)}.
\end{align}
In these expressions $\psi^{\dagger}_{n_{x},n_{y},n_{z}}$ is the 4-component creation operator for an electron at site $(n_{x},n_{y},n_{z})$, $(n_{x}^{c},n_{y}^{c},n_{z}^{c})$ is the inversion center at any lattice point, and $[\mathcal{I}] = \tau_{z}\sigma_{0}$ is the unitary inversion matrix which also squares to the identity and acts on the degrees of freedom within a unit cell. 
For the case of our helical model, recall that our example of Eq.~(\ref{eq:lattice_model_helical_sliding}) in Sec.~\ref{sec:promoted_helical_lattice_model} is spin-decoupled. 
Thus the $SU(2)$ gauge field given by $\vec{A} = (0,2\pi (q_{x}x+q_{z}z)\sigma_{z},0)$ to which the model is coupled acts effectively as oppositely oriented $U(1)$ magnetic fields for spin up and down electrons. 
We can thus define the unitary inversion operations squaring to identity up to a spin-dependent ($SU(2)$) gauge transformation
\begin{align}
    & \mathcal{I}_{n_{x}^{c},n_{y}^{c},n_{z}^{c}} \psi^{\dagger}_{n_{x},n_{y},n_{z},\sigma} \mathcal{I}_{n_{x}^{c},n_{y}^{c},n_{z}^{c}}^{-1} = \psi^{\dagger}_{2n_{x}^{c}-n_{x},2n_{y}^{c}-n_{y},2n_{z}^{c}-n_{z},\sigma} [\mathcal{I}]^{-1} e^{i 4\sigma\pi \left( q_{x}n_{x}^{c} + q_{z}n_{z}^{c} \right) \left( n_{y} - n_{y}^{c} \right)}  , \\
    & \mathcal{I}_{n_{x}^{c},n_{y}^{c},n_{z}^{c}} \psi_{n_{x},n_{y},n_{z},\sigma} \mathcal{I}_{n_{x}^{c},n_{y}^{c},n_{z}^{c}}^{-1} = [\mathcal{I}] \psi_{2n_{x}^{c}-n_{x},2n_{y}^{c}-n_{y},2n_{z}^{c}-n_{z},\sigma} e^{-i 4\sigma\pi \left( q_{x}n_{x}^{c} + q_{z}n_{z}^{c} \right) \left( n_{y} - n_{y}^{c} \right)},
\end{align}
where $\psi^{\dagger}_{n_{x},n_{y},n_{z},\sigma}$ is the 4-component creation operator for an electron at site $(n_{x},n_{y},n_{z})$ for a fixed spin $\sigma = \pm$ and $[\mathcal{I}] = \tau_{z}\mu_{0}$. 
Therefore, the 3D chiral (helical) HOTI coupled to a $U(1)$ ($SU(2)$) gauge field in our examples preserves inversion symmetry.

\section{\label{bigsec:chiral_sliding}Low energy bulk theory of chiral higher-order topological sliding modes}

In this section, we consider the low energy bulk theory of electrons in a 3D chiral HOTI minimally coupled to a $U(1)$ gauge field, shown in the main text and verify that it can capture several qualitative properties in numerical results of the lattice model calculation with $q_{y} = 0.02$.

As mentioned in the main text, the relevant low energy theory is 
\begin{align}
    H_{\text{bulk}} = m_{\text{bulk}} \tau_{z}\sigma_{0} + \tau_{x}\vec{p} \cdot \vec{\sigma} + \tau_{0} \vec{M} \cdot \vec{\sigma}. \label{eq:3D_chiral_HOTO_bulk}
\end{align}
Coupling this $H_{\text{bulk}}$ to a vector potential $\vec{A} = By \hat{z}$ such that $p_{z} \to p_{z} + By$, and defining 
\begin{align}
    a^{\dagger}_{k_{z}} = \frac{1}{\sqrt{2B}}\left( k_{z} + By - ip_{y} \right), \label{eq:3D_chiral_bulk_U1_ladder}
\end{align}
we can rewrite the Hamiltonian as 
\begin{align}
    H_{\text{bulk}}(k_{x},k_{z}) = m_{\text{bulk}} \tau_{z}\sigma_{0} + \tau_{x} \begin{bmatrix}
    \sqrt{\frac{B}{2}} \left(a_{k_{z}}+a^{\dagger}_{k_{z}} \right) & k_{x} -\sqrt{\frac{B}{2}} \left(a_{k_{z}}-a^{\dagger}_{k_{z}} \right) \\
    k_{x} +\sqrt{\frac{B}{2}} \left(a_{k_{z}}-a^{\dagger}_{k_{z}} \right) & -\sqrt{\frac{B}{2}} \left(a_{k_{z}}+a^{\dagger}_{k_{z}} \right)
    \end{bmatrix} + \tau_{0} \vec{M} \cdot \vec{\sigma}. \label{eq:3D_bulk_LL_Hamiltonian}
\end{align}

\begin{figure}[h]
\includegraphics[scale=0.4]{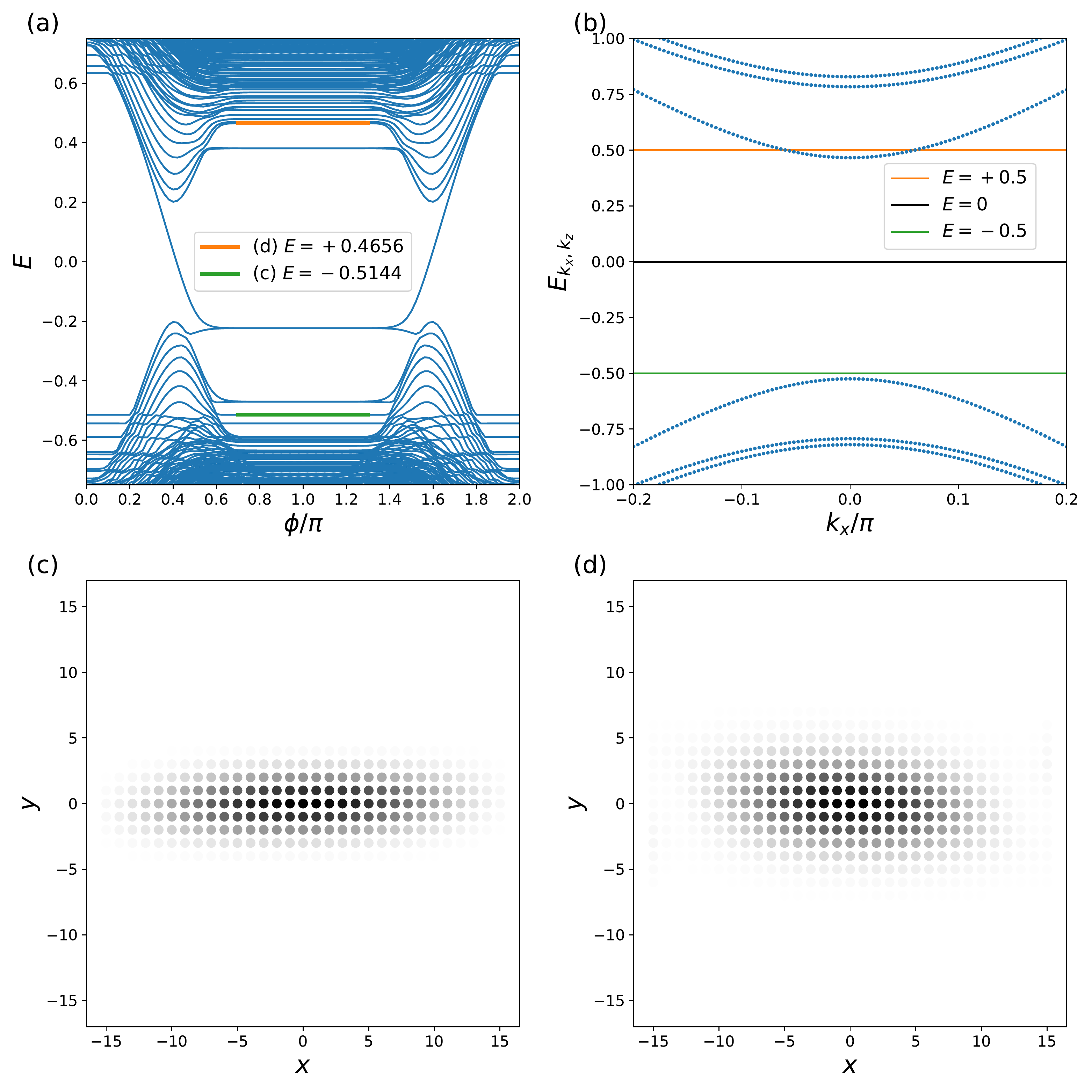}
\caption{(a) $\phi$-sliding spectrum of the 2D modulated system that can be promoted to a 3D chiral HOTI\cite{pozo2019quantization} coupled to a $U(1)$ gauge field with $\vec{q} = (0,q_{y})$ and $q_{y} = 0.02$. 
(b) Bulk Landau levels of Eq.~(\ref{eq:3D_bulk_LL_Hamiltonian}) for the low energy model of a 3D chiral HOTI with $m_{\text{bulk}} = 0.5$, $B = 0.2$ and $\vec{M} = (0.03,0.03,0.03)$. 
The band edges are given by $E = -0.5248$ and $0.4661$. 
The orange, black and green horizontal lines correspond to energies $E = +0.5$, $0$ and $-0.5$, respectively. 
(c) $\&$ (d) Probability distribution of bulk-confined modes in the flat bands of (a) at $\phi = \pi$ with energies $-0.5144$ and $0.4656$ marked by green and orange respectively.
We note that (c) and (d) share similar probability distributions, for example both are confined along $y$ and centered around the middle of the finite sample. This is consistent with the low energy theory prediction of energy-asymmetry in Eq.~(\ref{eq:3D_bulk_LL_Hamiltonian}) and (b). The darker (black) color in (c)--(d) implies higher probability density. 
In (c) and (d), the $x$- and $y$-coordinate both range from $-15, \ldots, +15$.}
\label{SM_chiral_bulk_low_energy}
\end{figure}

Fig.~\ref{SM_chiral_bulk_low_energy} (b) shows the numerically computed energy spectrum $E_{k_{x},k_{z}}$ for Eq.~(\ref{eq:3D_bulk_LL_Hamiltonian}) as a function of $k_{x}$. 
Note that $E_{k_{x},k_{z}}$ does not depend on $k_{z}$, giving rise to flat Landau levels (LLs) along $k_{z}$. 
We now relate this low energy bulk theory to the $\phi$-sliding spectrum of the 2D modulated system by identifying $k_{z}$ in Eq.~(\ref{eq:3D_bulk_LL_Hamiltonian}) with $\Delta \phi = \phi - \pi$, as we have done in the main text. 
To assist the following discussion, we also show again the $\phi$-sliding spectrum of the 2D modulated system which can be promoted to a 3D chiral HOTI coupled to a $U(1)$ gauge field with $\vec{q} = (0,q_{y})=(0,0.02)$ in Fig.~\ref{SM_chiral_bulk_low_energy} (a). 
The flat bands corresponding to bulk-confined modes in Figs.~\ref{SM_chiral_bulk_low_energy} (c) and (d) are marked in green ($E = -0.5144$) and orange ($E = +0.4656$).

First, the band edges for the valence and conduction in Fig.~\ref{SM_chiral_bulk_low_energy} (b) are at energy $E = -0.5248$ and $+0.4661$.
This indicates that the particle-hole symmetry in the bulk of a 3D chiral HOTI is generally broken upon $U(1)$ Landau quantization.
This is also reflected in the energy eigenvalues of the bulk flat bands  in the lattice model Fig.~\ref{SM_chiral_bulk_low_energy} (a), with probability density shown in Figs.~\ref{SM_chiral_bulk_low_energy} (c) and (d). 
We see that the energies ($E = -0.5144$ for Fig.~\ref{SM_chiral_bulk_low_energy} (c) and $E = +0.4656$ in Fig.~\ref{SM_chiral_bulk_low_energy} (d)) of these states are not symmetric about 0.

Second, the Hamiltonian in Eq.~(\ref{eq:3D_bulk_LL_Hamiltonian}) has eigenstates that are extended along $x$ and $z$ while confined along $y$. 
The confinement along $y$ can be understood through the definition of the ladder operator in Eq.~(\ref{eq:3D_chiral_bulk_U1_ladder}), which creates simple harmonic oscillator (SHO) states along $y$. 
Projecting the 3D eigenfunction to the 2D modulated system, only the probability distributions along $x$ and $y$ are physically meaningful. 
Therefore in 2D we expect to see flat bands along $k_{z}$, which is identified as $\Delta \phi  = \phi - \pi$ in the low energy model. The wave functions with these energies are confined along $y$ while extended along $x$. 
This is consistent with Figs.~\ref{SM_chiral_bulk_low_energy} (a) and (c)--(d). 
The above discussion shows that our low energy bulk theory in 3D does qualitatively explain the existence of bulk confined modes in 2D and their flat dispersion along the $\phi$-axis.

\section{\label{bigsec:helical_sliding}Low energy theory of helical higher-order topological sliding modes}

In this section, we construct various low energy theories for the dimensionally-promoted 3D helical HOTI coupled to an $SU(2)$ gauge field. 
We use these low energy models to explain corner, edge and bulk states in the 2D modulated system with helical sliding modes.

\subsection{2D corner modes $\leftrightarrow$ 3D hinge modes with $SU(2)$ gauge field}

In our promoted 3D model, different spin ($\vec{\sigma}$) subspaces are decoupled, see Sec.~\ref{sec:promoted_helical_lattice_model}. 
Therefore, all eigenstates are spin-polarized.
The 2D spin-polarized corner modes are the projection of the helical hinge modes in the 3D lattice. 
If we denote the third, synthetic dimension as $y$ and the corresponding crystal momentum as $k_{y}$, the low energy theory of the helical hinge mode will be
\begin{align}
    H_{\text{hinge}}= v_{F} \left( k_{y}\sigma_{z}' + 2\pi \left( q_{x}x_{\text{hinge}} + q_{z}z_{\text{hinge}} \right) \sigma_{0}' \right), \label{3D_helical_hinge_mode_Hamiltonian}
\end{align}
where we have minimally coupled the Hamiltonian $v_{F}k_{y}\sigma_{z}'$ to a $SU(2)$ gauge field 
\begin{align}
    \vec{A} =  (0,2\pi (q_{x}x+q_{z}z)\sigma_{z}',0). \label{SU2_A}
\end{align}
Notice that we denote our basis as $\vec{\sigma}'$. 
Although the eigenstates of $\sigma'_z$ have opposite spins, they might contain non-trivial orbital and sub-lattice textures. 
Also notice that we have $2\pi \left( q_{x}x_{\text{hinge}} + q_{z}z_{\text{hinge}} \right) \sigma_{0}'$ in Eq.~(\ref{3D_helical_hinge_mode_Hamiltonian}) instead of $2\pi \left( q_{x}x_{\text{hinge}} + q_{z}z_{\text{hinge}} \right) \sigma_{z}'$. 
Since as we replace $k_{y}$ by $k_{y} + 2\pi \left( q_{x}x_{\text{hinge}} + q_{z}z_{\text{hinge}} \right) \sigma_{z}'$ through the minimal coupling, we will have 
\begin{align}
   v_{F}k_{y}\sigma_{z}' = v_{F}\begin{bmatrix}
    k_{y} & 0 \\ 
    0 & -k_{y}
    \end{bmatrix} & \to v_{F}\begin{bmatrix}
    \left( k_{y} + 2\pi \left( q_{x}x_{\text{hinge}} + q_{z}z_{\text{hinge}} \right) \right) & 0 \\ 
    0 & -\left( k_{y} - 2\pi \left( q_{x}x_{\text{hinge}} + q_{z}z_{\text{hinge}} \right) \right)
    \end{bmatrix} \\
    & = v_{F}k_{y}\sigma_{z}' + v_{F} \cdot 2\pi \left( q_{x}x_{\text{hinge}} + q_{z}z_{\text{hinge}} \right)\sigma_{0}'.
\end{align}
In Eq.~(\ref{3D_helical_hinge_mode_Hamiltonian}), we have assumed that: 
(1) there is only one pair of helical hinge modes along this hinge,
(2) the magnitude of the group velocity is $v_{F}$, 
(3) the electron has charge $-1$, and 
(4) the fixed position of the hinge along $y$ is at $(x_{\text{hinge}},z_{\text{hinge}})$ which is set by the coordinate system. 
Eq.~(\ref{SU2_A}) again implies that the modulation wave vector $\vec{q} = (q_{x},q_{z})$ enters the definition of the $SU(2)$ gauge field. 
As we can see in Eq.~(\ref{3D_helical_hinge_mode_Hamiltonian}), if we tune $\vec{q}$, we are effectively shifting the hinge mode dispersion along the $k_{y}$-axis for spin up [down] electrons by an amount $-2\pi \left(q_{x}x_{\text{hinge}}+q_{z}z_{\text{hinge}}\right)$ [$+2\pi \left(q_{x}x_{\text{hinge}}+q_{z}z_{\text{hinge}}\right)$]. 
Since we use $q_{x}=0$ and $q_{z}\ne0$ in the main text for helical sliding modes, the following discussion will focus on this case. 
The generalization to other combinations of $q_{x}$ and $q_{z}$ follows the same procedure.

\begin{figure}[h]
  \includegraphics[scale=0.4]{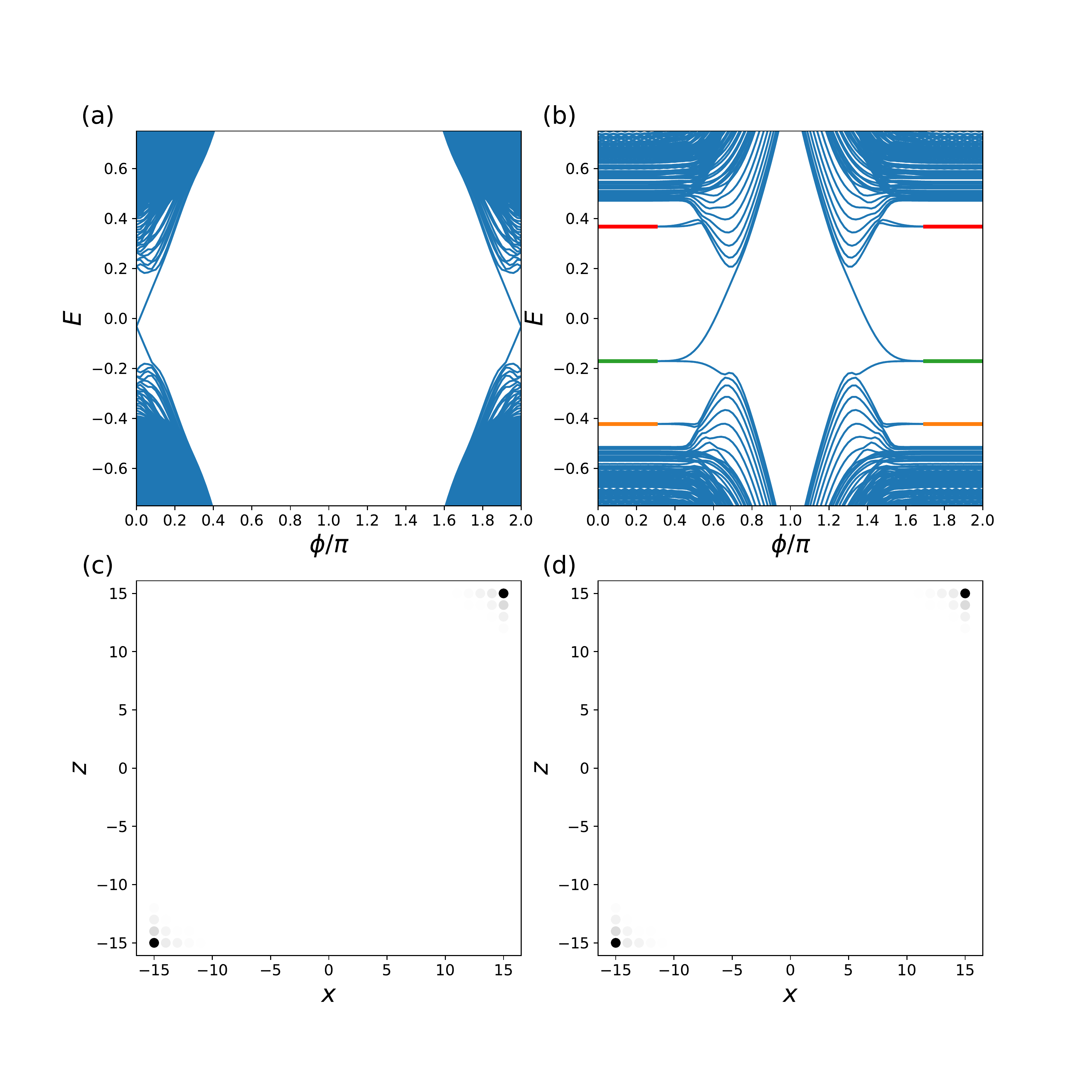}
  \caption{(a) $\phi$-sliding spectrum of Eq.~(35) in the main text with parameters $m_{1} = -3$, $m_{2} = 0.3$, $m_{3} = 0.2$, $m_{v_{1}} = -0.4$, $m_{v_{2}} = 0.2$, $v_{x}=v_{z}=u_{x}=u_{z} = 1$, $v_{y} = 2$, $v_{H} = 1.2$ and $\vec{q} = (0,q_{z})$ where $q_{z} = 0$. 
  This reduces to the $y$-rod band structure of a 3D helical HOTI\cite{Wieder_spin_decoupled_helical_HOTI} without $SU(2)$ gauge fields. 
  Notice the Rashba-like shifting of the surface band away from $\phi = 0$, near $E \approx \pm 0.2$. 
  (b) $\phi$-sliding spectrum with the same parameters as (a) but with $q_{z} = 0.02$. 
  We identify the flat dispersion in (b) marked by orange, green and red as $E^{-}_{\sigma',k_{y},n = 1}$, $E^{-}_{\sigma',k_{y},n =0}$ and $E^{+}_{\sigma',k_{y},n = 1}$ in Eq.~(\ref{wvfn_top}), respectively. 
  (c)$\&$(d) Summation of probability distribution of the doubly-degenerate corners modes related to each other by the $\mathcal{I}\mathcal{T}$-symmetry at $\phi = 0.6\pi$ and $1.4\pi$ with both energies equal to $-0.0124$. 
  The two corner modes in each of (c) and (d) are localized at inversion-related corners and have opposite spins. 
  The darker (black) color in (c)--(d) implies higher probability density. 
  In (c) and (d), the $x$- and $z$-coordinate both range from $-15,\ldots,+15$.}
  \label{SM_Fig_5}
\end{figure}

To connect the low energy theory of the $\phi$-sliding spectrum to the shifting of spin-polarized corner mode dispersion, we identified $k_{y}$ in Eq.~(\ref{3D_helical_hinge_mode_Hamiltonian}) as $\phi$, since the center of the $\phi$-sliding spectrum is at $\phi = 0$, as shown in Fig.~\ref{SM_Fig_5} (a) for $q_{z} = 0$. 
Upon projection to 2D, the position $(x_{\text{hinge}},z_{\text{hinge}})$ of the hinge along $y$ again becomes the position $(x_{\text{corner}},z_{\text{corner}})$ of the 2D corner. 
We thus obtain
\begin{align}
    H_{\text{corner}}= v_{F} \left( \phi \sigma_{z}' + 2\pi \left( q_{x}x_{\text{corner}} + q_{z}z_{\text{corner}} \right) \sigma_{0}' \right) \label{eq:H_helical_hinge_1}
\end{align}
from the main text. 
We now examine this low energy theory through numerical simulations. 
We will be focusing on how the doubly-degenerate gap-crossing bands corresponding to spin-polarized corner modes respond as we increase magnitude of $\vec{q}$.

We show in Fig.~\ref{SM_Fig_5} (b) the $\phi$-sliding spectrum for $q_{z} = 0.02$. 
Comparing Fig.~\ref{SM_Fig_5} (b) with Fig.~\ref{SM_Fig_5} (a), we see that as we turn on $q_{z}$, the two doubly-degenerate bands cross the gap with opposite slopes. 
To explain this, we notice that each of the doubly-degenerate gap-crossing states corresponds to a localized pair of modes at inversion-related corners with opposite spins, as they are related by the $\mathcal{I}\mathcal{T}$-symmetry (see main text Sec.~V) and shown in Figs.~\ref{SM_Fig_5} (c) and (d). 
In our coordinate system, the corner modes are localized at positions $(x_{\text{corner}},z_{\text{corner}})$ = $\pm(L/2,L/2)$, where $L=30$. 
The corresponding 3D helical hinge mode dispersion relations are
\begin{align}
    & H_{\text{hinge 1}}= v_{F} \left( k_{y}\sigma_{z}' - \pi q_{z}L \sigma_{0}' \right), \\
    & H_{\text{hinge 2}} = -v_{F} \left( k_{y}\sigma_{z}' + \pi q_{z}L \sigma_{0}' \right),
\end{align}
where inversion symmetry requires that the eigenstate of $H_{\text{hinge 1}}$ and $H_{\text{hinge 2}}$ with same spins have opposite group velocities. 
Identifying $k_{y}$ as $\phi$, we have that the Hamiltonians of the bands crossing the gap are given by
\begin{align}
    & H_{\text{corner 1}}= v_{F} \left( \phi \sigma_{z}' - \pi q_{z}L \sigma_{0}' \right), \label{eq:corner_1_SU2}\\
    & H_{\text{corner 2}} = -v_{F} \left( \phi \sigma_{z}' + \pi q_{z}L \sigma_{0}' \right). \label{eq:corner_2_SU2}
\end{align}
We then see that as we increase the $q_{z}$, the band in $H_{\text{corner 1}}$ along $\phi$ with slope $+v_{F}$ and spin $\uparrow$ moves in the same direction as the band in $H_{\text{corner 2}}$ with group velocity $+v_{F}$ and spin $\downarrow$, which is consistent with Fig.~\ref{SM_Fig_5} (b). 
In fact, a detailed comparison between Figs.~\ref{SM_Fig_5} (a) and (b) shows that the corner mode dispersion shifts along $\phi$ by $\pi q_{z}L \approx 0.6\pi $ for $q_{z} = 0.02$ and $L = 30$, implying that our low energy theory describes both the 2D corner modes and the corresponding 3D helical hinge modes. 
This confirms that we can tune the range of $\phi$ where the spin-polarized corner modes emerge from the bulk bands by modifying the periodicity of the modulation, as we have stated in the main text. 
Similar to the chiral sliding modes in the main text, when  $\pi q_{z}L > 2\pi$, the gap-crossing bands will be folded back within the range $\phi = [0,2\pi)$. 
This happens in Fig. 3 (a) of the main text, where $q_{z} = 0.11957$.

\subsection{\label{sec:surface_SU2}2D edge-confined modes $\leftrightarrow$ 3D surface modes with $SU(2)$ gauge field}

As stated in the main text, we may understand the flat bands corresponding to edge-confined modes using a low energy theory in the promoted 3D lattice. 
In this subsection we construct a surface theory minimally coupled to an $SU(2)$ gauge field to describe the projected 2D edge states in the modulated system.

Due to the shifted band edge of the surface bands in the corresponding 3D helical HOTI (see Fig.~\ref{SM_Fig_5} (a) and its caption), we may consider the surface theory by stacking a low energy Chern insulator with spin $\uparrow$ and its time-reversal counterpart, with a relative shift in momentum space. 
The low energy theory of a Chern insulator is given by $H_{\text{Chern}}=\vec{p}_{\parallel}\cdot \vec{\tau}_{\parallel}' - m\hat{n} \cdot \vec{\tau}'$, where $\vec{\tau}'$ is the basis describing spin $\uparrow$ electron with some orbital and sub-lattice textures, $\parallel$ is the parallel component along the surface, $\hat{n}$ is the surface normal vector and $m$ is the mass term. 
As shown in Eq.~(\ref{eq:B_SU2_1}), if $q_{x} = 0$, which is the case we consider in Fig.~\ref{SM_Fig_6} below, we have $\vec{B} = \vec{\nabla} \cross \vec{A} - i \vec{A} \cross \vec{A}$ parallel to $\hat{x}$. 
We then consider a surface theory on the $yz$-plane whose surface normal is $\hat{x}$. 
The corresponding surface Hamiltonian without an $SU(2)$ gauge field reads $H_{\text{surf}} = p_{y}\tau_{y}'\sigma_{0}'+p_{z}\tau_{z}'\sigma_{z}'+(\Delta k_{y})\tau_{y}'\sigma_{z}'-m\tau_{x}'\sigma_{0}'$, where $\vec{\sigma}'$ again denotes the spin degrees of freedom, and $\Delta k_{y}$ is a {\it real constant} denoting the shift of low energy surface band minima from $\vec{k}=\Gamma$. 
To facilitate the subsequent analysis, we perform a basis transformation through a $-2\pi/3$ radian rotation $U$ along the $[1,1,1]$ axis in the {\it orbital} space $\vec{\tau}'$ such that $U^{\dagger}(\tau_{x}',\tau_{y}',\tau_{z}')U = (\tau_{z}',\tau_{x}',\tau_{y}')$. 
The transformed surface Hamiltonian then reads
\begin{align}
    H_{\text{surf}} = p_{y}\tau_{x}'\sigma_{0}'+p_{z}\tau_{y}'\sigma_{z}'+(\Delta k_{y})\tau_{x}'\sigma_{z}'-m\tau_{z}'\sigma_{0}'.
    \label{eq:top_surf_SU2_helical_transformed}
\end{align}
Notice that this surface theory is spin-decoupled. 
In a general helical HOTI the spins are coupled. 
However, the general procedure will be the same: we first construct a low energy surface theory, and then couple it to the desired $SU(2)$ gauge field. 

We now couple Eq.~(\ref{eq:top_surf_SU2_helical_transformed}) to an $SU(2)$ gauge field $\vec{A} = (0,Bz \sigma_{z}',0)$ which produces a $SU(2)$ magnetic field $\vec{B} = (-B\sigma_{z}',0,0)$. 
Therefore opposite spins experience opposite magnetic fields. 
The minimally-coupled surface theory is
\begin{align}
    H_{\text{surf}} = \begin{bmatrix}
    \left(p_{y}+\Delta k_{y} + Bz \right)\tau_{x}' + p_{z}\tau_{y}' - m \tau_{z}' &  0 \\ 0 & \left(p_{y}-\Delta k_{y} - Bz \right)\tau_{x}' - p_{z}\tau_{y}' - m \tau_{z}'
    \end{bmatrix}, \label{eq:helical_surface_before_rewrite}
\end{align}
where we have assumed that both $B$ and $m$ are positive. 
Fourier transforming to replace $p_{y}$ by the wavenumber $k_{y}$, and defining spin- and $k_{y}$-dependent ladder operators
\begin{align}
    a^{\dagger}_{k_{y},\uparrow} = \frac{1}{\sqrt{2B}}\left(k_{y} + \Delta k_{y} + Bz - ip_{z} \right) \text{ and } a^{\dagger}_{k_{y},\downarrow} = \frac{1}{\sqrt{2B}}\left(k_{y} - \Delta k_{y} - Bz + ip_{z} \right), \label{eq:top_ladders_SU2}
\end{align}
we can rewrite Eq.~(\ref{eq:helical_surface_before_rewrite}) in each spin subspace ($\sigma'=\pm$) as
\begin{align}
    H_{\text{surf}}(k_{y},\sigma') = \begin{bmatrix}
    -m & \sqrt{2B} a_{k_{y},\sigma'}^{\dagger} \\
    \sqrt{2B} a_{k_{y},\sigma'} & m
    \end{bmatrix}.
\end{align}
We can solve for the eigenstates and energies to find
\begin{align}
    & \psi^{-}_{\sigma',k_{y},n=0} = e^{ik_{y}y}\begin{bmatrix}
    \ket{0,k_{y},\sigma'} \\ 0
    \end{bmatrix},\   E^{-}_{\sigma',k_{y},n=0} = -m, \nonumber \\
    & \psi^{-}_{\sigma',k_{y},n>0} = e^{ik_{y}y}\begin{bmatrix}
    \ket{n,k_{y},\sigma'} \\  \alpha_{-}(n)\ket{n-1,k_{y},\sigma'}
    \end{bmatrix},\   E^{-}_{\sigma',k_{y},n>0} = -\sqrt{m^{2} + 2Bn}, \nonumber \\
    & \psi^{+}_{\sigma',k_{y},n>0} = e^{ik_{y}y}\begin{bmatrix}
    \ket{n,k_{y},\sigma'} \\ \alpha_{+}(n)\ket{n-1,k_{y},\sigma'}
    \end{bmatrix},\   E^{+}_{\sigma',k_{y},n>0} = +\sqrt{m^{2} + 2Bn}, \nonumber \\
    & {\color{black}{\text{where }}} \alpha_{\pm}(n) = \frac{1}{\sqrt{2Bn}}\left(\pm \sqrt{m^{2} + 2Bn} + m \right). \label{wvfn_top}
\end{align}
Here $\sigma' = \pm$ is the spin quantum number, $k_{y}$ is the wavenumber along $y$, $n$ is an non-negative integer {\color{black}{labelling}} the $SU(2)$ LLs, and $\ket{n,k_{y},\sigma'}$ is the $n^{\text{th}}$ simple harmonic oscillator (SHO) eigenstate along $z$ defined using $a^{\dagger}_{k_{y},\sigma'}$. 
Notice that the coefficient $\alpha_{\pm}(n)$ does not depend on which spin sector we are considering. 
We now connect the above surface theory to the flat bands and corresponding wavefunctions in the $\phi$-sliding spectrum of the 2D modulated system by identifying $k_{y}$ as $\phi$ and $B$ as $2\pi q_{z}$. 
This is because in our 3D promoted lattice Eq.~(\ref{eq:lattice_model_helical_sliding}) and the numerical examples, we have $q_{x} = 0$ such that $\vec{A} =  (0,2\pi q_{z}z\sigma_{z}',0)$ produces an $SU(2)$ magnetic field with strength $2\pi q_{z}$. 
We will mainly focus on the two [one] flat bands in Fig.~\ref{SM_Fig_5} (b) with negative [positive] energy closest to the zero. 
These correspond to $E^{-}_{\sigma',k_{y},n \le 1}$ [$E^{+}_{\sigma',k_{y},n=1}$].

First, notice that the spectrum in Eq.~(\ref{wvfn_top}) breaks particle-hole symmetry in both spin sectors, as there are no $+m$ energy eigenvalues. 
This can be observed in Fig.~\ref{SM_Fig_5} (b) where there are no flat bands of edge-confined modes around $E\approx +0.2$ corresponding to $E = +m$. 
We thus identify the flat bands in Fig.~\ref{SM_Fig_5} (b) marked by orange, green and red as $E^{-}_{\sigma',k_{y},n = 1}$, $E^{-}_{\sigma',k_{y},n =0}$ and $E^{+}_{\sigma',k_{y},n = 1}$ in Eq.~(\ref{wvfn_top}), respectively.

Next, the probability densities $|\psi^{-}_{\sigma',k_{y},n=0}|^2$ and $|\psi^{\pm}_{\sigma',k_{y},n=1}|^2$ are respectively proportional to $\left|\varphi_{0,B}(z+(\Delta k_{y} + \sigma' k_{y})/B) \right|^{2}$ and $\left| \alpha_{\pm}(1) \right|^{2}\left|\varphi_{0,B}(z+(\Delta k_{y} + \sigma' k_{y})/B) \right|^{2} + \left|\varphi_{1,B}(z+(\Delta k_{y} + \sigma' k_{y})/B) \right|^{2}$, where $\varphi_{n,B}(z)$ is the $n^{\text{th}}$ SHO eigenstate along $z$. 
Notice that we have put explicit $B$-dependence on $\varphi_{n,B}(z)$ since the cyclotron frequency and the localization of wave functions depend on the field strength.
This implies that: 
(1) the probability density computed from $\psi^{-}_{\sigma',k_{y},n=0}$ has a pure Gaussian distribution along $z$, and 
(2) $\psi^{-}_{\sigma',k_{y},n=1}$ has a larger contribution from the SHO first excited state than $\psi^{+}_{\sigma',k_{y},n=1}$, since $\left| \alpha_{-}(1) \right|^{2} < \left| \alpha_{+}(1) \right|^{2}$ and we have assumed both $B$ and $m$ are positive. 
Figs.~\ref{SM_Fig_6} (a)--(c) show the 2D probability distribution at $\phi = 0$ for edge-confined modes in different LLs together with the insets showing the probability integrated over non-negative $x$-coordinates. 
While both Figs.~\ref{SM_Fig_6} (a) and (c) corresponds to $n = 1$ LL, Fig.~\ref{SM_Fig_6} (a) is from the negative energy branch and Fig.~\ref{SM_Fig_6} (c) is from the positive energy branch. 
Therefore Fig.~\ref{SM_Fig_6} (a) is more characteristic of the SHO first excited state than Fig.~\ref{SM_Fig_6} (c). 
In contrast, Fig.~\ref{SM_Fig_6} (b), being the $n=0$ LL wave function, shows a Gaussian probability distribution characteristic of the SHO ground state. 
We notice that, if we restrict ourselves to a single edge, the wave function in Fig.~\ref{SM_Fig_6} (a) contains a slightly asymmetric SHO first excited state. 
This is due to the complicated on-site and hopping energies in our model, which break extraneous symmetries, such that the wave function in Fig.~\ref{SM_Fig_6} (a) has nonzero penetration to bulk. 
This can be compared with Fig.~\ref{SM_Fig_6} (b) where the edge-confined modes are much more localized on the edges. 
Thus, the surface theory present above should be recognized as an effective theory we use to extract out qualitative properties of wave functions, and it suffices to identify the SHO first excited state character of Fig.~\ref{SM_Fig_6} (a) corresponding to $\psi^{-}_{\sigma',k_{y},n=1}$ with energy level $E^{-}_{\sigma',k_{y},n = 1}$ in Eq.~(\ref{wvfn_top}).

Third, the definition of the ladder operator in Eq.~(\ref{eq:top_ladders_SU2}) predicts that even when we have $\phi = 0$ (which corresponds to $k_{y} = 0$) the center of the SHO eigenstate will be shifted from the center of the coordinate system by a distance $|\Delta k_{y}/ B|$ along $z$. 
Recall that $\Delta k_{y}$ is the slight Rashba-like shift of the surface bands away from $\Gamma$ (see Fig.~\ref{SM_Fig_5} (a) and its caption). 
In the 2D modulated system, this is equivalent to saying that the edge-confined modes at $\phi = 0$ will not have wavefunctions centered at the middle of the edge along $z$. 
The probability distribution at $\phi = 0$ shown in Fig.~\ref{SM_Fig_6} (a) to (c) confirms this, as no states are centered around the middle of the edges.
Going further, we also expect from Eq.~(\ref{eq:top_ladders_SU2}) that the center of wave functions will be shifted along $z$ by $(-\Delta k_{y} - \sigma' k_{y})/B$ for a given $k_{y}$. 
Identifying $k_{y}$ as $\phi$ and $B$ as $2\pi q_{z}$, we deduce that the shifting of the edge-confined modes from the center of the edges is spin-dependent. 
The center of each probability distribution is given by $l_{\sigma'} = (-\Delta k_{y} - \sigma'  \phi)/(2\pi q_{z}) $. 
Notice that the edge-confined modes in Figs.~\ref{SM_Fig_6} (d) and (e) are shifted by $\approx \pm 2.5$ lattice constants comparing with Fig.~\ref{SM_Fig_6} (b). 
This is because Figs.~\ref{SM_Fig_6} (d) and (e) correspond to $\Delta \phi = \phi - 0 = 0.1\pi$, $q_{z} = 0.02$, and wave functions with opposite spins will be shifted in the opposite directions, according to the expression of $l_{\sigma'}$ given above and $a^{\dagger}_{k_{y},\sigma'}$ in Eq.~(\ref{eq:top_ladders_SU2}). 
To be more precise, we estimate the center of the wave functions in Figs.~\ref{SM_Fig_6} (b), (d) and (e) to be around $z \approx  1.5$, $4$ and $-1$, respectively.

Finally, similar to the chiral sliding modes in the 2D modulated system, we also have an additional degeneracies in the flat band states due to zone-folding (see Sec.~IV of the main text). 
Up to the degeneracy due to zone folding, the universal property we expect in these 2D helical sliding systems is that as we vary $\phi$, which corresponds to sliding of density waves (DWs) with spin-orbit coupled interactions, there will be flat bands together with edge-confined modes that are projected from the surface of a helical HOTI with $SU(2)$ Landau quantization. 
We have seen that the qualitative properties of the wave functions are all consistent with the low energy surface theory.

\begin{figure}[h]
\includegraphics[scale=0.4]{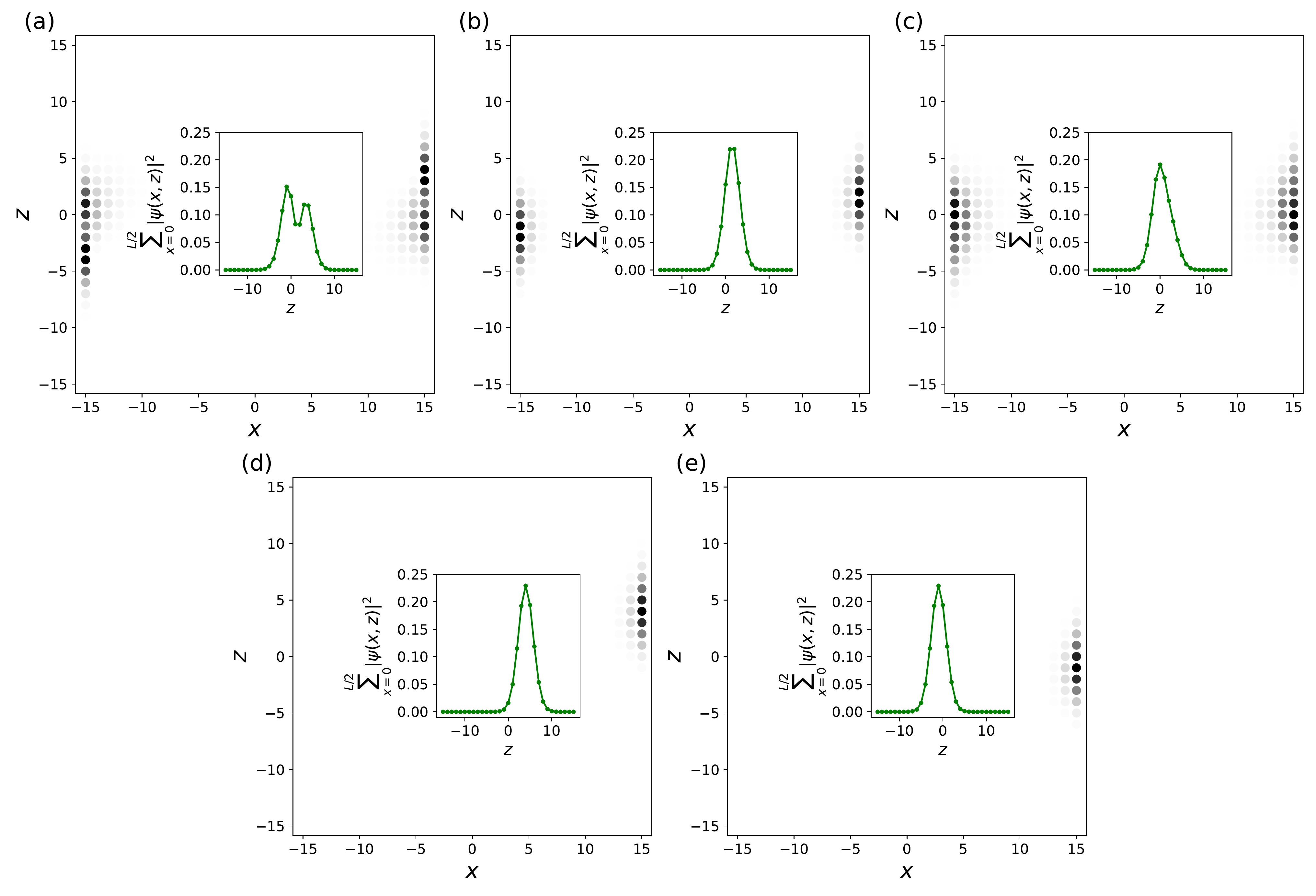}
\caption{(a)--(c) Average of the probability distribution for the four-fold degenerate edge-confined modes in the flat bands shown in Fig.~\ref{SM_Fig_5} (b). 
The four-fold degeneracy comes considering the two edges and two spins. 
(a)--(c) are edge-confined modes at $\phi = 0$ with $E = -0.4227$, $-0.1711$ and $0.3675$ which are marked orange, green and red in Fig.~\ref{SM_Fig_5} (b). 
The corresponding energy levels are $E^{-}_{\sigma^\prime,k_{y}=0,n=1}$, $E^{-}_{\sigma^\prime,k_{y}=0,n=0}$ and $E^{+}_{\sigma^\prime,k_{y}=0,n=1}$in Eq.~(\ref{wvfn_top}). 
(d)$\&$(e) Edge modes at $\phi = 0.1 \pi$ with $E = -0.1711$ confined at the right edge.
They correspond to $E^{-}_{\sigma^\prime,k_{y}=0.1\pi,n=0}$ with $\sigma^\prime = \pm$. 
We notice that in addition to (d) and (e) there are two other edge-confined modes with nearly identical energies localized on the left edge.
The darker (black) color in (a)--(e) implies higher probability density. 
The inset in (a)--(e) is the probability distribution integrated over non-negative $x$-coordinates, from $x= 0,\ldots, L/2$ where $L=30$. 
In (a)--(e), the $x$- and $z$-coordinate both range from $-15,\ldots, +15$.}
\label{SM_Fig_6}
\end{figure}

\subsection{2D bulk-confined modes $\leftrightarrow$ 3D bulk modes with $SU(2)$ gauge field}

Having accounted for the edge-confined modes in the 2D modulated system with helical sliding, we move on to consider the flat bands corresponding to bulk-confined modes. 
Similar to Sec.~\ref{sec:surface_SU2}, we will construct a low energy theory and couple it to an $SU(2)$ gauge field.

\begin{figure}[h]
  \includegraphics[scale=0.4]{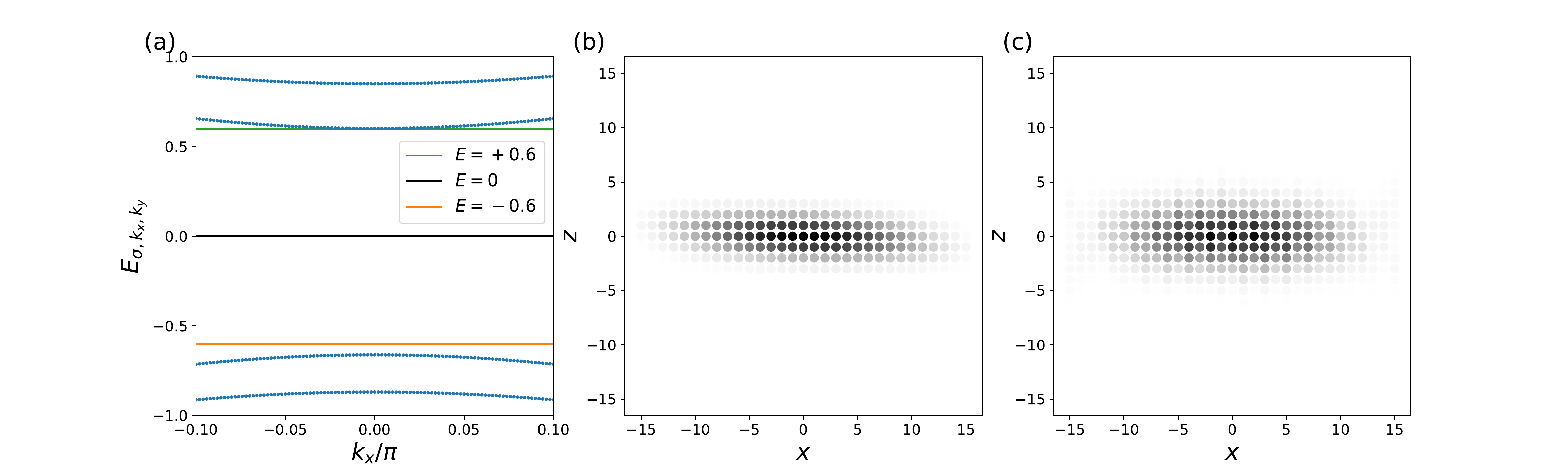}
  \caption{(a) Bulk $SU(2)$ Landau levels of Eq.~(\ref{eq:3D_bulk_SU2_LL_Hamiltonian}) for the low energy model of a 3D helical HOTI\cite{Wieder_spin_decoupled_helical_HOTI} with $m_{\text{bulk}} = 1$, $m_{2} = 0.3$, $m_{3} = 0.2$, $m_{v_{1}} = -0.4$, $m_{v_{2}} = 0.2$, $u_{x} = 1$, $v_{H}=1.2$, $u_{z} = 1$ and $B = 0.2$. 
  The band edges occur at $E = -0.6618$ and $0.601$. 
  The green, black and orange horizontal lines correspond to energy $+0.6$, $0$ and $-0.6$, respectively. 
  Notice that there is in general no particle-hole symmetry in the bulk $SU(2)$ LL spectrum of a 3D helical HOTI. 
  (b)$\&$(c) Probability distribution of bulk-confined modes in the flat band of Fig.~\ref{SM_Fig_5} (b) at $\phi = 0$ with energy $-0.5393$ and $0.4835$ respectively with similar wave function confinement. 
  These two modes are at the flat bulk continuum right below and above those flat bands  for {\it edge-confined} modes corresponding to $E^{-}_{\sigma',k_{y},n=1}$ and $E^{+}_{\sigma',k_{y},n=1}$ in Eq.~(\ref{wvfn_top}), which we have marked in orange and red in Fig.~\ref{SM_Fig_5} (b).
  We note that both (b) and (c) show modes confined along $z$, which is consistent with the ladder operators defined in Eq.~(\ref{eq:ladder_bulk_helical_SU2}). 
  However, the absolute value of energies for (b) and (c) are not same, indicating the spectrum is not particle-hole symmetric, and the minimal continuum model calculation in (a) captures this trend. 
  The darker (black) color in (b)--(c) implies higher probability density. 
  In (b) and (c), the $x$- and $z$-coordinate both range from $-15,\ldots, +15$.}
  \label{SM_Fig_7}
\end{figure}

We consider the Bloch Hamiltonian of the promoted 3D helical HOTI around the $\Gamma$ point (see Sec.~\ref{sec:promoted_helical_lattice_model}), which is
\begin{align}
    H_{\text{bulk}} = & m_{\text{bulk}} \tau_{z}\mu_{0}\sigma_{0} + m_{2} \tau_{z}\mu_{x}\sigma_{0} + m_{3} \tau_{z}\mu_{z}\sigma_{0} + m_{v_{1}}\tau_{0}\mu_{z}\sigma_{0} + m_{v_{2}} \tau_{0}\mu_{x}\sigma_{0} \nonumber \\
    & + u_{x} p_{x} \tau_{y}\mu_{y}\sigma_{0} + u_{z}p_{z}\tau_{x}\mu_{0}\sigma_{0} + v_{H}p_{y} \tau_{y}\mu_{z}\sigma_{z}. \label{eq:3D_helical_HOTI_bulk}
\end{align}
We have redefined several parameters in the original model in Sec.~\ref{sec:promoted_helical_lattice_model} for convenience. 
For example, $m_{\text{bulk}}$ is effectively $m_{1} +v_{x} + v_{y} + v_{z}$ from Eq.~(A1) 
in Ref.~\onlinecite{Wieder_spin_decoupled_helical_HOTI}. 
We now couple this $H_{\text{bulk}}$ to the $SU(2)$ vector potential $\vec{A} = Bz \tau_{0}\mu_{0}\sigma_{z} \hat{y}$, which is equivalent to Eq.~(\ref{SU2_A}) with $q_{x} = 0$. 
Therefore, the term $v_{H}p_{y} \tau_{y}\mu_{z}\sigma_{z}$ becomes
\begin{align}
    v_{H}p_{y} \tau_{y}\mu_{z}\sigma_{z} \to v_{H}\tau_{y}\mu_{z} \begin{bmatrix}
    p_{y} + Bz & 0 \\ 0 & -(p_{y}-Bz)
    \end{bmatrix},
\end{align}
where the matrix acts in spin space. 
Fourier transforming along $x$ and $y$ and defining the $k_{y}$- and spin($\sigma = \pm$)-dependent ladder operators
\begin{align}
    a^{\dagger}_{\sigma,k_{y}} = \frac{1}{\sqrt{2B}}\left( \sigma k_{y} + Bz - ip_{z} \right), \label{eq:ladder_bulk_helical_SU2}
\end{align}
we can rewrite Eq.~(\ref{eq:3D_helical_HOTI_bulk}) coupled to $\vec{A} = Bz \tau_{0}\mu_{0}\sigma_{z} \hat{y}$ in different spin sectors as
\begin{align}
    H_{\sigma,k_{x},k_{y}} &= m_{\text{bulk}} \tau_{z}\mu_{0} + m_{2} \tau_{z}\mu_{x} + m_{3}\tau_{z}\mu_{z} + + m_{v_{1}} \tau_{0}\mu_{z} + m_{v_{2}}\tau_{0}\mu_{x} \nonumber \\
    & + u_{x}k_{x}\tau_{y}\mu_{y} + v_{H}\sqrt{\frac{B}{2}} \left( a_{\sigma,k_{y}} + a^{\dagger}_{\sigma,k_{y}} \right)\tau_{y}\mu_{z} -i u_{z} \sqrt{\frac{B}{2}}\left( a_{\sigma,k_{y}} - a^{\dagger}_{\sigma,k_{y}} \right)\tau_{x}\mu_{0}. \label{eq:3D_bulk_SU2_LL_Hamiltonian}
\end{align}
In Fig.~\ref{SM_Fig_7} (a) we show the numerically computed spectrum $E_{\sigma,k_{x},k_{y}}$ as a function of $k_{x}$. 
Note that $E_{\sigma,k_{x},k_{y}}$ does not depend on $\sigma$ and $k_{y}$, giving rise to flat Landau levels as a function of $k_{y}$. 
We relate this low energy bulk theory to the $\phi$-sliding spectrum of the 2D modulated system by identifying $k_{y}$ in Eq.~(\ref{eq:3D_bulk_SU2_LL_Hamiltonian}) with $\phi$. 
We now examine the consequences of this correspondence.

First, examining the band edges for the valence and conduction bands in Fig.~\ref{SM_Fig_7} (a) we notice that there is no particle-hole symmetry in the low energy spectrum of helical HOTI with $SU(2)$ Landau quantization.
This is also reflected in the energy eigenvalues and eigenstate probability distribution for the flat bands in Fig.~\ref{SM_Fig_5} (b) corresponding to bulk-confined modes shown in Figs.~\ref{SM_Fig_7} (b) and (c), which are in the bulk flat continuum directly below and above the edge flat dispersion marked in orange and red in Fig.~\ref{SM_Fig_5} (b). 
The figure caption of Fig.~\ref{SM_Fig_7} gives the asymmetric energies of the two bulk states in Figs.~\ref{SM_Fig_7} (b) and (c).

Second, we note that the Hamiltonian in Eq.~(\ref{eq:3D_bulk_SU2_LL_Hamiltonian}) has eigenstates that are extended along $x$ and $y$ while confined along $z$. 
The confinement along $z$ can also be understood through the definition of the spin-dependent ladder operators in Eq.~(\ref{eq:ladder_bulk_helical_SU2}), which create SHO states along $z$. 
Projecting this 3D wave function to the 2D modulated system, only the probability distribution along $x$ and $z$ is preserved. 
Therefore in 2D we expect to see flat bands along $k_{y}$--which is identified as $\phi$--and with corresponding wave functions confined along $z$ while extended along $x$. 
This is consistent with Figs.~\ref{SM_Fig_7} (b) and (c). 
The above discussion shows that our low energy bulk theory in 3D qualitatively explains the existence of bulk confined modes in 2D. 
The universal property of these types of 2D helical sliding systems, even with larger $|\vec{q}|$ presented in the main text, is that there will be flat bands as we vary $\phi$ with bulk-confined modes that are projected from the bulk of helical HOTI with $SU(2)$ Landau quantization.

\section{\label{sec:numerical_method}Numerical methods to identify non-trivial Chern insulating layers in 3D Weyl-CDW systems}

In this section, we describe the numerical methods we used to identify layers (hybrid Wannier functions) and corresponding Bloch states in our 3D Weyl-CDW model. 
In the following, we use the term $xy$-slab to denote a sample of 3D Weyl-CDW system infinite along the $x$ and $y$ directions and finite along $z$ with size $L_{z}$. 
In addition, we use $y$-rod to denote a sample of 3D Weyl-CDW system infinite along $y$ and finite along $x$ and $z$ with size $L_{x} \times L_{z}$. 
In this section, we will present the numerical methods based on Berry phase, Berry curvature, and hybrid Wannier functions to identify the non-trivial bands in the $xy$-slab for $q = 1/5$, $L_{z} = 25 $, $t_{x} = -t_{y} = t_{z} = 1$, $m=2$, $2|\Delta|=0.75$ with $\phi = 0$ and $\phi = \pi$ in the main text. 
The numerical methods can also be applied to systems such as incommensurate CDWs with $q = \tau /4$ where $\tau = (1+\sqrt{5})/2$ is the golden ratio (see main text).

As mentioned in the main text, we can understand the existence of quantum anomalous Hall (QAH) surface states by viewing the $xy$-slab as composed of a layered stack of Chern insulators (CIs). 
When we cut the $xy$-slab into the corresponding $y$-rod, those states in the $xy$-slab with non-zero Chern number will each contribute a number of chiral edge modes, depending on the Chern number of the bands. 
These chiral edge modes collectively form the QAH surface states. The number of chiral edge modes in the $y$-rod on each edge is equal to the Chern number of occupied bands in the corresponding $xy$-slab. Specifically, the low energy theory in 4D predicts that topologically non-trivial bands in the $xy$-slab have a layered structure in position space, in which non-trivial bands will be localized along $z$ and separated by some amount of lattice constants\cite{NicoDavidAXI2,dynamical_axion_insulator_BB}. 
In the following, we will use a combination of Berry phase, Berry curvature and hybrid Wannier function calculations to identify those non-trivial bands in the $xy$-slab. 
The goal is to find non-trivial bands whose probability distributions along $z$ match the probability distribution of chiral edge modes (or QAH surface states) in the $y$-rod {\color{black}{shown in Figs.~4 and 5 in the main text}}.

\begin{figure}[ht]
  \includegraphics[width=\columnwidth]{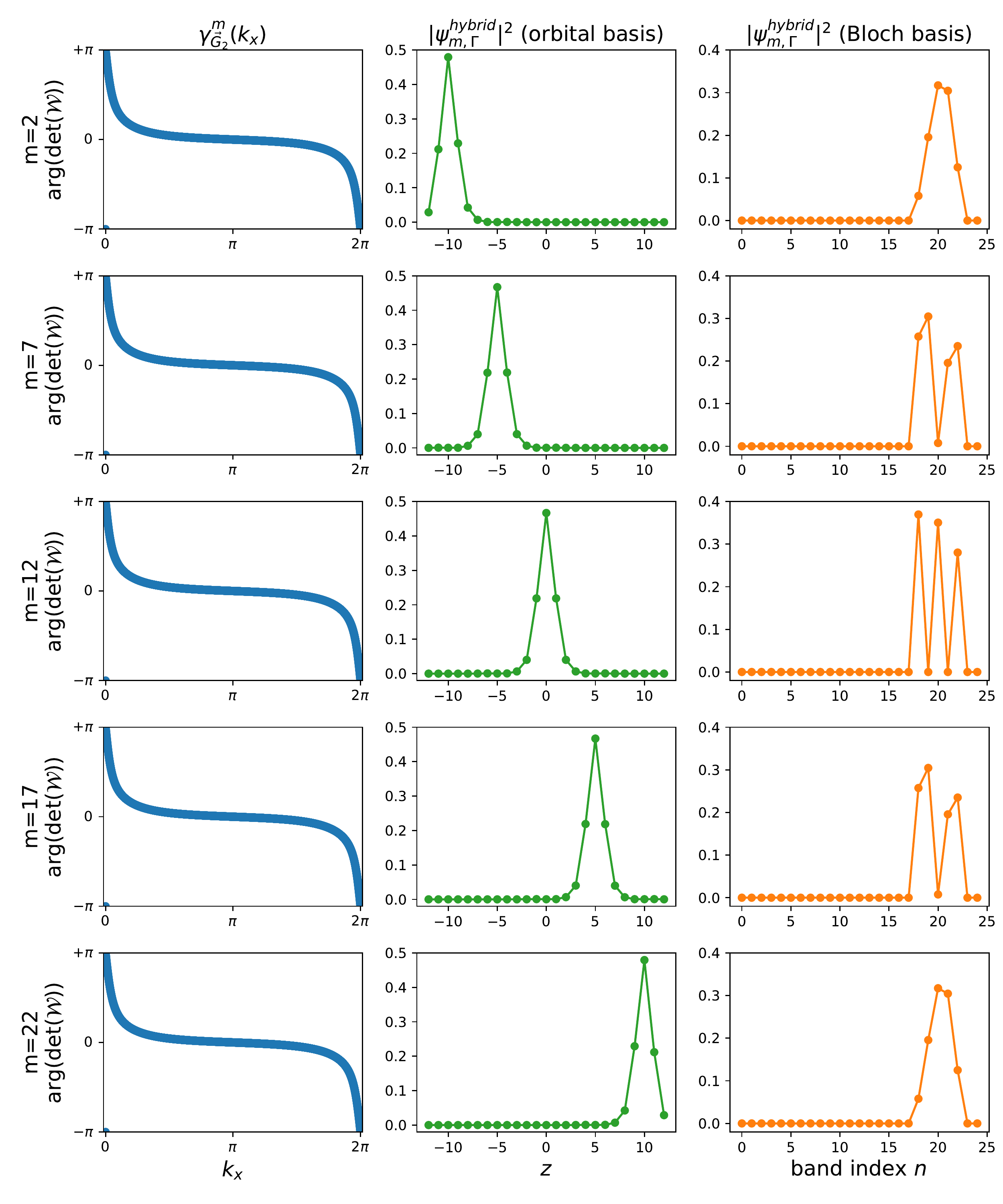}
  \caption{Berry phase winding of non-trivial hybrid Wannier bands (first column), hybrid Wannier functions at $\vec{k} = \Gamma$ decomposed in terms of orbital (second column), and Bloch bases (third column) for an $xy$-slab of the Weyl-CDW model with $t_{x}=-t_{y}=t_{z}=1$, $m =2$, $2|\Delta|=0.75$, $q = 1/5$, $L_{z} =25$ and $\phi = 0$. 
The $z$-coordinate ranges from $-12, \ldots, +12$. 
There are 5 non-trivial hybrid Wannier bands for the occupied states, each of them displaying $-2\pi$ winding of $\gamma^{m}_{\vec{G}_{2}}(k_{x})$ as $k_{x}$ goes from $0$ to $2\pi$, such that the Chern number for each of the CI layers is $C = -1$. 
From the second column, we can deduce that the non-trivial states are localized at $z \approx -10$, $-5$, $0$, $+5$, $+10$. 
From the third column, where we plot $|c^{n}_{m,\Gamma}|^{2}$, we can deduce that the non-trivial Bloch band indices at $\vec{k} = \Gamma$ are $ n =18,\ldots, 22$ (the first band is labelled by $n=0$).}
  \label{hwf_winding_decomposition_orbitals_bloch_Q_0.4pi_Lz_25_D_0.75_phi_0}
\end{figure}

The first step of our analysis is to compute the hybrid Wannier bands for the occupied states in the $xy$-slab. 
This involves diagonalizing the position operator $z$ projected into the space of occupied bands at each $\vec{k}$-point of the 2D Brillouin zone (BZ)\cite{PythTB,Marzari2012}. 
Notice that, as we have finite size along $z$, the position operator $z$ is well-defined. 
We thus obtain hybrid Wannier bands, which are the eigenstates of the projected $z$ operator as a function of $k_x$ and $k_y$. 
In our calculation we find that for $q = 1/5$, $t_{x}=-t_{y}=t_{z}=1$, $m = 2$, $2|\Delta|=0.75$, $\phi = 0$ and $L_{z} = 25$ the hybrid Wannier bands are non-degenerate.
This allows us to compute the Berry phase integrated along $k_{y}$ as a function of $k_{x}$ for each hybrid Wannier band, as shown in the first column of Fig.~\ref{hwf_winding_decomposition_orbitals_bloch_Q_0.4pi_Lz_25_D_0.75_phi_0}. 
We denote such Berry phases as $\gamma^{m}_{\vec{G}_{2}}(k_{x})$, where $m$ indexes the hybrid Wannier band, $\vec{G}_{2}$ indicates that we are integrating along $k_{y}$, and $k_{x}$ denotes the functional dependence on $k_{x}$. 
Due to the non-degeneracy of the hybrid Wannier bands, the Berry phases $\gamma^{m}_{\vec{G}_{2}}(k_{x})$ coincide with the eigenvalues of the $\hat{y}$-directed non-Abelian slab Berry phase (Wilson loop) $\text{arg}(\mathcal{W})$\cite{wieder2018axion,Wilson_2,Yu11,Alexandradinata16}. 
From the negative winding of $\gamma^{m}_{\vec{G}_{2}}(k_{x})$ shown in the first column of Fig.~\ref{hwf_winding_decomposition_orbitals_bloch_Q_0.4pi_Lz_25_D_0.75_phi_0}, we find that each of the $m = 2$, $7$, $12$, $17$ and $22$ hybrid Wannier bands (the first hybrid Wannier band is labelled by $m=0$) carries Chern number $C = -1$. 
Importantly, at $\Gamma$ point, the hybrid Wannier functions for these 5 hybrid Wannier bands are localized at $z_\Gamma \approx -10$, $-5$, $0$, $5$ and $10$, see the second column of Fig.~\ref{hwf_winding_decomposition_orbitals_bloch_Q_0.4pi_Lz_25_D_0.75_phi_0} and the second column of Table~\ref{tab:hwfc_table}. 
This is consistent with our deduction in the main text that the CIs will be separated by $1/q$ along $z$, which is $5$ lattice constants in this case. 
This also supports our identification that this system can be viewed as layered stack of CIs.
The discrepancy between $z_\Gamma$ and $\langle z\rangle$ (reported in the third column in Table~\ref{tab:hwfc_table}), the average of the hybrid Wannier center over all momenta, can be attributed to the fact that our present model has nonzero coupling along the $z$-direction, and hence nonzero coupling between CI layers. 

\begin{figure}[ht]
  \includegraphics[scale=0.4]{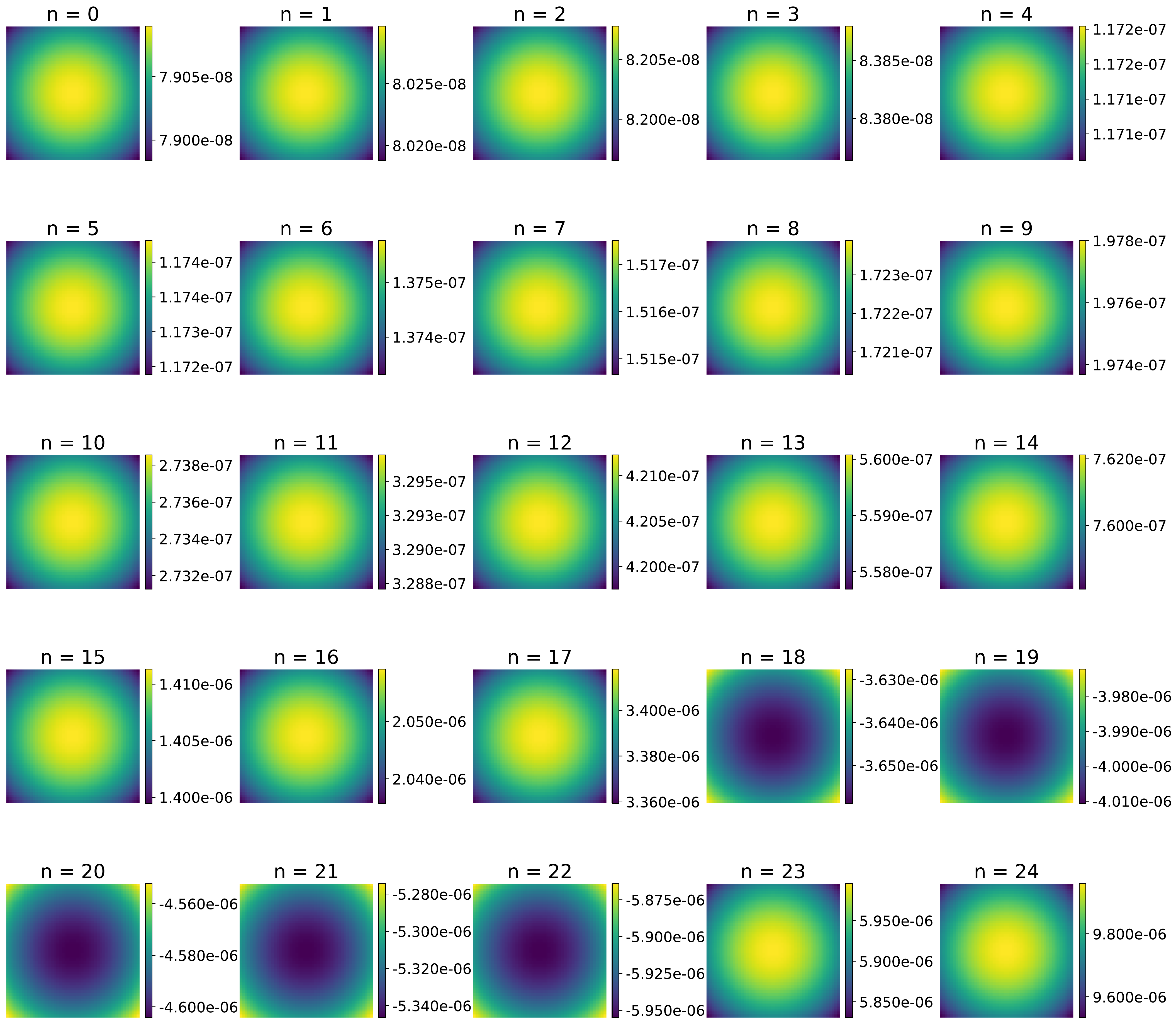}
  \caption{The Berry curvature around $\vec{k} = \Gamma$ for the 25 valance Bloch bands (not hybrid Wannier bands) of the Weyl-CDW model for an $xy$-slab with $t_{x}=-t_{y}=t_{z}=1$, $m =2$, $2|\Delta|=0.75$, $q = 1/5$, $L_{z} =25$ and $\phi = 0$.
   All the figures have $k_{x}$ and $k_{y}$ in the range range $2\pi [-0.003,0.003]$.
    We notice that the Bloch bands with indices $n = 18,\ldots, 22$ have negative Berry curvature around $\vec{k} = \Gamma$ contributing to the total Chern number $C = -5$ of this $xy$-slab, leading to $G_{xy} = -5 e^{2}/h$.}
  \label{Berry_curvature_around_Gamma_Q_0.4pi_Lz_25_D_0.75_phi_0}
\end{figure}

For those hybrid Wannier bands with non-zero winding of the Berry phase, we compute the decomposition of hybrid Wannier functions in terms of Bloch states at $\vec{k} = \Gamma$: 
\begin{align}
	\psi^{\text{hybrid}}_{m,\Gamma} = \sum_{n=1}^{N_{\text{occ}}} c_{m,\Gamma}^{n}  \psi^{\text{Bloch}}_{n,\Gamma}, \label{eq:wvfn_decomposition}
\end{align}
where $N_{\text{occ}}$ is the number of occupied bands. 
The values of $n$ for those non-zero $ c_{m,\Gamma}^{n} $ then directly tell us the index of the topologically non-trivial Bloch bands (not hybrid Wannier bands) at $\vec{k} = \Gamma$. 
Hereafter, we will use $m$ to denote hybrid Wannier band indices and $n$ to denote Bloch band indices.
In our specific examples, the decomposition of $\psi^{\text{hybrid}}_{m,\Gamma}$ in the Bloch basis where $m$ is the index of non-trivial hybrid Wannier bands are given in the third column of Fig.~\ref{hwf_winding_decomposition_orbitals_bloch_Q_0.4pi_Lz_25_D_0.75_phi_0}. 
We can then identify that the Bloch band with index $n = 18,\ldots, 22$ at $\vec{k} = \Gamma$ are non-trivial. 
Importantly, we notice that for the hybrid Wannier bands at $\Gamma$ with $m \ne 2$, $7$, $12$, $17$ and $22$, the decomposition coefficient $c^{n}_{m,\Gamma}$ are zero for $n = 18,\ldots, 22$, showing that the hybrid Wannier bands at $\Gamma$ with $m = 2$, $7$, $12$, $17$ and $22$ are truly spanned only by Bloch bands with $n = 18,\ldots, 22$. 
These states contribute to the chiral QAH surface states. 
Although the band structure of $xy$-slab is complicated with various band entanglements, we can track the topologically non-trivial bands by starting from the non-trivial bands at $\vec{k} = \Gamma$. 
Whenever we encounter band crossing, we choose to proceed along the direction where there is no discontinuity of the bands. 
For example, we can identify the non-trivial valence bands at $\vec{k} = \Gamma$ and $\vec{k} = (0,\pi) = Y $, as shown in Fig.~\ref{slab_band_and_wvfn_Q_0.4pi_Lz_25_D_0.75_phi_0_and_pi} marked by orange and green respectively. 
We also plot the  probability distribution of these non-trivial states in Fig.~\ref{slab_band_and_wvfn_Q_0.4pi_Lz_25_D_0.75_phi_0_and_pi}. 
The probability distribution along $z$ for these non-trivial states are exactly the same as the QAH zero modes for the corresponding $y$-rod, which we have shown in the main text, and the hybrid Wannier function at $\Gamma$ for $m = 2$, $7$, $12$, $17$ and $22$, shown in the middle column of Fig.~\ref{hwf_winding_decomposition_orbitals_bloch_Q_0.4pi_Lz_25_D_0.75_phi_0}.
Furthermore, we have computed the Berry curvature of each Bloch band (not hybrid Wannier band) around the $\Gamma$ point of the $xy$-slab in Fig.~\ref{Berry_curvature_around_Gamma_Q_0.4pi_Lz_25_D_0.75_phi_0}. 
Those non-trivial bands with band index $n = 18,\ldots, 22$ show negative values of Berry curvature around $\Gamma$, which is distinct from the other Bloch bands having positive values of Berry curvature around $\Gamma$.
This negative Berry curvature contributes to the total Chern number $C = -5$ of occupied bands. 
This is consistent with $G_{xy}(\phi = 0) = -5 e^{2}/h$, as we stated in the main text.

\begin{figure}[ht]
  \includegraphics[width=\columnwidth]{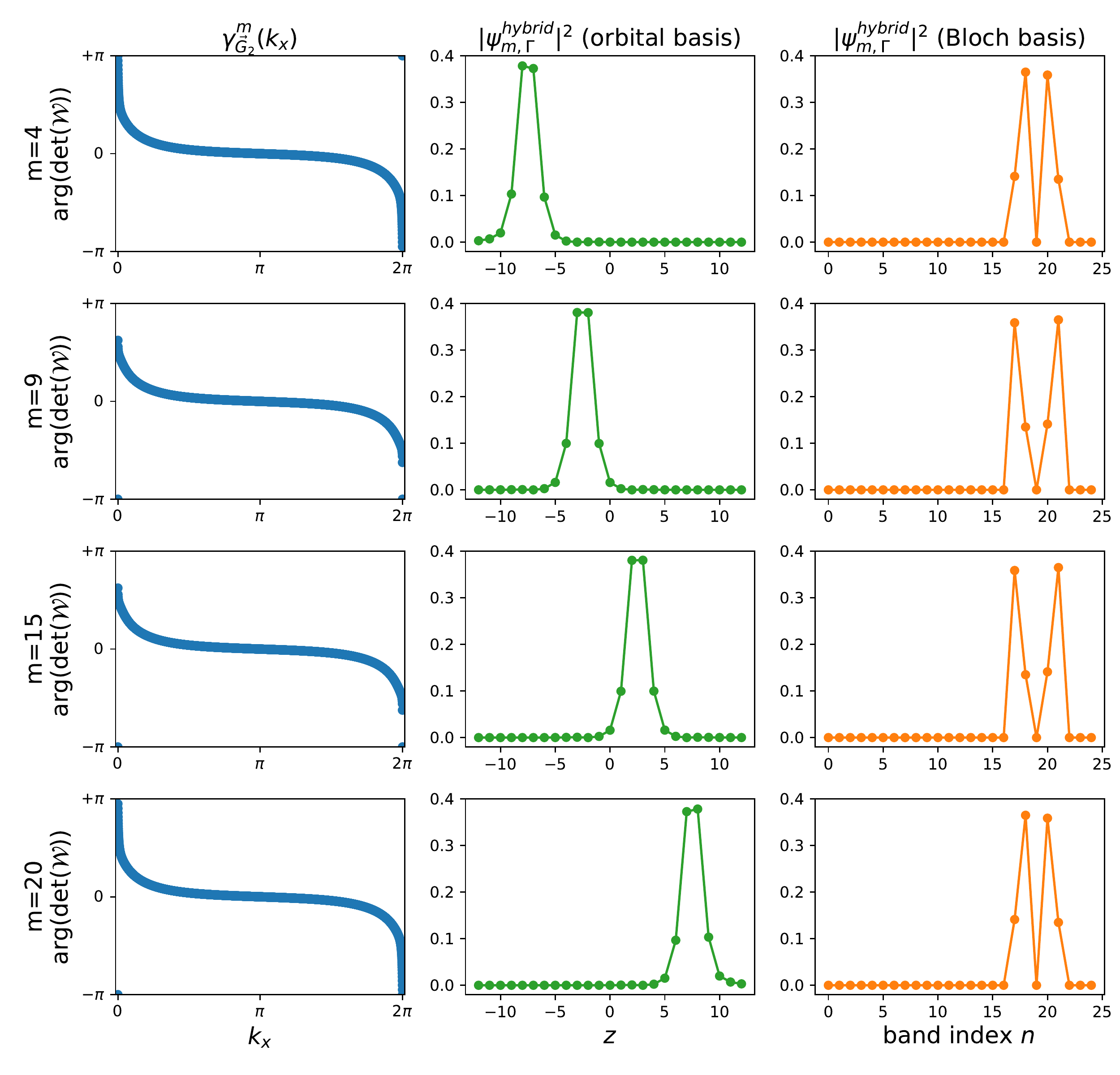}
  \caption{Berry phase winding of non-trivial hybrid Wannier bands (first column), hybrid Wannier functions at $\vec{k} = \Gamma$ decomposed in terms of orbital (second column), and Bloch basis (third column) for an $xy$-slab of the Weyl-CDW model with $t_{x}=-t_{y}=t_{z}=1$, $m =2$, $2|\Delta|=0.75$, $q = 1/5$, $L_{z} =25$ and $\phi = \pi$.
The $z$-coordinate ranges from $-12, \ldots, +12$. 
There are 4 non-trivial hybrid Wannier bands for the occupied states, each of them displaying $-2\pi$ winding of $\gamma^{m}_{\vec{G}_{2}}(k_{x})$ as $k_{x}$ goes from $0$ to $2\pi$, such that the Chern number for each of the CI layers is $C = -1$. 
From the second column we can deduce that the non-trivial states are localized at $z \approx -7.5$, $-2.5$, $+2.5$, $+7.5$. 
From the third column we can deduce that the non-trivial Bloch band indices at $\vec{k} = \Gamma$ are $n = 17$, $18$, $20$ and $21$ (the first band is labelled by $n=0$).}
  \label{hwf_winding_decomposition_orbitals_bloch_Q_0.4pi_Lz_25_D_0.75_phi_pi}
\end{figure}

\begin{figure}[ht]
  \includegraphics[scale=0.4]{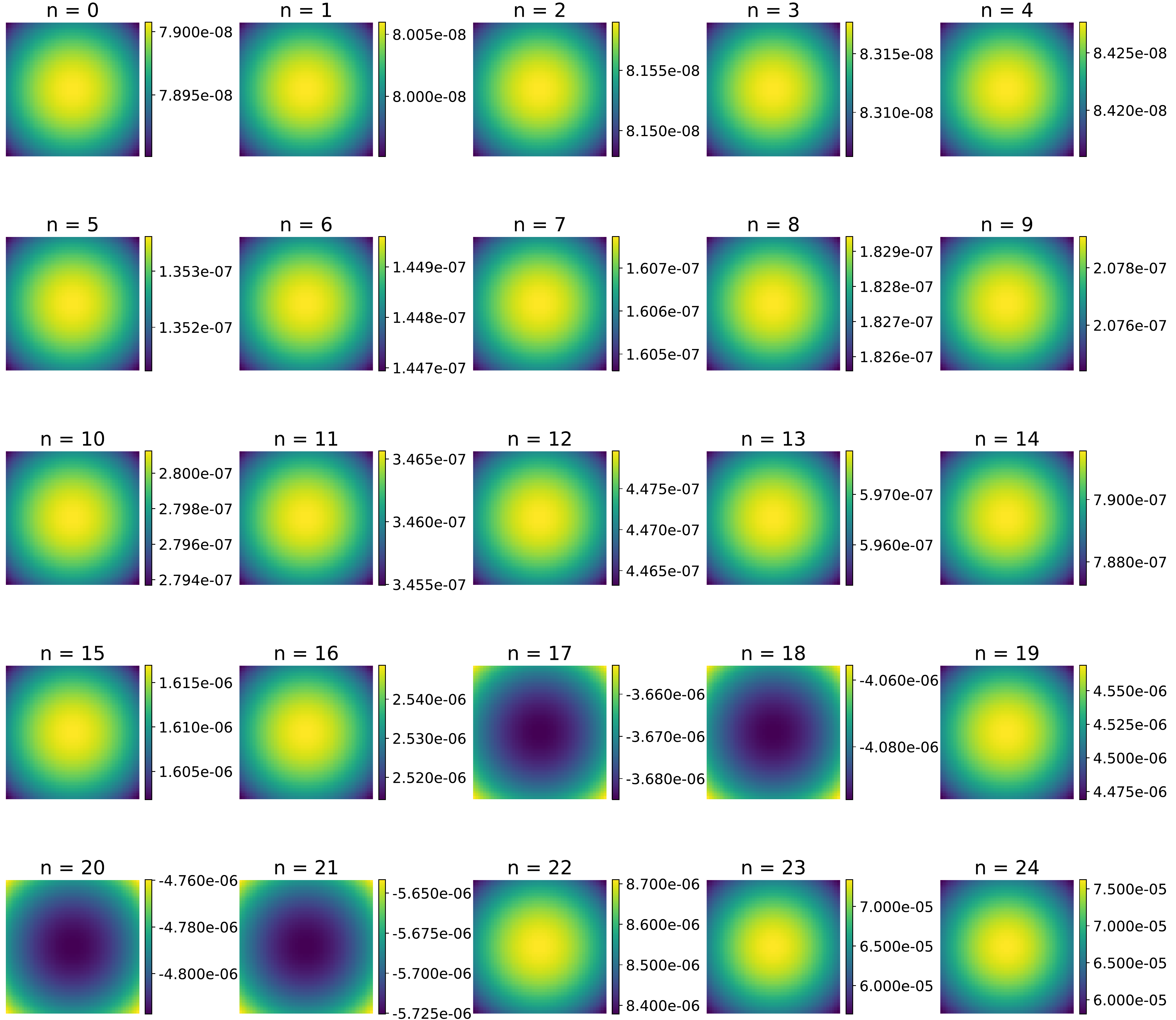}
  \caption{The Berry curvature around $\vec{k} = \Gamma$ for the 25 valance Bloch bands (not hybrid Wannier bands) for the $xy$-slab of the Weyl-CDW with $t_{x}=-t_{y}=t_{z}=1$, $m =2$, $2|\Delta|=0.75$, $q = 1/5$, $L_{z} =25$ and $\phi = \pi$. 
  All the figures have $k_{x}$ and $k_{y}$ in range $2\pi [-0.003,0.003]$. 
  We notice that the Bloch bands with indices $n=17$, $18$, $20$ and $21$ have negative Berry curvature around $\vec{k} = \Gamma$ contributing to the total Chern number $C = -4$ of this $xy$-slab, leading to $G_{xy} = -4 e^{2}/h$.}
  \label{Berry_curvature_around_Gamma_Q_0.4pi_Lz_25_D_0.75_phi_pi}
\end{figure}

\begin{figure}[ht]
  \includegraphics[scale=0.6]{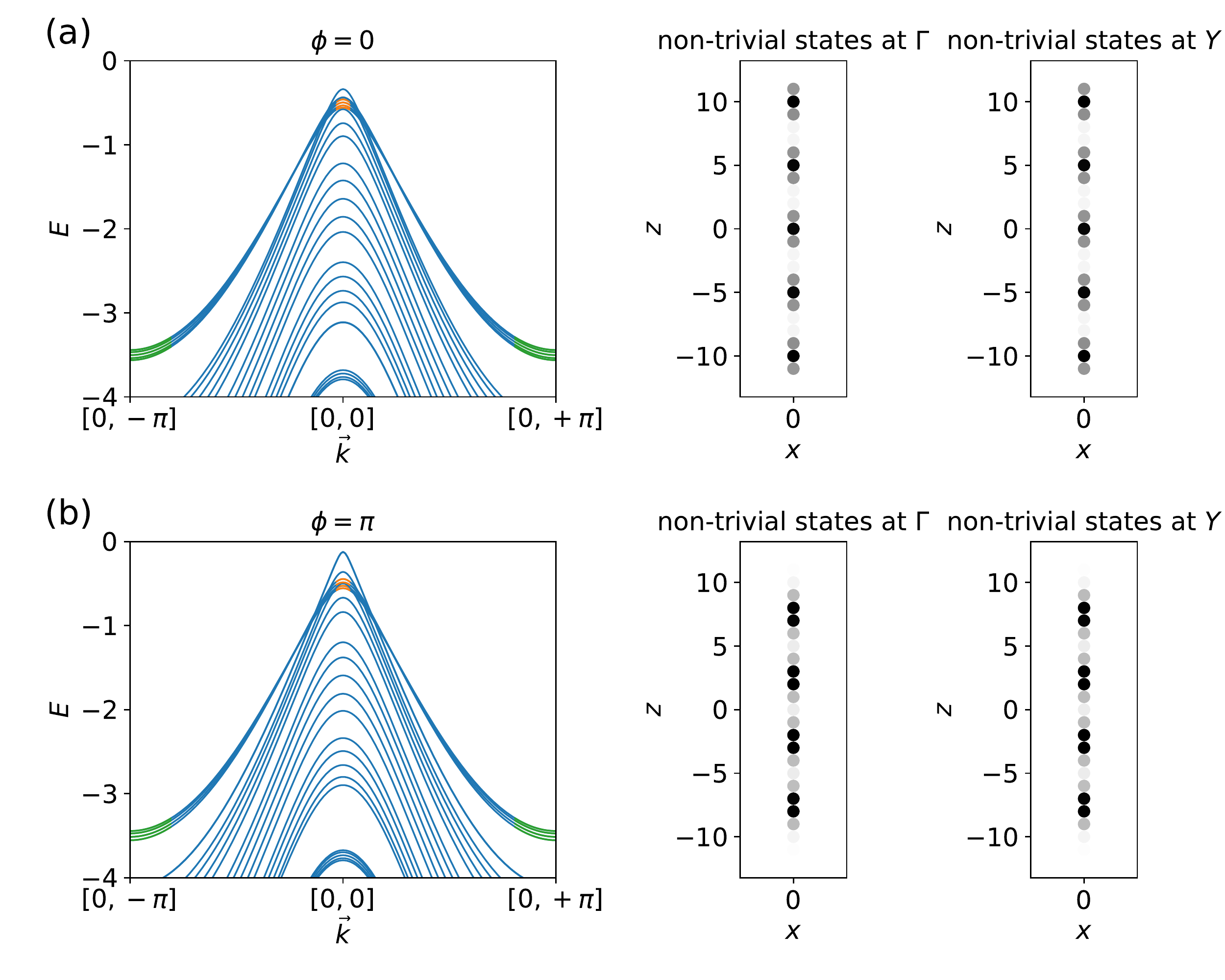}
  \caption{The $xy$-slab valence band structure of the Weyl-CDW model plotted along the path $-Y \to \Gamma \to Y$ with $t_{x}=-t_{y}=t_{z}=1$, $m =2$, $2|\Delta|=0.75$, $q = 1/5$, $L_{z} =25$, (a) $\phi = 0$ and (b) $\phi = \pi$.
   The (a) 5 and (b) 4 non-trivial bands around $\vec{k}=\Gamma$ and $Y$ are marked by orange and green. 
   The summation of the probability distribution for the (a) 5 and (b) 4 non-trivial states at $\vec{k}=\Gamma$ and $Y$ are also plotted on the right of each $xy$-slab band structure for (a) $\phi = 0$ and (b) $\pi$. 
 The $z$-coordinate ranges from $-12,\ldots, +12$. 
 As the non-trivial states at $\vec{k} = \Gamma$ and $\vec{k} =Y$ have exactly same probability distribution along $z$, they lie in the same Landau level index $n$ (see Sec.~VI of the main text) subspace. 
 This also confirms that our identification of non-trivial bands is consistent.}
  \label{slab_band_and_wvfn_Q_0.4pi_Lz_25_D_0.75_phi_0_and_pi}
\end{figure}

The above process can be also applied to the case with $\phi = \pi$. 
We also identify the non-trivial bands for the $xy$-slab, and plot their probability distribution at $\vec{k} = \Gamma$ and $\vec{k} = Y$ in Fig.~\ref{slab_band_and_wvfn_Q_0.4pi_Lz_25_D_0.75_phi_0_and_pi}. 
Again, the probability distribution along $z$ for these non-trivial states are exactly the same as the QAH zero modes for the corresponding $y$-rod, which we have shown in the main text, and the hybrid Wannier functions at $\Gamma$ for $m = 4$, $9$, $15$ and $20$, shown in the second column of Fig.~\ref{hwf_winding_decomposition_orbitals_bloch_Q_0.4pi_Lz_25_D_0.75_phi_pi}. 
The similar analysis of Berry phase for hybrid Wannier bands, wave function decomposition of Eq.~(\ref{eq:wvfn_decomposition}) and Berry curvature around $\vec{k} = \Gamma$ are also shown in Fig.~\ref{hwf_winding_decomposition_orbitals_bloch_Q_0.4pi_Lz_25_D_0.75_phi_pi} and Fig.~\ref{Berry_curvature_around_Gamma_Q_0.4pi_Lz_25_D_0.75_phi_pi}. 
Importantly, we can see that in the second column of Fig.~\ref{hwf_winding_decomposition_orbitals_bloch_Q_0.4pi_Lz_25_D_0.75_phi_pi} and the fifth column of Table~\ref{tab:hwfc_table}, the hybrid Wannier functions at $\Gamma$ corresponding to the non-trivial bands are localized at $z_\Gamma \approx -7.5$, $-2.5$, $2.5$ and $7.5$. 
As above, we attribute the discrepancy between $z_\Gamma$ and $\langle z \rangle$ (the average of the hybrid Wannier center over the Brillouin zone, reported in column six of Table~\ref{tab:hwfc_table}) to the nonvanishing interlayer coupling in our model. 
This is again consistent with our stacked layer identification. 
As mentioned in the main text, as we vary $\phi$ by $\pi$, each of the CIs will be shifted by $-\Delta \phi / (2\pi q) = -\pi/(2\pi q) = -1/(2q)$, which in our current example is equal to $-2.5$ lattice constants. 
Comparing this with the second column of Fig.~\ref{hwf_winding_decomposition_orbitals_bloch_Q_0.4pi_Lz_25_D_0.75_phi_0}, which shows that the non-trivial CIs are at $z \approx -10$, $-5$, $0$, $5$ and $10$, we can see that the stacked layers in Fig.~\ref{hwf_winding_decomposition_orbitals_bloch_Q_0.4pi_Lz_25_D_0.75_phi_pi} can be identified as those in Fig.~\ref{hwf_winding_decomposition_orbitals_bloch_Q_0.4pi_Lz_25_D_0.75_phi_0} with a shift of $-2.5$ lattice constant along $z$. 
We notice that in this case, where we have used $q = 1/5$, $t_{x}=-t_{y}=t_{z}=1$, $m = 2$, $2|\Delta|=0.75$, $\phi = \pi$ and $L_{z} = 25$, the hybrid Wannier bands are nearly degenerate at $\Gamma$ for $m = 9$, $10$ and $m = 14$, $15$, which result in a dramatic (though still continuous) change of $\gamma^{m}_{\vec{G}_{2}}(k_{x})$ for $m = 9$ and $15$ around $k_{x} = 0$, as shown in the first column of Fig.~\ref{hwf_winding_decomposition_orbitals_bloch_Q_0.4pi_Lz_25_D_0.75_phi_pi}. 
Nevertheless, we are able to separate the hybrid Wannier bands.

We conclude this section by making several remarks. 
First, since the above process does not involve computing the band structure in a 3D BZ, there is no problem in generalizing the above method to 3D Weyl-CDW systems with incommensurate density waves with modulation wave vector along $z$. 
We have carried out the same numerical analysis for $q = \tau / 4$ (where $\tau = (1+\sqrt{5})/2$ is the golden ratio).
The results are shown in the main text in which we can identify non-trivial states corresponding to CI layers and these states have exactly the same probability distribution along $z$ as the QAH zero modes in the $y$-rod at $k_{y} = 0$. 
We shall in here briefly summarize some numerical results for $q = \tau / 4$, $t_{x} = -t_{y} = t_{z} = 1$, $m=2$, $2|\Delta|=2$: 
(1) The $xy$-slab Hall conductances with $L_{z} = 21$ are $G_{xy}(\phi = 0)=-9 e^{2}/h$ and $G_{xy}(\phi = \pi)=-8 e^{2}/h$\cite{dynamical_axion_insulator_BB}. 
(2) There are 9 and 8 non-trivial hybrid Wannier bands in the $xy$-slab for $\phi = 0$ and $\pi$, respectively. 
Each of these non-trivial hybrid Wannier bands carries Chern number $C = -1$. 
(3) Similar to Fig.~\ref{hwf_winding_decomposition_orbitals_bloch_Q_0.4pi_Lz_25_D_0.75_phi_0} and Fig.~\ref{hwf_winding_decomposition_orbitals_bloch_Q_0.4pi_Lz_25_D_0.75_phi_pi}, we have identified 9 and 8 non-trivial Bloch bands of the $xy$-slab using non-zero values of $c^{n}_{m,\Gamma}$ in Eq.~(\ref{eq:wvfn_decomposition}). 
The Berry curvature of these 9 and 8 non-trivial Bloch bands around $\Gamma$ are negative, which is distinct from other Bloch bands having positive values of Berry curvature around $\Gamma$, and contribute to $G_{xy}(\phi = 0)=-9 e^{2}/h$ and $G_{xy}(\phi = \pi)=-8 e^{2}/h$. 
This is similar to the cases of $q = 1/5$ in Fig.~\ref{Berry_curvature_around_Gamma_Q_0.4pi_Lz_25_D_0.75_phi_0} and Fig.~\ref{Berry_curvature_around_Gamma_Q_0.4pi_Lz_25_D_0.75_phi_pi}. 
Secondly, the above procedure to filter out non-trivial CI layers might break down when we have strong interlayer coupling, causing complicated hybridization between layer states. 
We emphasize again that the hybrid Wannier center $\left\langle z \right\rangle$ averaged over the 2D Brillouin zone for a fixed non-trivial hybrid Wannier band can differ slightly from the hybrid Wannier center $z_{\Gamma}$ at $\vec{k} = (k_{x},k_{y})=(0,0)=\Gamma$ where our low energy theory works. 
For example, as shown in Table~\ref{tab:hwfc_table}, for $\phi = 0$, the hybrid Wannier band with index $m = 2$ has $z_{\Gamma} = -9.9245$ while $\left\langle z\right\rangle = -9.9997 \pm 0.0026$. 
This can be viewed as the consequence of hybridization between layers. 
We expect that if the layers are completely decoupled from each other, there will be no difference between $z_{\Gamma}$ and $\left\langle z \right\rangle$. 
Notice that for $\phi = 0$ the inversion center of the $xy$-slab is at $z = 0$, which pins both $z_{\Gamma}$ and $\left\langle z \right\rangle$ to $z = 0$ for the $m=12$ hybrid Wannier band. 
This is consistent with Refs.~\onlinecite{song2017,MTQC}, where it was emphasized that the distinction between QAH and oQAH states is due to the presence or absence of a non-trivial CI layer exactly at the inversion center $\left\langle z \right\rangle = 0$, respectively. 
We have also verified this for the incommensurate case of $q=\tau /4$ where $\tau = (1+\sqrt{5})/2$, which indicates that the difference of a non-trivial CI layer at the inversion center $\left\langle z \right\rangle=0$ between QAH and oQAH states holds for generic CDW wave vectors. 
Our method can be viewed as a way to map a system with interlayer coupling to another system with decoupled CI layers.

\begin{table}[h]
\centering
\begin{tabular}{|c|c|c||c|c|c|}
\hline
$m$ for $\phi = 0$ & $z_{\Gamma}$ & $\left\langle z\right\rangle$ & $m$ for $\phi = \pi$ & $z_{\Gamma}$ & $\left\langle z\right\rangle$ \\
\hline
2 & $-9.9245$ & $-9.9997 \pm 0.0026$ & 4 & $-7.5467$ & $-7.9963 \pm 0.0167$\\
\hline
7 & $-4.9972$ & $-5.0 \pm 0.0001$ & 9 & $-2.5023$ & $-2.9963 \pm 0.0176$ \\
\hline
12 & $0.0$ & $0.0 \pm 0.0$ & 15 & $2.5023$ & $2.9963 \pm 0.0176$ \\
\hline
17 & $4.9972$ & $5.0 \pm 0.0001$ & 20 & $7.5467$ & $7.9963 \pm 0.0167$ \\
\hline
22 & $9.9245$ & $9.9997 \pm 0.0026$ &  &  &  \\
\hline
\end{tabular}
\caption{Hybrid Wannier centers along $z$ for different hybrid Wannier bands with non-zero winding of $\gamma^{m}_{\vec{G}_{2}}(k_{x})$ and $q = 1/5$.
$m$ denotes the non-trivial hybrid Wannier band index. 
The first and last three columns summarize the results for $\phi = 0$ and $\phi = \pi$, respectively. 
There are $5$ and $4$ non-trivial hybrid Wannier bands for $\phi = 0$ and $\pi$ with $m = 2$, $7$, $12$, $17$, $22$ and $m = 4$, $9$, $15$, $20$, respectively.
$z_{\Gamma}$ is the hybrid Wannier center along $z$ for the hybrid Wannier functions at $\vec{k} = (k_{x},k_{y}) = (0,0) = \Gamma$.
$\left\langle z \right\rangle$ is the hybrid center along $z$ averaged over the 2D Brillouin zone with grid size $100 \times 100$. 
We also report the standard deviation of $\left\langle z \right\rangle$ in the same column.}
\label{tab:hwfc_table}
\end{table}

\clearpage

\bibliography{refs}